\RequirePackage{fix-cm}
\documentclass[smallextended]{svjour3}       
\usepackage{hyperref}
\usepackage{bbm}
\usepackage[utf8]{inputenc}
\usepackage{graphicx}
\usepackage[usenames]{color}
\usepackage[table]{xcolor}
\usepackage{soul}
\usepackage{textcomp}
\usepackage{multicol}
\usepackage{multirow}
\usepackage{xspace}
\usepackage{mathtools}
\usepackage[T1]{fontenc}
\usepackage{cancel}
\usepackage{amsmath} 
\usepackage{amssymb}

\usepackage{algorithm}
\usepackage{float}
\usepackage{graphicx} 
\usepackage{graphics}
\usepackage[noend]{algpseudocode}
\usepackage{algorithmicx}
\usepackage{amsmath}
\usepackage{amssymb}
\usepackage{mathtools}
\usepackage{color}

\algrenewcommand\algorithmicrequire{\textbf{Input:}}
\algrenewcommand\algorithmicensure{\textbf{Output:}}
\algrenewcommand{\algorithmiccomment}[1]{\hfill$\{$\textit{#1}$\}$}
\newcommand{\lines}[2]{(lines #1-#2)}
\newcommand{\sline}[1]{(line #1)}

 \newtheorem{fact}{Fact} 

\def\G{\mbox{$\mathcal{G}$}} %
\def\V{\mbox{$\mathcal{V}$}} %
\def\E{\mbox{$\mathcal{E}$}} %

\def\A{\mbox{$\mathcal{A}$}}  
\def\D{\mbox{$\mathcal{D}$}}  

\newcommand{\argmax}{\operatornamewithlimits{argmax}}
\newcommand{\argmin}{\operatornamewithlimits{argmin}}

\newcommand{\olf}[1]{C(#1)} 
\newcommand{\ActiveSet}[1]{\mu(#1)} 
\def\TargetSet{TS} 
\def\symbOF{aDC} 

 \newcommand{\myalgo}{\textsf{ADITUM}\xspace}  
 \newcommand{\algo}[1]{\textsf{#1}}
 
\begin{document}

\title{Monotone Submodular Diversity functions for Categorical Vectors with Application to Diversification of Seeds for Targeted Influence Maximization  }
\titlerunning{Categorical Diversification of Seeds for Targeted Influence Maximization}

\author{Antonio Caliò \and Andrea Tagarelli} 

\institute{A. Caliò \at Dept. Computer Engineering, Modeling, Electronics, and Systems Engineering, University of Calabria, 87036 Rende (CS), Italy. 
\email{a.calio@dimes.unical.it}
\and 
A. Tagarelli \at Dept. Computer Engineering, Modeling, Electronics, and Systems Engineering, University of Calabria, 87036 Rende (CS), Italy. 
\email{andrea.tagarelli@unical.it}
}

\date{Initially conceived: October 2018. First article-version: February 1, 2019. Last update: September 11, 2019}

\maketitle

\begin{abstract}
Embedding diversity into knowledge discovery tasks is of crucial importance to enhance the meaningfulness of the mined patterns with high-impact aspects related to novelty, serendipity, and  ethics.   
 Surprisingly, in the classic problem of  influence maximization   in social networks,  relatively little study has been devoted to diversity  and its integration into  the objective function of an influence maximization  method. 
 
 In this work, we propose the integration of a side-informa\-tion-based notion of seed diversity   into the objective function of a targeted influence maximization problem. Starting from the assumption that side-information is available at node level in the general form of categorical attribute values, 
 we design a class of monotone   submodular functions specifically conceived for  determining the diversity within  a set of  categorical profiles associated with  the seeds to be discovered.   This allows us to develop an efficient scalable approximate  method, with a constant-factor guarantee of optimality. More precisely, 
 we formulate the \textit{attribute-based diversity-sensitive targeted  influence maximization} problem   under the state-of-the-art reverse influence sampling framework,  and we develop a method, dubbed \myalgo, that  ensures a  $(1-\frac{1}{e}-\epsilon)$-approximate solution under the general triggering diffusion model.  
 We experimentally evaluated \myalgo on five  real-world networks, including comparison with  methods that exploit numerical-attribute-based diversity and topology-driven diversity in influence maximization.  
\end{abstract}

%

%
%

%
%
%

\section{Introduction}

Online social networks (OSNs) have become 
the most profitable channel for ``viral marketing'' purposes. 
 In this regard, a classic optimization problem in OSNs is   \textit{influence maximization} (IM), i.e.,   to discover a set of initial influencers, or \textit{seeds}, that can maximize the spread of information through the network~\cite{Kempe:2003,Chenbook}.  
 In most practical scenarios, companies want to tailor their advertisement strategies in order to address only selected OSN-users as potential customers. 
This is the perspective adopted in the context
of \textit{targeted IM} (e.g.,~\cite{kbtim,Lu:2015:CCC:2850578.2850581,M.thai1,M.thai2,8326536}), 
 which is also the focus of this work.   


Besides trying to maximize the spread of information (e.g., advertising of a product), which is directly related to an a-priori specified budget, i.e., the number of seeds, a further yet less explicit issue in (targeted) IM is in the attempt of maximizing the ``potential'' of the selected seeds to influence, or engage, the users  in the network. We believe that  such a  kind of potential can be well-explained in terms of \textit{diversity} that may characterize the  seeds.   %
   Intuitively, influencers that have different  ``features'' (e.g.,   age, gender, socio-cultural aspects, preferences)    bring unique opinions, experiences, and perspectives to bear on the influence propagation process. 
   As a consequence, \textit{seed users that have more different characteristics are more likely to maximize their strategies to  engage the target users}. 
   Also, from a different view, 
   favoring diversity has important ethical implications in  choosing the seeds as well as the target users. 

Surprisingly, despite diversity has been recognized as a key-enabling dimension in  data science (e.g., to improve user satisfaction in content recommendation based on novelty and serendipity), 
 relatively few studies have considered diversity in the context of (targeted) IM problems.
  One of the earliest attempt is provided by Bao et al.~\cite{BaoCZ13}, which extends the Independent  Cascade model to account for the structural diversity of nodes' neighborhood, however without addressing an optimization problem. 
   Other works have studied relations between diversity and    spreading ability, but focusing on  a single node  in a network~\cite{HuangLCC13}.  
    Node diversity into the IM task has been first introduced by Tang et al.~\cite{6921625}.  They 
    consider numerical attributes reflecting user's preferences on some predefined categories (e.g., movie genres) to address a generic IM task.  
    In~\cite{8326536}, we originally define  an IM problem that is both targeted and diversity-sensitive, which  
     however,     only considers specific notions of diversity that are driven by the topology of the information diffusion graph.

{\bf Contributions.\ } 
We aim to advance research on IM by introducing a targeted IM problem that accounts for side-information-based  diversity of the seeds to be identified.  Our contributions are summarized as follows.
\vspace{-1mm}
\begin{itemize}
\item
We propose \ the {\em  {\bf A}ttribute-based {\bf DI}versity-sensitive}  \  {\em{\bf T}argeted Infl{\bf{U}}ence {\bf M}axi\-mization} problem, dubbed  \myalgo.\footnote{Latin term for \textit{access}, \textit{admission}, \textit{audience}.}    Our notion of diversity  assumes that nodes in the network are   associated with side-information in the form of a schema of categorical attributes and corresponding   values. 

\item
We   provide different definitions of diversity that are able to   reflect  the variety in the amount and type of categorical values that characterize the seeds being discovered. Remarkably, we design a class of nondecreasing monotone and submodular functions for categorical diversity, which also has the nice property of allowing incremental computation of a node's marginal gain when added to  the current seed set.  
To the best of our knowledge, we are the first to propose a formal systematization of approaches and functions for determining   node-set diversity in   influence propagation and related problems in information networks. 
    
\item
 We design our solution to the \myalgo problem under the Reverse Influence Sampling (RIS) paradigm~\cite{doi:10.1137/1.9781611973402.70,Tang:2014:IMN:2588555.2593670} and recognized as the  state-of-the-art approach for IM problems. 
 One challenge that we address   is   revisiting   the RIS framework     to deal with  both the targeted nature and the diversity-awareness of the \myalgo problem. 
 
\item
  We develop the \myalgo algorithm, which
 returns a $(1-1/e-\epsilon)$-approxima\-tion with at least $1-1/n^l$ probability in $O((k+l)(|\E|+|\V|) \log |\V| /\epsilon ^2)$ time, under the triggering model, a general diffusion model adopted by most existing work.  

\item
We experimentally evaluated \myalgo on publicly available network data\-sets, three of which were used in a user engagement context, one in community interaction,   and the other one in recommendation. 
This choice was mainly motivated by the opportunity of comparing our \myalgo with the aforementioned   methods in~\cite{6921625} and~\cite{8326536}.
\end{itemize}

{\bf Plan of the paper.\ }
The rest of this paper is organized as follows. 
Section~\ref{sec:related} briefly discusses related work on targeted IM and diversity-aware IM. 
Section~\ref{sec:problem-statement} formalizes the information diffusion context model, the objective function, and the optimization problem under consideration. 
Section~\ref{sec:diversity} presents our study on monotone and submodular diversity functions for the categorical data modeling the profiles of nodes in a network. 
Section~\ref{sec:framework} describes our proposed approach and algorithm for the \myalgo problem. 
Sections~\ref{sec:eval} and \ref{sec:results} contain our experimental evaluation methodology and results, respectively. 
In Section~\ref{sec:conclusions}, we provide our conclusions and pointers for future research.

\section{Related work}
\label{sec:related}

The foundations of   IM as an optimization problem, initially posed by Kempe et al. in their seminal work~\cite{Kempe:2003}, 
 rely on two main findings, namely the intractability of the problem in its two sources of complexity (i.e., given the budget $k$ and a diffusion model, to discover a size-$k$ seed set that maximizes the expected spread, and to estimate   the expected spread of the final activated node-set) and the possibility of designing an approximate greedy solution with theoretical guarantee provided that the activation function is nondecreasing monotone and submodular. 
Upon the findings in the breakthrough work by Borgs et al.~\cite{doi:10.1137/1.9781611973402.70}, Tang. et al.~\cite{Tang:2014:IMN:2588555.2593670}     proposed a randomized algorithm, \algo{TIM}/\algo{TIM+}, that can perform orders of  magnitude faster than the greedy one, overcoming   the bottleneck in the computation of the expected spread   by exploiting a \textit{reverse sampling} technique. 
 Since then, other methods have followed, such as  IMM \cite{IMM}, BCT \cite{M.thai1}, TipTop \cite{M.thai2}. 
Also,  \cite{Lu:2015:CCC:2850578.2850581} generalizes the theoretical results 
in~\cite{doi:10.1137/1.9781611973402.70,Tang:2014:IMN:2588555.2593670} 
  to any diffusion model with an equivalent 
 live-edge model of the   diffusion graph. 
 %
In the following, we focus our discussion on targeted IM approaches,  while for broader and more complete  views on the IM topic, the interested reader can refer to recent surveys, such as \cite{lifan,annappa,Peng+18}.

{\bf Targeted influence maximization.\ } 
Research on targeted IM has also  gained attention in recent years.  
 A  query processing problem for distinguishing specific users from others is considered in~\cite{LeeChung15}. 
  In the keyword-based targeted IM  method proposed in~\cite{kbtim}, the target nodes are identified as those having preferences (i.e., keywords) in common with a certain advertisement. 
   In~\cite{Devotion}, the targeted IM problem is studied in the context of user engagement, whereby  a node is regarded as target on the basis of its social capital. 
    The RIS-based BCT method is proposed in~\cite{M.thai1}, whereby  
    each node is associated with a  cost (i.e., the effort required to engage a node as a  seed), and a   benefit  score (i.e., the profit resulting from its involvement in the  propagation).

A few studies  focus on the special case of a single selected target-node~\cite{GuoZZCG13,YangHLC13,GulerVTNZYO14}. By contrast, more general targeted IM methods, like ours,  
aim  at maximizing the probability of activating a target set of arbitrary size 
 by discovering a seed set which is neither fixed and singleton 
  nor has  constraints related to the topological closeness to a fixed initiator. 
  
Other approaches  incorporate information on the users' profiles into the diffusion process or into the influence probability estimation. 
In~\cite{LagnierDGG13}, a family of probabilistic diffusion models is proposed to exploit   vectors of features representing the content of   information to be diffused and the profile of users. 
%
In~\cite{ZhouZC14}, the independent cascade model is adapted  
to accommodate   user preferences, which are learned from a set of users' documents labeled with topics.    
  In the conformity-aware cascade model~\cite{LiBSC15}  
   the influence probability from node $u$ to node $v$ is computed based on a sentiment analysis approach and  proportionally to the product of $u$'s influence and $v$'s conformity, where the latter refers to the inclination of a node  to be influenced by others. 
 %
 User activity, sensitivity, and affinity  are considered in~\cite{Deng+16} to define node features, which are then used to adjust the influence between any two users. 

 A further perspective that can be regarded as related to targeted IM consists in   exploiting network structures to  drive the seed selection.   
In~\cite{Suman+19}, a  budget constraint on the cumulative cost of the seeds to be selected is divided among available communities, then seeds are selected inside each community based on some centrality measure.  
 %
 Community structure is also exploited in the three-phase greedy approach proposed in~\cite{PHG}. 
 Yet, coreness is used in~\cite{8031449} for estimating nodes' influence and developing  a simulated annealing based algorithm for IM. 
%
Note that the aforementioned works, 
besides \textit{discarding any  diversity notion}, are concerned with the development of  heuristics for IM  while we are interested in   designing a solution with approximation guarantee.

{\bf Diversity-aware influence maximization}.
Diversity notions have been considered in several research fields,   such as web searching, recommendation, and information spreading (e.g.,~\cite{Santos:2015:SRD:2802186.2802187,Wu:2016:RMC:2885506.2700496,BaoCZ13,HuangLCC13}).   
  However, a relatively little amount of work has been devoted to   integrating  diversity in the objective function of IM problems. 
 Tang et al.~\cite{6921625} proposed the first study on    diversity-aware IM, where a linear combination of the expected  spread function and a numerical-attribute-based diversity is  maximized by   means on heuristic search strategies, defined upon
classic centrality measures. 
%
%
%
 In~\cite{8326536}, we formulated the topology-driven diversity-sensitive targeted IM problem, dubbed \algo{DTIM}, with an emphasis on maximizing the social engagement of a given network. The provided solution, built upon the Simpath method~\cite{Simpath}, supports only  the Linear Threshold model. 
%
 It should be noted that, although the optimization problem presented in this work is similar to the one  in~\cite{8326536}, 
 here we provide different formulation and algorithmic solution  than the earlier ones, since unlike \algo{DTIM} (i) \myalgo builds  on state-of-the-art approximation methods  for IM, and (ii) it is designed to handle  different notions of attribute-based diversity. 
In Sects.~\ref{sec:eval}--\ref{sec:results} 
we present a comparative evaluation with the methods in~\cite{6921625} and~\cite{8326536}.

\section{Problem statement}
\label{sec:problem-statement}

{\bf Representation model.\ }  
Given a social network graph  $\G_0=\langle \V, \E \rangle$, with set of nodes  $\V$ and set of edges $\E$, let  $\G = \G_0(b, t) = \langle \V, \E, b, t \rangle$ be a directed weighted graph representing the \textit{information diffusion} context associated with $\G_0$, with 
$b:\E \rightarrow (0,1]$ 
 edge weighting function, 
and $t: \V \rightarrow (0,1]$ 
 node weighting function. 
       
   Function  $t$ determines  the status of each node as \textit{target}, i.e.,   a node toward which the information diffusion process is directed. Given a user-specified threshold $\tau_{TS} \in  [0,1]$, we define the \textit{target set}  $\TargetSet$ for $\G$    as:  
   $$\TargetSet  = \{ v \in \V  |  t(v)\geq  \tau_{TS}\}.$$

 \vspace{-1mm}   
 Function $b$ corresponds to the parameter of the \textit{Triggering}  model~\cite{Kempe:2003}, which in line with several existing studies on IM is also adopted here as  information diffusion model. 
 Under this model, each node chooses a random subset of its neighbors as \textit{triggers}, where the choice of triggers for a given node is independent of the choice for all other nodes. If a node $u$ is inactive at a given time and a node   in its trigger set becomes active, then $u$ becomes active at  the subsequent time. Notably, Triggering   has an equivalent interpretation as   ``reachability via live-edge paths'', such that an edge $(u, v)$ is designated as live when $v$ chooses $u$ to be in its trigger set. Therefore,  $b(u,v)$ represents the probability that edge $(u, v)$ is live. 
 Linear Threshold   and Independent Cascade~\cite{Kempe:2003} are special cases of Triggering  with particular  distributions of   trigger sets. 
 
 Note also that function $b$ and $t$  are usually defined as data-driven. We will discuss possible instances of both  functions   in Sect.~\ref{sec:settings}.  
     %

{\bf Objective function.\ }
 The objective function of our targeted IM problem is comprised  of two functions. 
The first one, denoted as $C(\cdot)$, is determined as   the cumulative amount of the scores associated with the target nodes that are activated by the seed set $S$. 
Following the terminology in~\cite{8326536}, we call this function social capital, or simply \textit{capital}, which is defined as 
\begin{equation}\label{eq:DC}
\olf{\ActiveSet{S}}=\sum \limits_{v \in  \ActiveSet{S}\cap \TargetSet } t(v) 
\end{equation}
where $\ActiveSet{S}$  denotes the set of nodes that are   active at the end of the diffusion   starting from $S$. 



The second term in our objective function, denoted as $div(\cdot)$,   is   introduced to determine   the   \textit{diversity} of the nodes in any subset of $\V$. 
%
%
%
%
As previously mentioned, our approach is to measure node diversity in terms of a-priori knowledge provided in the form of symbolic values corresponding to a predetermined set of \textit{categorical  attributes}.  In Section~\ref{sec:diversity}, we provide a class of diversity functions for categorical datasets. 

We now formally define our proposed problem of targeted IM,   {\em  {\bf A}ttribute-based {\bf DI}versity-sensitive {\bf T}argeted Infl\-{\bf{U}}ence {\bf M}aximization} (\myalgo).

\begin{definition} 
\label{def:tim} 
{\em (}\textsc{Attribute-based Diversity-sensitive  Targeted Influence Maximization}{\em )}  
Given a diffusion graph $\G = \langle \V, \E, b, t \rangle$,   a budget $k$, and a threshold  $\tau_{TS}$, find a   set $S \subseteq \V$ with $|S| \leq k$ of seed-nodes such that 
\begin{equation}\label{eq:problem}
S = \argmax_{S' \subseteq \mathcal{V} \ s.t. \ |S'| \leq k} \alpha \times \olf{\ActiveSet{S'}} + (1-\alpha)\times div(S')
\end{equation}
where $\alpha \in [0,1]$ is a smoothing parameter that controls the weight of capital $\olf{\cdot}$ w.r.t diversity $div(\cdot)$. 
\hfill~\qed
\end{definition}

The problem in Def.~\ref{def:tim} preserves the NP-hard complexity of the IM problem. 
 However, as for the classic IM problem, if we are able to design an objective function for which  
the natural \textit{diminishing property} holds, then the output of a greedy solution provides a $(1-1/e-\epsilon)$-approximation guarantee w.r.t. the optimal solution.  
 To  this aim, we need to ensure that   Eq.~(\ref{eq:problem}) is a linear combination of two monotone and submodular functions. 
  Here we point out that  monotonicity and submodularity of the capital function  $C(\cdot)$ was previously demonstrated in~\cite{8326536}. In the next section, we provide our definitions of $div(\cdot)$.

\section{Monotone and submodular diversity functions for a set of categorical tuples} 
\label{sec:diversity}
We assume that nodes in the social network graph $\G_0=\langle \V, \E \rangle$  are associated with side-information in the form of symbolic values that are valid for a predetermined set of categorical attributes, or \textit{schema}, $\A  = \{A_1, \ldots, A_m \}$. 
For each $A \in \A$, we denote with $dom_A$ its domain, i.e., the set of admissible values known for $A$, and with $dom$ the union of attribute domains. Moreover,   we define  $val_A: \V \mapsto dom_A$ 
 as a function that associates a node with   a value of $A$. 
For any $S \subseteq \V$, we will also use symbols $dom_A(S)$ and $dom(S)$ to denote the subset of values in $dom_A$, resp. $dom$, that are associated with nodes in $S$.  

Given the schema $\A$, we will refer to the categorical tuple   associated to any $v \in \V$ as the \textit{profile} of node $v$, and to   the categorical dataset   for all  nodes in $\V$ as the \textit{profile set} of $\V$. 
We will use symbol $\A[v]$ to denote the profile of $v$ and symbol $\D_S$ to denote the profile set of nodes in $S \subseteq \V$. Note that $\D_S$ is a multiset such that $\D_S = \bigcup_{v \in S} \A[v]$, and  any $\A[v]$ is generally regarded as a sparse vector,  as it could contain  \textit{missing values} for some attributes; i.e.,   
if we denote with $\bot$ a missing attribute value,   
$\A[v] = \langle val_{A_1}(v) \vee \bot, \ldots, val_{A_m}(v) \vee \bot\rangle$. 
 Moreover, we will use symbol $|\A[v]|$ to denote the actual length of $\A[v]$ as the number of attribute values contained in the profile. 

\vspace{-2mm} 
 \paragraph*{\bf General requirements. \ }
Given our setting of an information diffusion graph $\G = \G_0(b, t) = \langle \V, \E, b, t \rangle$  associated with  $\G_0$, 
 here we define a class of functions $div$ that, for any $S \subseteq \V$ with associated $\D_S$,    satisfy the following requirements:
 \begin{itemize}
 \item $div(S)$  defines a notion of   diversity of nodes in $S$ w.r.t. their categorical representation given in $\D_S$; 
 \item $div(S)$ must be   \textit{nondecreasing monotone and submodular}; hereinafter, we will use the more simple  term  ``monotone and submodular''. 
 \item for any $v \in \V \setminus S$, the marginal gain $div(S \cup \{v\}) - div(S)$ should    be computed efficiently.
 \item $div(S)$ should be     \textit{meaningful}, in terms of ability in capturing the subtleties underlying the variety of node profiles according to their categorical attributes and values.   
 \end{itemize}

\vspace{-4mm}
\subsection{Challenges in defining  set diversity functions} 
\label{sec:negativefunctions}
\vspace{-1mm}
Before providing our definitions of diversity functions in Sects.~\ref{sec:diversity:attributeprojection}--\ref{sec:diversity:class}, 
here we mention some of the negative outcomes that were drawn by an attempt of devising  apparently simple and intuitive  approaches based on \textit{attribute-wise} functions as well as based on \textit{profile-wise} functions, eventually demonstrating their unsuitability as diversity functions for the task at hand, as they do not satisfy one or more of the above listed general requirements.
 %

Let us begin with attribute-wise functions.   
Given  $A \in \mathcal{A}$ and $S \subseteq \V$, one simple approach  would be to compute the \textit{number  of unique values} admissible for   $A$ that occur   in $\D_S$, normalized by the size of $S$;  however, this coarse-grain function is not only unable  to characterize the variety of nodes in terms of repetitions of the different values of the attribute under consideration, but also it is not nondecreasing  monotone since it decreases by adding nodes with identical values of the attribute. 
 The   desired properties of monotonicity and submodularity could be satisfied by  just counting the number of unique values of attribute in $\D_S$, however at the cost of a further worsening in meaningfulness, thus obtaining  a useless notion of diversity.  
 
 An alternative approach would be to aggregate \textit{pairwise distances} of the node profiles w.r.t.  a given attribute. For instance, we could count the  (normalized) number of mismatchings over each pair of nodes in a set; however,  it is easy to prove that the derived function will not be  submodular in general. 

Let us now extend to calculating    pairwise distances  of the node profiles over the entire schema. In this regard, we could consider a widely-applied measure for computing the distance between two sequences of symbols, namely \textit{Hamming distance}. However, for different varying set-size-based normalization schemes, this might result in a function that is not submodular or even not monotone.  
Alternatively, we could consider  a standard statistic for   dissimilarity of finite sample sets, namely \textit{Jaccard distance}. This is defined as the complement of Jaccard similarity, that is, for any two sets, substracting from 1 the ratio between the size of the intersection and the size of the union of the sets. (In our context, a sample set corresponds to a categorical tuple, i.e., a node profile.) Again,  the resulting function will not ensure submodularity.  
The interested reader can refer to the \textbf{\em Appendix} for analytical details of the aforementioned  functions and relating examples that show their unsuitability as nondecreasing monotone submodular diversity functions. 

Please note that, in \textbf{\em Appendix}, we  also report    {\em Proofs} for   the main theoretical results that will be presented next in Sections~\ref{sec:diversity:attributeprojection}--\ref{sec:diversity:class}.

\vspace{-1mm}
\subsection{Attribute-wise  diversity} 
\label{sec:diversity:attributeprojection}
\vspace{-1mm}
In this section, we discuss the first of our proposed diversity functions, which is \textit{attribute-wise}.  
 We consider a notion of diversity of nodes  that builds on the variety in the amount and type of categorical values that characterize the nodes in a selected set. In particular, we consider a linear combination of the contributions the various attributes provide to the diversity of nodes in a set.  
 
 \begin{definition}
 \label{def:diversity}
 Given a set of categorical attributes $\mathcal{A} = \{A_1, \ldots, A_m \}$ and associated profile set $\D$ for the nodes in a  graph $\G_0=\langle \V, \E \rangle$, we define the  {\em attribute-wise  diversity} of  any set $S \subseteq \V$ as: 
\begin{equation}
\label{eq:diversity}
div(S) = \sum_{j=1..m} \omega_j \ div_{A_j}(S)
\end{equation}
where  $div_{A_j}(S)$ evaluates the diversity of nodes in $S$ w.r.t. attribute $A_j$, 
and $\omega$'s are   real-valued coefficients in $[0,1]$, which sum up to 1 over $j=1..m$.    
\hfill~\qed
\end{definition}

  To  meet    the monotonicity,  submodularity, meaningfulness and efficiency requirements,  we provide the following attribute-specific set  diversity   function.  
 
%
%
%
%
%

\begin{definition}
\label{def:divA}
Given a categorical attribute $A$, with domain of values $dom_A$, and node set $S \subseteq \mathcal{V}$, we define the \textit{attribute-specific set diversity} for $S$ as: 
\begin{equation}
\label{eq:divA}
div_A(S) = \sum_{a \in dom_A(S)} \sum_{i=1}^{n_a} \frac{1}{i^{\lambda}} 
\end{equation}
where $n_a$   is the number of nodes in $S$ that have value $a$ for     $A$, and $\lambda \geq 1$.  
\hfill~\qed
\end{definition}
 
One nice property of the function in Eq.~(\ref{eq:divA}) is that the contribution of a node to the set diversity, i.e., \textit{the node's marginal gain} can be    determined at constant time, thus without recomputing the set diversity from scratch.  
This holds based on the following fact.

\begin{fact} 
The marginal gain of adding a node $v$ to $S$ is equal to 
$$\sum_{j=1..m} \omega_j \sum_{a \in dom(A_j) \ \wedge \ a \in \mathcal{A}[v]}  (n_a+1)^{-\lambda},$$
 where $n_a$   is the number of nodes in $S$ that have value $a$ for     $A$, and $\lambda \geq 1$.
 \end{fact}



\begin{proposition} 
The attribute-wise   diversity function defined in Eq.~(\ref{eq:diversity}) is monotone and submodular.
\end{proposition}
%

\begin{lemma} \label{lemma:max_div_value}
Given a set $S$ and a categorical attribute $A$,   \
 consider $M_A = \max_{a \in dom_A(S)} n_a$ and 
    $m_A = \min_{a \in dom_A(S)} n_a$.   For any $$S = \argmax_{S' \subseteq \mathcal{V} \ s.t. \ |S'| \leq k} div_A(S'),$$ it holds that   $M_A - m_A <= 1.$
\end{lemma}

 We also observed that the \textit{theoretical maximum} value reach\-ed by Eq.~(\ref{eq:diversity}) depends only 
on the budget $k$, as provided by the following result.  
 
\begin{proposition} 
\label{def:diversity_max_possible_value}
 Given the set of categorical attributes $\mathcal{A} = \{A_1, \ldots, A_m \}$,  $m$-real valued coefficients $\omega_j \in [0,1]$ ($j=[1..m]$), and a budget $k$, the theoretical maximum value     for Eq.~(\ref{eq:diversity}) is function of $k$ and determined as ($d_j \triangleq |dom_{A_j}|$):
\begin{equation}
\label{eq:diversity_max_possible_value}
div^*[k] = \sum_{j=1}^{m} \omega_j \left(d_j \sum_{i=1}^{k/d_j} \frac{1}{i^{\lambda}} +  \frac{k~\mathrm{mod}~d_j}{\big(1+ \frac{k}{d_j}\big)^{\lambda}} \right)
\end{equation}
\end{proposition}

\vspace{-2mm}
  \subsection{Distance-based diversity}
\label{sec:diversity:hball}
\vspace{-1mm}
In Sect.~\ref{sec:negativefunctions}, we  showed  that   an aggregation    by sum of the profile-wise Hamming distances  does not generally ensure submodularity or even monotonicity.  
Given the profiles of   two nodes $u,v$,  the \textit{Hamming distance}    is defined as:
\begin{equation}\label{eq:hamming}
dist^H(u,v) =  \sum_{j=1}^m \mathbbm{1}[val_{A_j}(u) \neq val_{A_j}(v)], 
\end{equation}
 where $\mathbbm{1}[\cdot]$ denotes the indicator function.\footnote{For any nodes $u$ and $v$, we assume that if either $u$'s or $v$'s profile  is not associated with a value in the domain of $A_j$ (i.e., missing value for $A_j$), with $j=1..m$, then the indicator function will be evaluated as 1.}
%
%
%
%

To design a set-function that satisfies both 
the properties of monotonicity and submodularity, we borrow the notion of \textit{Hamming ball} introduced in~\cite{hammingball}, i.e., a set of objects each having a Hamming distance from a selected object-center at most equal to a predefined threshold, or \textit{radius}. 
Our  definition of Hamming ball for a given node in the network takes also into account the \textit{influence range} 
of the node, i.e., all the nodes reachable starting from the node at the center of  the ``ball''. Formally, given     $v \in \V$ and a positive  integer $\xi$, we define the Hamming ball  as:
\begin{equation}
\mathcal{B}_v^{\xi} = \{ u \mid  u \in \mathrm{IR}(v) \wedge  dist^H(u,v) \leq \xi \},
\end{equation}
where $\mathrm{IR}(v) \subseteq \V$ denotes the set of nodes $u$ for which there exists 
a path connecting $v$ to $u$. 
 %
 %
 Restricting the Hamming balls to the center's influence range 
  is   beneficial in terms of efficiency, but also 
  licit   since only the Hamming balls that are meaningful
in  an influence spread scenario might be considered.

\begin{definition}
\label{def:hamming-div}
 Given a set of categorical attributes $\mathcal{A} = \{A_1, \ldots, A_m \}$ and associated profile set $\D$ for the nodes in a  graph $\G_0=\langle \V, \E \rangle$ and a radius $\xi$,  we define the {\em Hamming-based  diversity} of any $S \subseteq \V$ as:
\begin{equation}
\label{eq:hamming-div}
div(S) = \bigl\vert \bigcup\limits_{v \in S} B^{\xi}_v \bigr\vert
\end{equation} 

\vspace{-4mm}
\hfill~\qed
\end{definition} 

Intuitively, as similar nodes have overlapping Hamming balls,
by taking the union in Eq.~(\ref{eq:hamming-div}) we implicitly force
the selection of seeds so that nodes are as different
as possible from each other. 
In fact, this 
 eventually leads an extension of the Hamming
ball given by the individual balls associated with every selected seed. 
%
 %
 Moreover,  one nice effect of accounting for the influence reachable set
 in computing the Hamming balls, is that we inherently favor the selection
of nodes with   higher connectivity, since having a ``large'' Hamming ball  also implies   a large influence range, which is a particularly valuable aspect 
for our problem. 

The above defined function has the   property of allowing an incremental   computation of the marginal gain of any node. 

\begin{fact} 
\label{fact:hamming-inc}
The marginal gain of adding a node $u$ to $S$, with $u$ having 
 Hamming ball $B^{\xi}_u$, is equal to $\mid B^{\xi}_u \setminus B^{\xi}_S \mid$,
where $B^{\xi}_S =   \cup_{v \in S} B^{\xi}_v$.
\end{fact} 

\begin{proposition} 
The Hamming-based diversity function defined in Eq.~(\ref{eq:hamming-div}) is monotone and submodular.
\end{proposition}

\vspace{-2.5mm}
\subsection{Entropy-based diversity}
\label{sec:diversity:entropy}
\vspace{-1.5mm}
Diversity for categorical data can   naturally be associated with notions of heterogeneity, or variability, for discrete random variables, such as entropy and Gini-index. 
  Unfortunately, it is easy to note that such measures cannot be used to define a monotone submodular function of diversity as long as they are evaluated on any discrete random variable      whose    sample space (i.e., set of admissible values) corresponds to the categorical content of $\D_S$, for any $S \subseteq \V$. 
   For instance, if we describe each node-profile, resp. each attribute-value, in $\D_S$ by means of a vector whose generic entry represents the frequency of that profile, resp. attribute-value, then the entropy for the correponding probability mass function does not even preserve monotonicity for any $T \supseteq S$. 
  
Nonetheless, it is known that entropy is monotone and submodular if defined for a \textit{set of discrete random variables}~\cite{Fujishige78}. 
  Given a collection $\mathcal{X} = \{X_i\}_{i=1..|\mathcal{X}|}$ of discrete random variables, for the entropy function $H: 2^{\mathcal{X}} \mapsto [0, +\infty)$ it holds that  $H(\mathcal{X}_S) \leq H(\mathcal{X}_T)$ and that $H(\mathcal{X}_S,X) - H(\mathcal{X}_S) \geq H(\mathcal{X}_T,X) - H(\mathcal{X}_T)$, with $\mathcal{X}_S \subseteq \mathcal{X}_T \subseteq \mathcal{X}$ and $X \in \mathcal{X}, X \notin \mathcal{X}_T$. 
 Hence, one question  here becomes how to suitably define the variables over  $\D_S$, for any $S \subseteq \V$.  We next provide an intuitive definition valid in our context.
 %
 
\begin{definition}
 Given any $S \subseteq \V$, we define a set $\mathcal{X}_S = \{X_i\}_{i=1..|S|}$  of discrete random variables associated with the profiles of nodes in $S$, where for each $v_i \in S$, $X_{i}:  dom  \mapsto \{0,1\}$,   such that $dom$ is equipped with a probability function that assigns each $a \in dom$ with its relative frequency in $\D$, and   $X_i$ takes the value 1 if   $a$ is contained in $\A[v_i]$, 0 otherwise.
 \hfill~\qed
\end{definition} 
 
  By definition, the entropy of a set of $n$ discrete random variables is the joint entropy $H(X_1, \ldots, X_n) = \mathbb{E}[-\log P(X_1,$ $\ldots, X_n)]$. 
 %
  This  can   be rewritten in terms of conditional entropy through a \textit{chain rule} for discrete random variables~\cite{CoverThomas2006}: 
$$H(X_1, \ldots, X_n)   = H(X_1) + H(X_2|X_1) + \ldots +H(X_n|X_{n-1}, \ldots, X_1).$$ 
 That is, the entropy of a collection of random variables is the sum of the conditional entropies. In particular, given  three variables, 
  it holds that: 
 \begin{equation*}
 \begin{split}
     H(X_1, X_2, X_3) & = H(X_1) + H(X_2, X_3|X_1) \\
     & = H(X_1) + H(X_2|X_1) + H(X_3|X_2,X_1) \\
     & =  H(X_1,X_2) + H(X_3|X_2,X_1).
 \end{split}
 \end{equation*}

It should also be noted that a sequence of random variables can be considered as a single vector-valued random variable, therefore the joint probability distribution $p(\mathcal{X})$ can also be seen   as the  probability distribution $p(\mathbf{X})$ of  the random vector $\mathbf{X}=[X_1, \ldots, X_n]$. This naturally reflects as well on the  computation of  the conditional entropy of a variable given a sequence of random variables. 

 \begin{definition}
\label{def:entropy-div}
 Given a set of categorical attributes $\mathcal{A} = \{A_1, \ldots, A_m \}$ and associated profile set $\D$ for the nodes in a  graph $\G_0=\langle \V, \E \rangle$,   we define the {\em entropy-based  diversity} of any $S \subseteq \V$ as:
\begin{equation}
\label{eq:entropy-div}
div(S) =  H(X_1, \ldots, X_{|S|}) = \sum_{i=1}^{|S|} H(X_i | \mathbf{X}^{<i}),  
\end{equation}
where $\mathcal{X}_S = \{X_i\}_{i=1..|S|}$ is the set   of discrete random variables corresponding to nodes in $S$,   
$\mathbf{X}^{<i}$ denotes  the vector of variables $X_1, \ldots, X_{i-1}$, and 
\begin{equation}
\begin{split}
H(X_i|\mathbf{X}^{<i}) & = -\!\!\sum_{x \in \{0,1\}^{i-1}} p(\mathbf{X^{<i}}\!=\!x)   \\
 & \times \sum_{x_i \in \{0,1\}} p(x_i|\mathbf{X}^{<i}\!=\!x) \log p(x_i|\mathbf{X}^{<i}\!=\!x)  \\
 & = - \sum_{x \in \{0,1\}^{i-1}} p(\mathbf{X}^{<i}\!=\!x) \times H(X_i|\mathbf{X}^{<i}\!=\!x). \nonumber
\end{split}
\end{equation}

\vspace{-4mm}
\hfill~\qed
\end{definition}

In the above equation, note that the enumeration of 0-1 tuples of length $i$ is only limited to the  joint variable combinations   corresponding to the attribute-values occurring in $\D$, whereas for all other attribute-values $a'$ not in $\D$, the same tuple of all zeros is associated with the sum of probabilities of $a'$ in $\D$. 

The following fact states that the entropy-based diversity    function allows for an incremental   computation of a node's marginal gain. 

\begin{fact} 
The marginal gain of adding a node $v$ to $S$ is equal to the conditional entropy $H(X_{|S|+1} \ |\  \mathbf{X}^{<|S|+1})$.
 \end{fact}

\begin{proposition} 
The entropy-based diversity function defined in Eq.~(\ref{eq:entropy-div}) is monotone and submodular.
\end{proposition}
%

 \vspace{-2mm}
 \subsection{Class-based diversity}
\label{sec:diversity:class}
We now introduce a subclass of diversity functions which differs from the ones previously described in that it exploits a-priori knowledge  on  a grouping of the node profiles. This might be particularly relevant in scenarios where we are interested in distinguishing the nodes based on a coarser grain than their individual profiles. An available organization of the profiles into categorically-cohesive groups could reflect some predetermined equivalence classes of the profiles w.r.t. a given schema of attributes $\A$.  (This in principle also includes the opportunity of defining profile groups based on the availability of a \textit{community structure} over the set of nodes in the network.)
 
A simple yet efficient approach to measure diversity based on the exploitation of profile groups is to cumulate the \textit{selection rewards} for  choosing  nodes with a profile that belongs to any given class. 

\begin{definition}
\label{def:partition-div}
 Given a set of categorical attributes $\mathcal{A} = \{A_1, \ldots, A_m \}$ and associated profile set $\D$ for the nodes in a  graph $\G_0=\langle \V, \E \rangle$,   we define the {\em class-based  diversity} of any $S \subseteq \V$ as:
\begin{equation}
\label{eq:partition-div}
div(S) = \sum_{l=1..h}  \mathrm{f}(\sum\limits_{v_j \in C_l \cap S} r_j) 
\end{equation}

\noindent
where $\mathcal{C} = \{C_1, \ldots, C_h\}$  is a partition of $\D$ (i.e., $\bigcup_{l=1}^h C_l  = \D$, and $C \cap C' = \emptyset$, for each $C,C' \in \mathcal{C}$, with $C \neq C'$),  
$\mathrm{f}: \mathbb{R} \mapsto \mathbb{R}$ is any non-decreasing concave function, and $r_j >0$   is the selection reward for $v_j \in \V$.  
\hfill~\qed
\end{definition} 


The effect of $\mathrm{f}$ is that repeatedly selecting  nodes of  the 
 same class yields increased diminishing gains for the previously selected nodes. 
 In fact, since $\mathrm{f}$ is nonnegative concave and $\mathrm{f}(0)\geq 0$, $\mathrm{f}$ is also \textit{subadditive} on $\mathbb{R}^+$, i.e., 
 $ 
 \sum_{x_i=0}^{+\infty} \mathrm{f}(x_i) \geq  \mathrm{f}(\sum_{x_i=0}^{+\infty} x_i).
 $
 Therefore, adding (to the set $S$ being discovered) a node from a different class is preferable in terms of marginal gain than adding a node from an already covered class. 
Example  instances of $\mathrm{f}(x)$ are $\sqrt(x)$ and $\log(1+x)$, but  any other non-decreasing concave function can in principle be adopted.   
 We now provide the lower bound and upper bound of Eq.~(\ref{eq:partition-div}) when the logarithmic function is adopted. 

\begin{proposition}
Given a budget $k$  and $h$ classes, 
the function in Eq.~(\ref{eq:partition-div}), equipped with $\mathrm{f}(x)=\log(1+x)$, with $r_j = 1, \forall v_j \in \V$, achieves the minimum value of $\log(1+k)$ when all $k$ nodes belong to the same class (i.e.,  1 class  covered),  and the maximum value of $k$ when all $k$ nodes belong to different classes (i.e.,  $k$ classes  covered). 
\end{proposition}
%

Again, the above defined function enables an incremental   computation of the marginal gain of any node. 
 
 \begin{fact} 
The marginal gain of adding a node $v$ to $S$, with $v$ belonging to class $C_l$, is equal to $\log(1+ r/R_l)$, where $r$ is the reward of adding $v$ and $R_l$ is one plus the sum of rewards of nodes in $S$ that belong to class $C_l$.  
 \end{fact} 
 

\begin{proposition} 
The partition-based diversity function defined in Eq.~(\ref{eq:partition-div}) is monotone and submodular.
\end{proposition}

\section{A RIS-based framework for the ADITUM problem}
\label{sec:framework}
We develop our framework for the  \myalgo problem based on  the \textbf{R}everse \textbf{I}nfluence \textbf{S}ampling (RIS) para\-digm first introduced in~\cite{doi:10.1137/1.9781611973402.70} and recognized as the  state-of-the-art approach for IM problems. 
 %
  %

%

The breakthrough study by Borgs et al.~\cite{doi:10.1137/1.9781611973402.70}  overcomes the limitations of a greedy, Monte Carlo based, approach to IM  by proposing 
a novel solution  based on the two following concepts. 

Given the diffusion graph $\G$ with node set $\V$ and edge set $\E$, let $G$ be an instance of $\G$ obtained by removing each edge   $e \in \E$ with probability $1-p(e)$. The \textit{reverse reachable set} (RR-Set) rooted in $v$ w.r.t. $G$ contains all the nodes reachable from $v$ in a backward fashion. 
 %
A \textit{random RR-Set} is any RR-Set generated on an instance $G$, 
for a node  selected uniformly at random from $G$. 

The key idea of the RIS framework  is that the more a node $u$ appears in a random RR-Set rooted in $v$, the higher the probability that $u$, if
selected as seed node, will activate $v$. 
The design of the RIS framework   follows a \textit{two-phase schema}~\cite{doi:10.1137/1.9781611973402.70}: 
(1) Generate a certain number of random RR-Sets, 
and (2) Select as seeds the $k$ nodes that cover the most RR-Sets. (The latter step can be solved by using any greedy algorithm for the Maximum Coverage problem.) 

Based on  RIS, Tang et al.~\cite{Tang:2014:IMN:2588555.2593670} developed the \algo{TIM} and \algo{TIM+} algorithms  
that achieves  $(1-1/e-\epsilon)$-approxi\-mation
with at least $1-|\V|^{-l}$ probability (by default $l=1$) in time 
$O((k+l)(|\E|+|\V|) \log |\V| /\epsilon ^2)$. 
 \algo{TIM}/\algo{TIM+} works in two major stages: \textit{parameter estimation} and \textit{seed selection}. The first stage   aims at deriving a lower-bound for the maximum expected spread a size-$k$ seed set can achieve, from which depends the number $\theta$ of random RR-Sets that must be generated in the second stage; the latter essentially coincides with the second phase
of the RIS method.\footnote{\algo{TIM+} aims to improve upon \algo{TIM} by adding an intermediate step between parameter estimation and node selection, which  heuristically refines $\theta$ into a tighter lower bound of the maximum expected influence of any size-$k$ node set. Also, in~\cite{IMM}, IMM is introduced to further speed up \algo{TIM+}.}   
  The effectiveness of \algo{TIM}/\algo{TIM+} is explained by Lemma 2 provided in~\cite{Tang:2014:IMN:2588555.2593670}, which  states that, if   $\theta$ is sufficiently
large, the fraction of random RR-Sets covered by any
seed set $S$ is a good and unbiased estimator of the average node-activation probability. 

 \vspace{-2mm}
\subsection{Proposed approach}
\label{sec:algo:revisiting}
 \vspace{-2mm}
Our proposed RIS-based framework follows the typical two-phase  schema,  however it originally embeds both the targeted nature and the diversity-awareness in an influence maximization   task. To accomplish this, we revise the two-phase schema as follows.  
%
  
%


\paragraph*{\bf Parameter estimation. \ } 
%
%
%

We want to understand how much capital can be captured from a size-$k$ seed set. Therefore, to compute the number $\theta$ of RR-Sets, we need to identify a  lower-bound on the maximum capital score. 

We select a node $v$ as the root of an RR-Set with 
probability $p(v) \propto t(v)$. Since we are interested  in the activation
of the target nodes only, we set $p(v) = \frac{t'(v)}{\mathcal{T}_{TS}}$, where $\mathcal{T}_{TS} = \sum_{u \in TS} t(u)$,  
and $t'(v) = t(v)$ if $v \in TS$ and $t'(v) = 0$  otherwise. 
 %
 We leverage on the \algo{TIM+} procedures \textsl{KPTEstimation} and \textsl{RefineKPT}, in order to estimate 
  a lower-bound
for the expected spread achieved by any optimal seed set of size $k$. 
More specifically, the first procedure generates a small number of RR sets upon which it provides an initial approximation that it is further improved 
by the second procedure.

We borrowed these procedures from \algo{TIM+} as  
 our capital function is contingent on the activation process, thus we still need to have an unbiased estimator for the spread function. 
In fact,  any target node will contribute in terms of capital  as long as it has been activated starting from the seed set.   
The lower-bound on the expected spread allows us to derive 
a lower-bound on the average activation probability, from which we   
compute the \textit{expected capital} score of a seed set as 
\vspace{-1.5mm}
\begin{equation} \label{eq:exp_capital_estimation}
\mathbb{E}[C(S)] = \mathcal{T}_{TS} (\mathbb{E}[\mu(S)])/|\V|. 
\end{equation}
 Above, the rightmost term is the average fraction of
total capital score, denoted by $\mathcal{T}_{TS}$, the seed set $S$ is able to capture. 
Moreover, since every random RR-Set is rooted in a target node, the aforementioned Lemma 2~\cite{Tang:2014:IMN:2588555.2593670} ensures that 
 $\mathbb{E}[\mu(S)]/|\V|$ is very close
to the average activation probability of the target nodes.

\paragraph*{\bf Seed selection. \ } 
Once   all $\theta$ RR-Sets are computed, this stage is in charge of detecting   the $k$ seed sets. To this end, we need also to account for a notion of set-diversity to   choose  the candidate seeds.  
 The selection of   
best seeds is accomplished in a greedy fashion, one seed at a time. A node $v$ is associated with a  linear combination of (i)  the \textit{node's capital score}, obtained by  summing  the target  scores of    the roots of the RR-Sets to which  $v$ belongs and that are not already covered by seeds, and (ii) the \textit{node's diversity score}, which    corresponds to the node's marginal gain for the diversity function w.r.t. the  current seed set.  
%



\begin{algorithm}[t!]
\small
\caption{{\bf A}ttribute-based {\bf DI}versity-sensitive {\bf T}argeted Infl\textbf{U}ence {\bf M}aximization (\myalgo)}
\label{alg:ris-dtim}
\begin{algorithmic}[1]
\Require A diffusion graph $\mathcal{G} = \langle \mathcal{V}, \mathcal{E}, b, t \rangle$ based on triggering model $\mathcal{M}$, a budget $k$, a target selection threshold $\tau_{TS} \in [0,1]$, a smoothing
parameter $\alpha \in [0,1]$. 
\Ensure Seed set $S$ of size $k$.
\vskip 0.1em
\State $TS \gets \{v \mid t(v) \geq \tau_{TS} \}$\Comment{Select the target nodes}
\State Compute $\theta$ by using \algo{TIM+} procedures \textsl{KPTEstimation} and \textsl{RefineKPT}
\State $\mathcal{R} \gets \emptyset$
\For{$i \gets 1$ \textbf{to} $\theta$}
\State $R \gets \textsl{computeRandomRRSet}(TS, \mathcal{M}, i)$ 
\State $\mathcal{R} \gets \mathcal{R} \cup \{ R \}$
\EndFor
\State $S \gets \textsl{buildSeedSet}(\mathcal{R},k,\alpha)$\Comment{Seed Selection stage}\\
\Return $S$
\vskip 0.2cm
\Procedure{\textsl{computeRandomRRSet}}{$TS,\mathcal{M}, id$}
\State $R \gets \emptyset$\Comment{Initialize the RR-Set}
\State Select node $r \in TS$ as root, with probability $p(r) \propto t(r)$
\State $R.id  \gets id, R.root \gets r$\Comment{Associate   id and   root to the RR-Set}
\State Add to $R$ the nodes that can reach $r$  according to live-edge model of $\mathcal{M}$ \\
\Return $R$
\EndProcedure
\vskip 0.2cm
\Procedure{\textsl{buildSeedSet}}{$\mathcal{R}, k, \alpha$}
\State $q \gets \emptyset$\Comment{Priority queue for lazy-greedy  optimization}
\For{$v \in \V$}
\State $v.pushedC \gets \sum_{R \in \mathcal{R}(v)} c(root(R))$
\State $v.pushedD \gets \textsl{marginalGainInDiversity}(v, \emptyset)$  
\State $q.add(\langle (\alpha \times v.pushedC + (1-\alpha) \times  v.pushedD), v, 0 \rangle)$
\EndFor

\State $S \gets \emptyset, CS \gets \emptyset$
\Repeat
\State $\langle \symbOF\_val, v, it \rangle \gets q.removeFirst()$
\If{$it = |S|$}
\State $S \gets S \cup \{v \}, CS \gets CS \cup \mathcal{R}(v)$
\Else

\For{$R \in \mathcal{R}(v) \cap CS$}	
	\State $v.pushedC \gets v.pushedC - t(root(R))$
	\State Remove $R$ from $\mathcal{R}(v)$
\EndFor
\State $v.pushedD \gets \textsl{marginalGainInDiversity}(v, S)$ 
\State $q.add(\langle (\alpha \times v.pushedC + (1-\alpha) \times  v.pushedD), v, |S| \rangle)$
\EndIf
\Until{$|S| = k$}\\
\Return $S$
\EndProcedure
\end{algorithmic}
\end{algorithm}

{\bf Remarks.} 
The objective function we seek to maximize is a linear combination 
of two main quantities: the expected capital and the diversity of the seeds. 
 Note that there is a key difference between these two measures: the former is defined globally over the whole network, while the latter is limited to the seed nodes, namely the solution itself. 

Our approach hence reflects this inherent interplay between capital and diversity.
In fact, the sampling procedure in the first stage corresponding to the  parameter estimation, is driven by only the capital score --
there are no seeds upon which the diversity must be assessed -- whereas the diversity aspect comes into play only during the process of  seed set formation, thus it drives the discovery of the seeds.

  \vspace{-2mm}
\subsection{The ADITUM algorithm}
\label{sec:algo:aris}

Algorithm~\ref{alg:ris-dtim} shows the pseudocode of our implementation of \myalgo. 
%
The algorithm  starts by identifying the target nodes \sline{1}, then
it infers  the number $\theta$ of RR-Sets to be computed, according to \algo{TIM+} subroutines of estimation and refinement of $KPT$, i.e., the mean of the expected spread of possible size-$k$ seed sets \sline{2}. In lines 4-6 the $\theta$ RR-Sets are generated by invoking the \textsl{computeRandomRRSet} procedure \lines{4}{6}. 
in $\mathcal{R}$.  
The procedure \textsl{buildSeedSet} eventually returns the size-$k$ seed set \lines{7}{8}. 
 %
 In the following, we provide details about the two procedures.

\vspace{1mm}
Procedure \textsl{computeRandomRRSet} starts
 by sampling no\-de $r$ as the root of $R$ from a  distribution of probability proportional to the target-node scores  \sline{11}.
Each RR-Set is associated with an integer identifier and the root node  \sline{12} --- this information is needed  since the capital   associated with a set is given by the target score of its root.
Finally, an instance of the influence graph $G \sim \mathcal{G}$ is computed according to the live-edge model related to $\mathcal{M}$, then all the nodes that can reach $r$ in $G$
are inserted in the RR-Set   to be returned.

\vspace{1mm}
Procedure \textsl{buildSeedSet}
 exploits a priority queue $q$, which is initialized \sline{16} to store triplets comprised of:   value of the linear combination of capital and diversity,    node and    iteration   to which the value  refers to. The triplets are ordered by decreasing values of capital-diversity combination.
 For each node $v$, its  capital score is computed by summing   the target  score of all   nodes that are roots of an RR-Set $v$ belongs
to \sline{18}. Moreover, the $v$'s diversity score is computed as its marginal gain  for the $div$ function w.r.t. the  current seed set \sline{19}; in particular, since the latter is initialized as empty, the initial $v$'s diversity score equals 1 (according to Eqs.~(3--4) of the main paper).
 Once all the scores are computed, the procedure starts to select the seeds, by getting at each iteration the best triplet from the queue \sline{23}: if the choice is done at  iteration $it$ equal to the number of nodes currently in the seed set \sline{24}, then $v$ is
inserted in $S$, and all sets covered by $v$ are stored in
$CS$; otherwise, all the score are to be   recomputed.
 By denoting with  $\mathcal{R}(v)$  the set of     random RR-Set containing  $v$,  the $v$'s capital score is decreased  by the target   score   of
each node $r$ that  is root of an already covered RR-Set (i.e., a set in  $\mathcal{R}(v) \cap CS$) \sline{28}, and this set is also  removed  from $\mathcal{R}(v)$ \sline{29}.
The diversity score needs also to be recomputed, finally the updated triplet is inserted  into the priority queue  \lines{30}{31}.
The procedure loop ends when the   desired size $k$ is reached for the seed set \sline{32}.

\begin{proposition}
\myalgo   runs in $O((k+l)(|\E|+|\V|)$ $\log |\V| /\epsilon ^2)$ time and returns  a $(1-1/e-\epsilon)$-approximate solution 
with at least $1-|\V|^{-l}$ probability.
\end{proposition}

\section{Evaluation methodology}
\label{sec:eval}


\begin{table*}[t!] 
\caption{Summary of evaluation network data.}
\label{tab:graph:properties}
\scalebox{0.84}{
\begin{tabular}{|c||c|c|c|c|c|c|c|c|}
\hline 
network & \#nodes & \#edges & avg.    & avg.    & clust.   & assort. & \#sources & \#sinks \\
         &        &         & in-degree   & path length & coeff. & & &  \\
\hline\hline
FriendFeed & 493\,019 & 19\,153\,367 & 38.85 & 3.82 & 0.029 & -0.128 & 41\,953 &   292\,003 \\
\hline
GooglePlus & 107\,612 & 13\,673\,251 & 127.06 & 3.32 & 0.154 & -0.074 & 35\,341 &  22 \\
\hline 
Instagram & 17\,521 & 617\,560 & 35.25 & 4.24 & 0.089 & -0.012 & 0 & 0  \\
\hline
MovieLens & 943 & 229\,677 & 243.5 & 1.87 & 0.752 & -0.323 & 1 & 1   \\
\hline
Reddit & 11\,224 &  91\,924 &  8.18 & 4.11  & 0.083 & -0.072 & 0 & 0  \\
\hline

\end{tabular}
}
\end{table*}

\subsection{Data}
\label{sec:data}
 \vspace{-1.5mm}
We used five real-world OSN datasets, namely  
\textit{FriendFeed}~\cite{8326536},
\textit{GooglePlus}~\cite{8326536},  \textit{Instagram}~\cite{8326536},   \textit{MovieLens}~\cite{6921625}, and 
\textit{Reddit}~\cite{kumar2018community}. 
 Table~\ref{tab:graph:properties} shows  main  statistics about the evaluation networks.  
It should be emphasized that we came to our  choice of the datasets because of  the following reasons: 
\begin{itemize}
\item \textit{reproducibility}, i.e., all of the networks are publicly available; 
\item \textit{diversification} of the evaluation scenarios, which include user engagement and item recommendation;
\item \textit{continuity} w.r.t. previous studies; 
\item \textit{fair comparative evaluation}, i.e., we based our choice also in relation of the competing methods include in our evaluation, so to enable a fair comparison between them and our \myalgo.
\end{itemize} 

FriendFeed, GooglePlus, and Instagram network datasets refer to OSNs previously studied in a  \textit{user engagement} scenario, which has been recognized as an important case in point for demonstrating  targeted IM tasks~\cite{8326536}. 
 For each of these networks, the meaning of any directed edge $(u,v)$ is that user $v$ is ``consuming'' information received from $u$ (e.g., $v$ likes/comments/rates a $u$'s media post).  
No side information is originally provided with such datasets, therefore we synthetically  generated the user profiles as follows:  
  Given $m$  categorical attributes, each with $n_i$  admissible values ($i=1..m$), we associated  each user with a set of values sampled from  either uniform or exponential (with $\lambda=1$) distribution. We set $m=n_i=10$.
    We used these datasets also for comparison with \algo{DTIM}.
%

Originally used for \textit{movie recommendation}, MovieLens is associated with a  (user, movie-genre) rating matrix   storing the number of movies each user rated for each genre, at any given time over a predefined observation period.  
 This dataset was previously included in the evaluation of our competitor \algo{Deg-D}. 
 To enable   \myalgo to work on MovieLens, we   mapped 
 each genre to an attribute, with    unique rating-values as corresponding attribute-values.  
%
 %
 The MovieLens  network was built so to have users as  nodes  and any directed edge $(u,v)$ is drawn if user $u$ rated \textit{first} at least 10 movies in common with $v$   (timestamps are available in the original data). 
 %
%
%

  Reddit network represents the directed connections between two subreddits, i.e., communities on the Reddit platform.  Each connection refers  to a post in the source community that links to a post in the target community. 
 From the original network, we kept only the connections for which the source post is explicitly positive towards the target post, and finally extracted the largest strongly connected component to overcome sparsity issues. 
 Reddit connections are also rich in terms of numerical attributes associated with each source post, which include both lexical and sentiment information. We selected 11 attributes which appeared to be the most informative for influence propagation reasons.\footnote{We selected the \textsc{POST\_PROPERTIES}  attributes corresponding to the following identifiers: 19, 20, 21, 43, 44, 45, 46, 51, 52, 53, 66. }
%
 To generate the profile of each node (community), we grouped the posts by community and summed up the scores for each attribute; finally, the values of each attribute were discretized through a 10-quantile binning scheme. 
 

\vspace{-2.5mm}
\subsection{Settings}
\label{sec:settings}
\vspace{-1.5mm}
We considered \myalgo instantiations with each of the   definitions of diversity proposed in Sect.~\ref{sec:diversity}. Hereinafter, we will use notations $div^{(AW)}$, $div^{(H)}$, $div^{(E)}$, $div^{(C)}$ to refer to the attribute-wise, Hamming-, entropy-, and class-based definitions, respectively. When using $div^{(H)}$, we set 
the radius $\xi$ of the Hamming balls within $\{1,3,5\}$. 
 We experimentally varied the setting of \myalgo parameters: 
the seed-set size $k$, within [5..50], 
the smoothing parameter $\alpha$, from 0 to 1 with step 0.1, 
and  the target selection threshold $\tau_{TS}$; the latter was controlled in terms of percentage  of top-values from the target score distribution, thus we selected target sets corresponding to the top-$\{5\%,10\%,25\%\}$. 
We   used the default $\epsilon=0.1$ for the approximation-guarantee in 
 the parameter estimation phase. 
   Concerning the edge weighting function ($b$) and the node weighting  function ($t$), we devised the following settings:
 
\vspace{-1mm}
\begin{description}
\item[(S1)]
The first setting  refers to the basic, non-targeted setting adopted in~\cite{6921625}, i.e.,  $b(u,v)=1/n_v$, with $n_v$ number of $v$'s in-neighbors, and $t(u)=1$, for all $u,v$ in $\V$. We used this setting for MovieLens evaluation.  

\item[(S2)] 
The second setting refers to Reddit, for which  the  influence weights are set to be proportional to the amount of interactions between communities: 
 for any  two nodes $u$ and $v$,   $b_{uv} = P_{uv}/P_v$, where $P_{uv}$ is the number of posts of $u$ towards $v$, and $P_v$ is the total number of posts having $v$ as target.  
 The node weighting function is here simply defined as the in-degree function, in order to mimic a scenario of influence targeting as corresponding to communities that are highly popular in terms of post recipients.

\item[(S3)] 
The third setting refers to a user engagement scenario and applies to FriendFeed, GooglePlus and Instagram, which were previously used on that context~\cite{8326536}. 
 User engagement  is addressed as a topology-driven   task for encouraging silent users, a.k.a. ``lurkers'',   to return their acquired social capital, through a more active participation to the community life. Note that such users are    effective members of an OSN, who are not actively involved in tangible content production and sharing with other users in the network, but rather they are information consumers. 
  Given this premise,   in~\cite{8326536}    a specific instance of targeted IM is developed such that   lurkers are regarded as the target of the diffusion process. Therefore, the user engagement task becomes: Given a budget $k$,  to  find a  set of $k$ nodes that are capable of maximizing the likelihood of ``activating'' (i.e., engaging) the target lurkers.
  In this context, the two  weighting functions rely on a pre-existing solution of a \textit{lurker} \textit{ranking} algorithm  applied to the social graph. 
  The intuition is as follows (the interested reader is referred  to \cite{8326536} for analytical  details about the above functions):  
For any node $v$,  the node weight  $t(v)$ indicates the status of $v$ as lurker, such as the higher the lurker ranking score of $v$  the higher should be $t(v)$; for any edge $(u,v)$, the weight $b(u,v)$  is computed to measure how much   node $u$ has contributed to the $v$'s lurking score calculated by the lurker ranking algorithm, which resembles a measure of ``influence'' produced by   $u$ to  $v$. 
\end{description}

\vspace{-2mm} 
\subsection{Competing methods}
\vspace{-1mm} 
 The closest methods to \myalgo are \algo{DTIM}~\cite{8326536} and \algo{Deg-D}~\cite{6921625}. 
  As previously mentioned, 
  \algo{DTIM} addresses targeted IM, but it considers  topology-driven notions of diversity only; conversely, \algo{Deg-D} utilizes side-information-based diversity, however it assumes a numerical representation of node attributes and the addressed problem is not  targeted. 
 We next provide details on the objective function of \algo{Deg-D} and  \algo{DTIM}. 

 The objective function in \algo{DTIM}~\cite{8326536} shares the capital term with \myalgo, which is however combined with a diversity term defined as $\sum_{s \in S} \sum_{v \in TS} div_v(s)$, i.e., as the sum of diversity scores that each seed has in relation with each of the target nodes, where $div_v(\cdot)$ is either the global topology-driven or the local topology-driven diversity function~\cite{8326536}.

\algo{Deg-D}~\cite{6921625} follows a simple greedy scheme to  maximize the objective function
$(1-\gamma) \sum_{u \in S} deg(u) + \gamma D(S)$, 
where $deg(u)$  denotes the out-degree of node $u$, while $D(S)$ represents 
the diversity of the set $S$, whose value is given by:
%
$D(S) = \sum_{m=1}^{M} f(\sum_{u \in S} \omega_{um} \times g(u))$,
where $M$ denotes a given number of types of external information, 
$\gamma$ is a smoothing parameter, 
$\omega_{um} \in [0,1]$ is a real-valued coefficient expressing the preference of node $u$ toward type  $m$, 
$f$ denotes any nondecreasing concave function (with default form set to   $f(x)=\log(1+x)$),  
whereas $g$ is a function defined   for each node $u$,  either as $g(u)=1$ or $g(u)=deg(u)$; the two different definitions of $g$ lead to the variants named  \algo{Deg-DU} and \algo{Deg-DW}, respectively. 
Note that, compared to $\alpha$ in \myalgo, $\gamma$ in \algo{Deg-D} has an opposite role, 
therefore we set $\gamma = 1-\alpha$ in all the experiments.
%
 %
 Moreover, 
 \algo{Deg-D}  requires a numeric   vector of size $M$ to be associated with each node.

 To enable a comparison with \algo{DTIM}, we integrated its  global  
 topology-driven diversity function  into our RIS-based framework, following the guidelines provided in~\cite{8326536}. 
As concerns   \algo{Deg-D}, we also had to account for the different (i.e., numerical) representation of side-information by \algo{Deg-D}. Thus, we devised two settings: 
\begin{itemize}
\item 
 Integration of the \textit{uniform} and \textit{weighted}  functions, i.e.,   \algo{Deg-DU} and \algo{Deg-DW}, resp.,  into our RIS-based framework, upon numerical representation of nodes' attributes; 
 \item
  Comparison of the two methods:  \myalgo upon categorical representation derived from a numerical representation of nodes' attributes vs. \algo{Deg-DU} and \algo{Deg-DW} upon normalized numerical representation.
  \end{itemize}
  %

 \section{Experimental results}
\label{sec:results}

\paragraph*{\bf Goals.\ \ } 

We pursued  four main goals of experimental evaluation, around which we organize the presentation of our results.   
First, we want to assess the significance of the estimation of capital produced by \myalgo (Sect.~\ref{sec:results:capital}).  
Second,  we want to understand the effect of the three 
proposed  definitions of diversity on  the solutions provided by \myalgo  (cf. Sect.~\ref{sec:results:diversity}). 
Third, we analyze   the sensitivity of \myalgo   w.r.t. its various parameters and the attributes' distributions (Sect.~\ref{sec:results:seedset}).  
Fourth, we comparatively evaluate   \myalgo with the  competing methods  
\algo{DTIM}  and \algo{Deg-D}  (Sects.~\ref{sec:results:comparison_DTIM} and~\ref{sec:results:comparison_Tang}).
%


\vspace{-2mm}
\subsection{Capital estimation}
\vspace{-2mm}
\label{sec:results:capital}
To begin with, we    analyzed the correctness of the RIS-based estimation of  the capital captured by the seeds discovered by \myalgo, which refers to   Eq.~\ref{eq:exp_capital_estimation}.  
To this purpose, we compared the \myalgo capital estimation (with $\alpha=1$)
with the capital scores obtained from a Monte Carlo simulation ($10\,000$ runs). 

\begin{figure}[t]
\centering 
\begin{tabular}{@{\hskip -2.4mm}c@{\hskip -2.3mm}c@{\hskip -2.3mm}c@{\hskip -2.3mm}c}
\includegraphics[width=.27\linewidth]{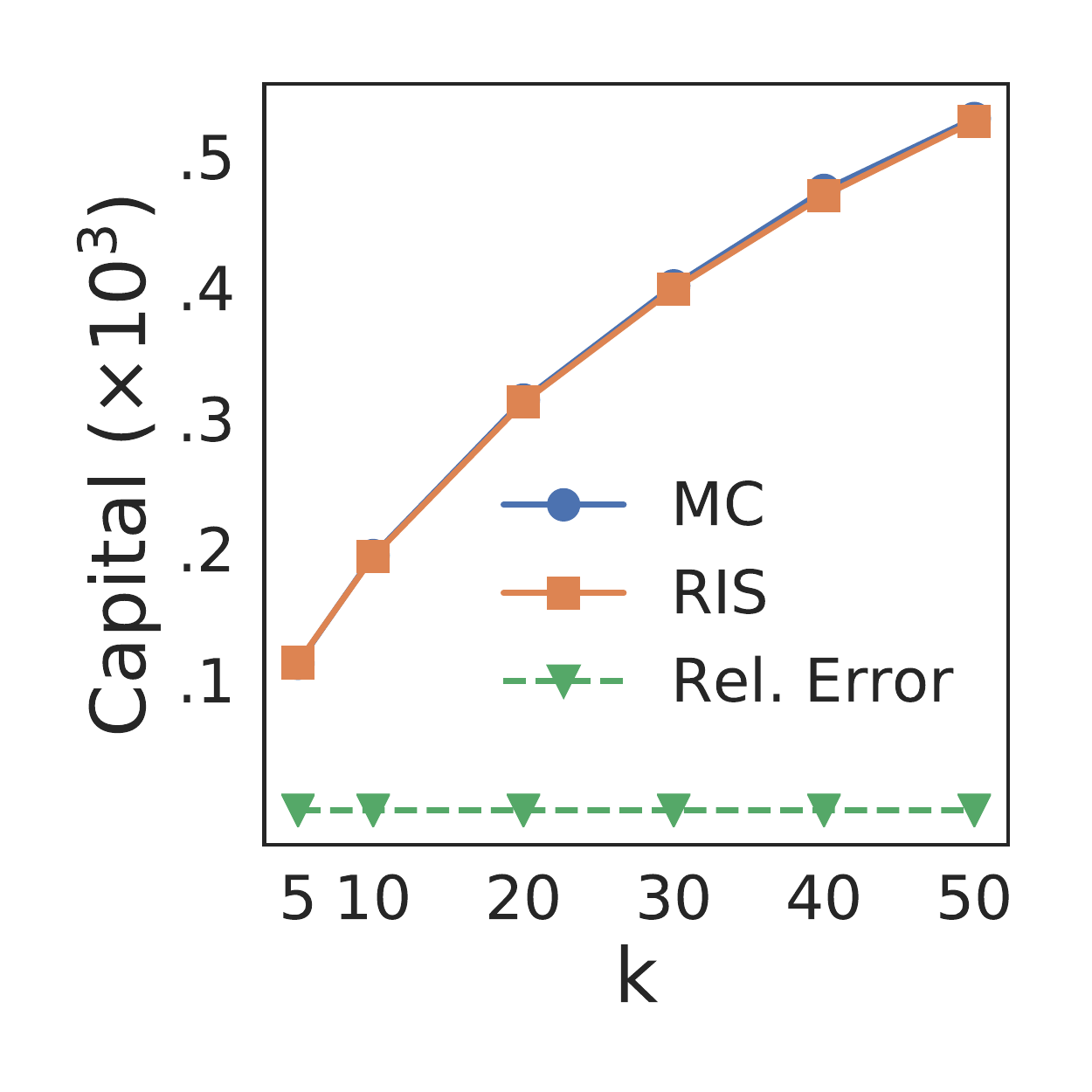} & 
\includegraphics[width=.27\linewidth]{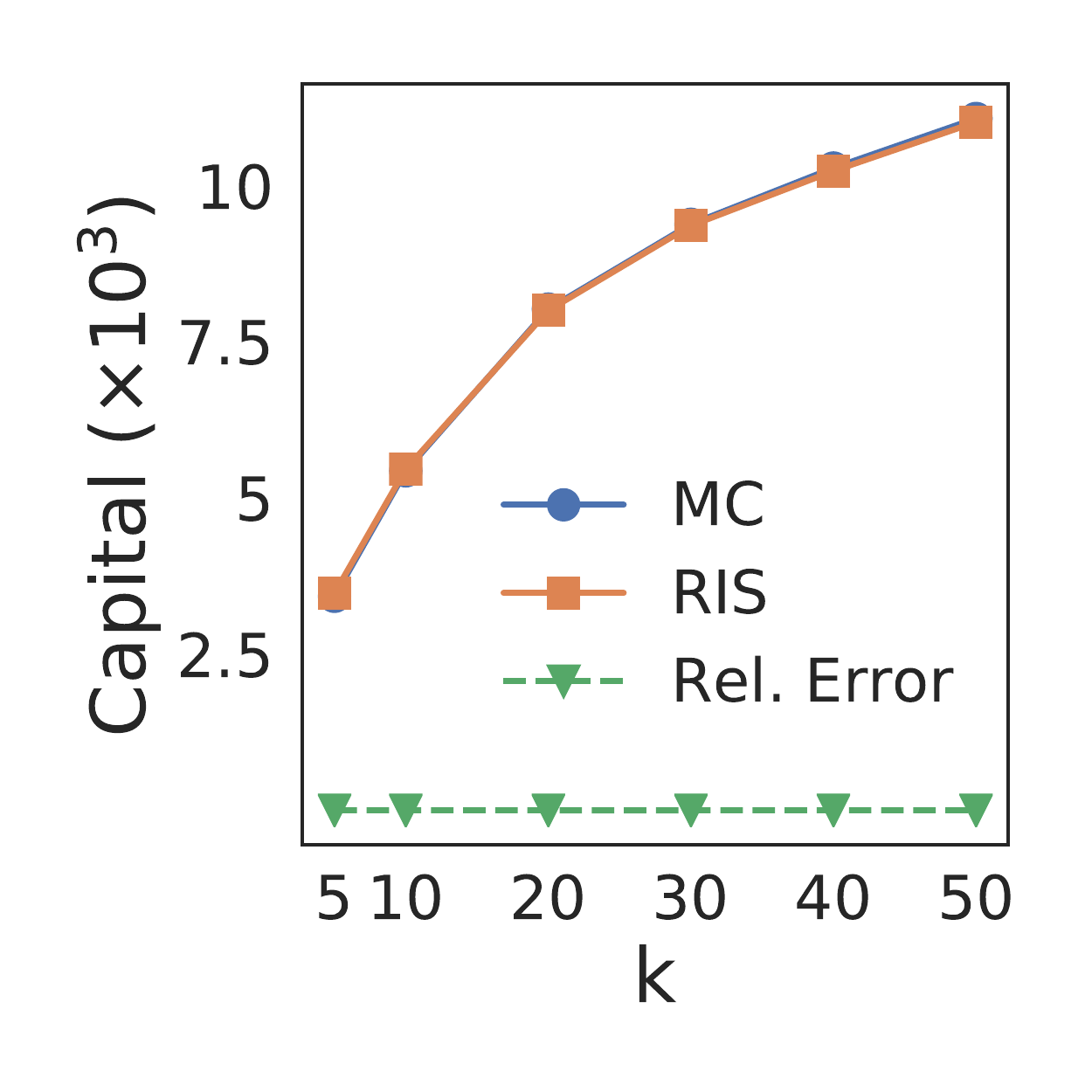} &
\includegraphics[width=.27\linewidth]
{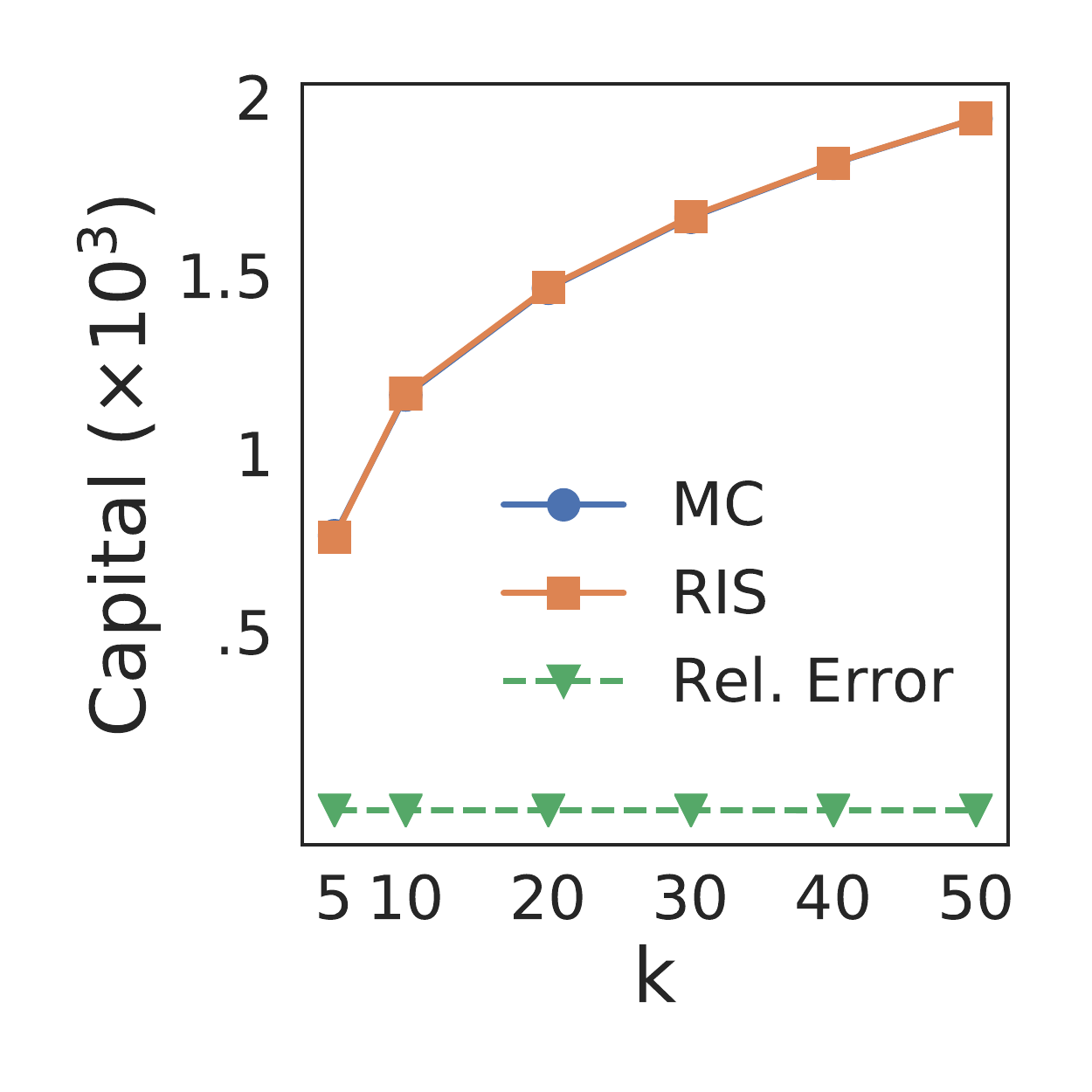} &
\includegraphics[width=.27\linewidth]
{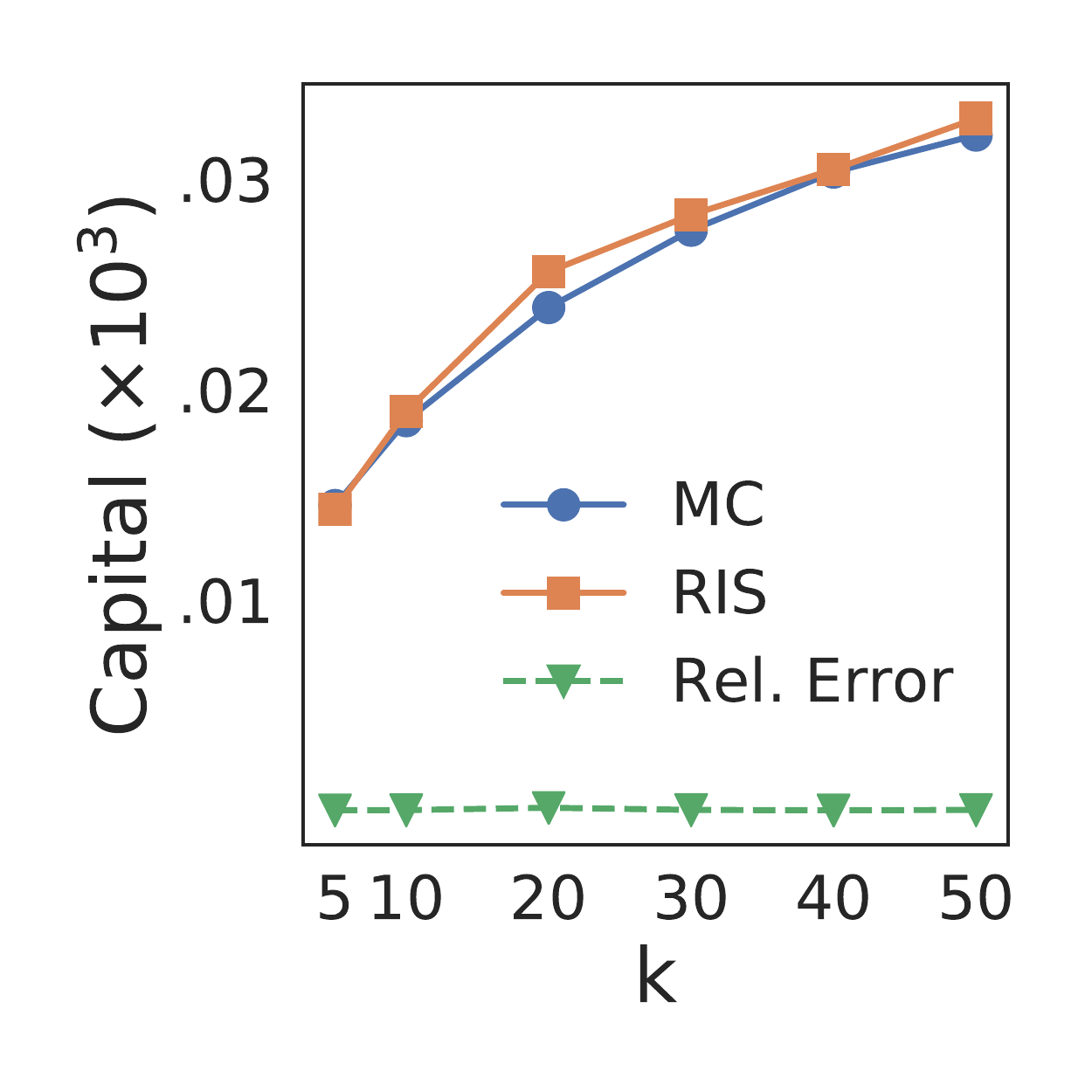} \\
(a) Instagram  & (b) FriendFeed  &
 (c) GooglePlus  & (d) Reddit \\
\end{tabular}
\caption{Capital estimation for seed sets obtained by \myalgo: RIS-based estimation by \myalgo vs.  estimation by Monte Carlo simulations, with top-25\%  target selection.}
\label{fig:estimation}
\end{figure}

 As shown in Fig.~\ref{fig:estimation}, for top-25\% target selection and varying $k$,  the two capital estimations are practically identical (i.e., relative error almost zero), even for higher $k$. The same   holds  for other settings of target selection.    
This     confirms the correcteness of the RIS-based estimation of capital in \myalgo.  
 
 
\begin{figure}[t]
\centering 
\begin{tabular}{@{\hskip -3mm}c@{\hskip 0mm}c@{\hskip -2mm}c@{\hskip -2mm}c}
\includegraphics[width=.26\linewidth, height=3cm]{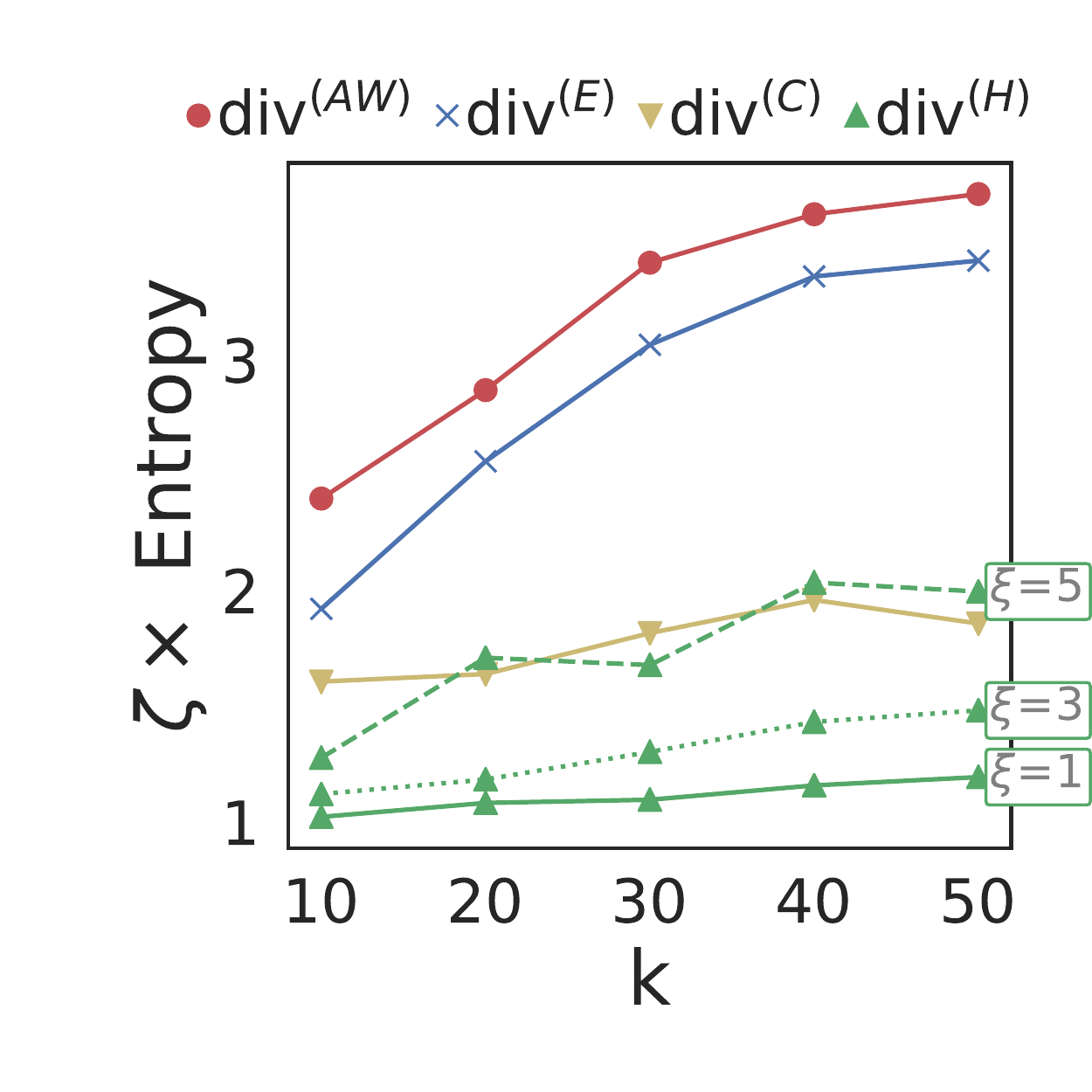} & 
\includegraphics[width=.27\linewidth, height=3cm]
{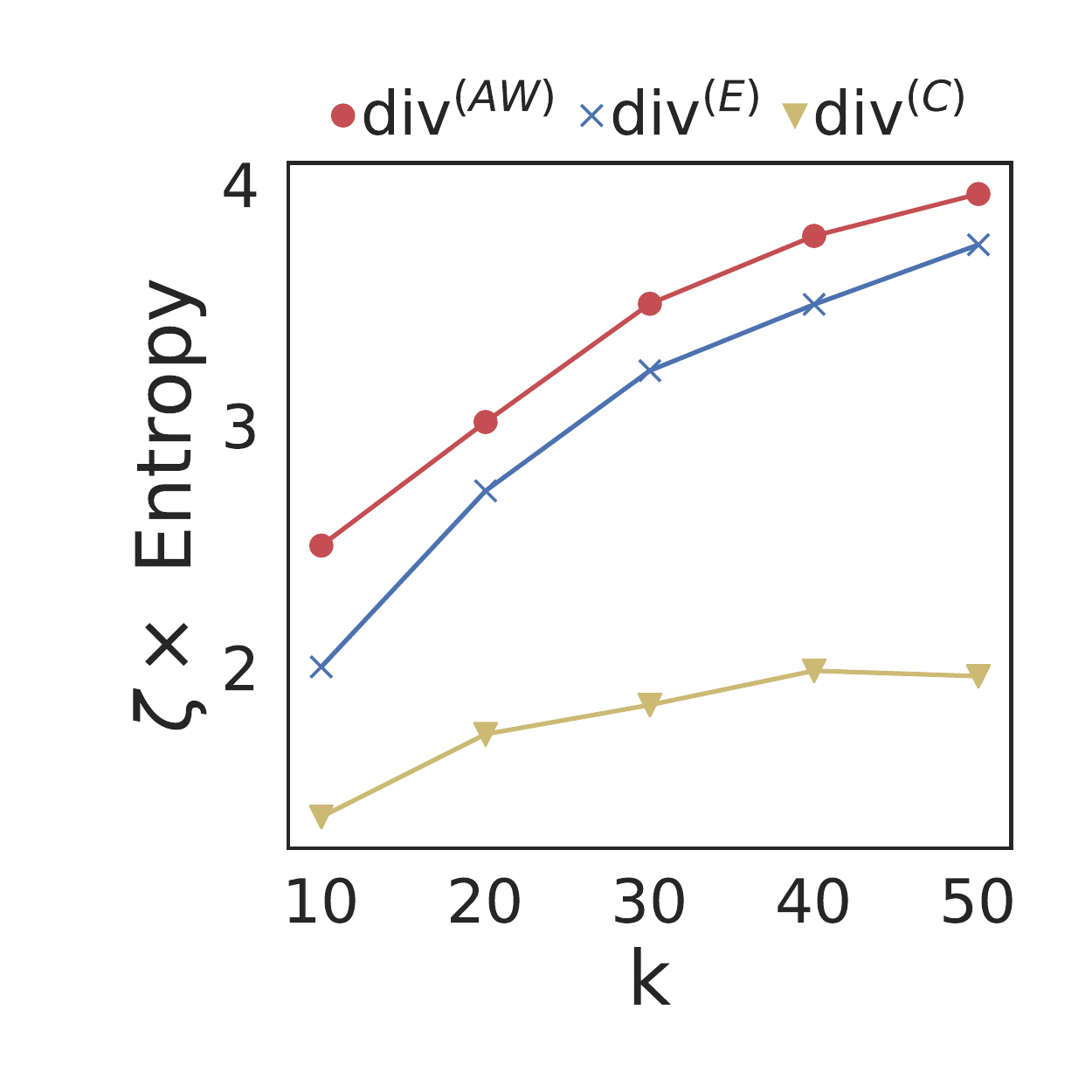} &
\includegraphics[width=.27\linewidth, height=3cm]
{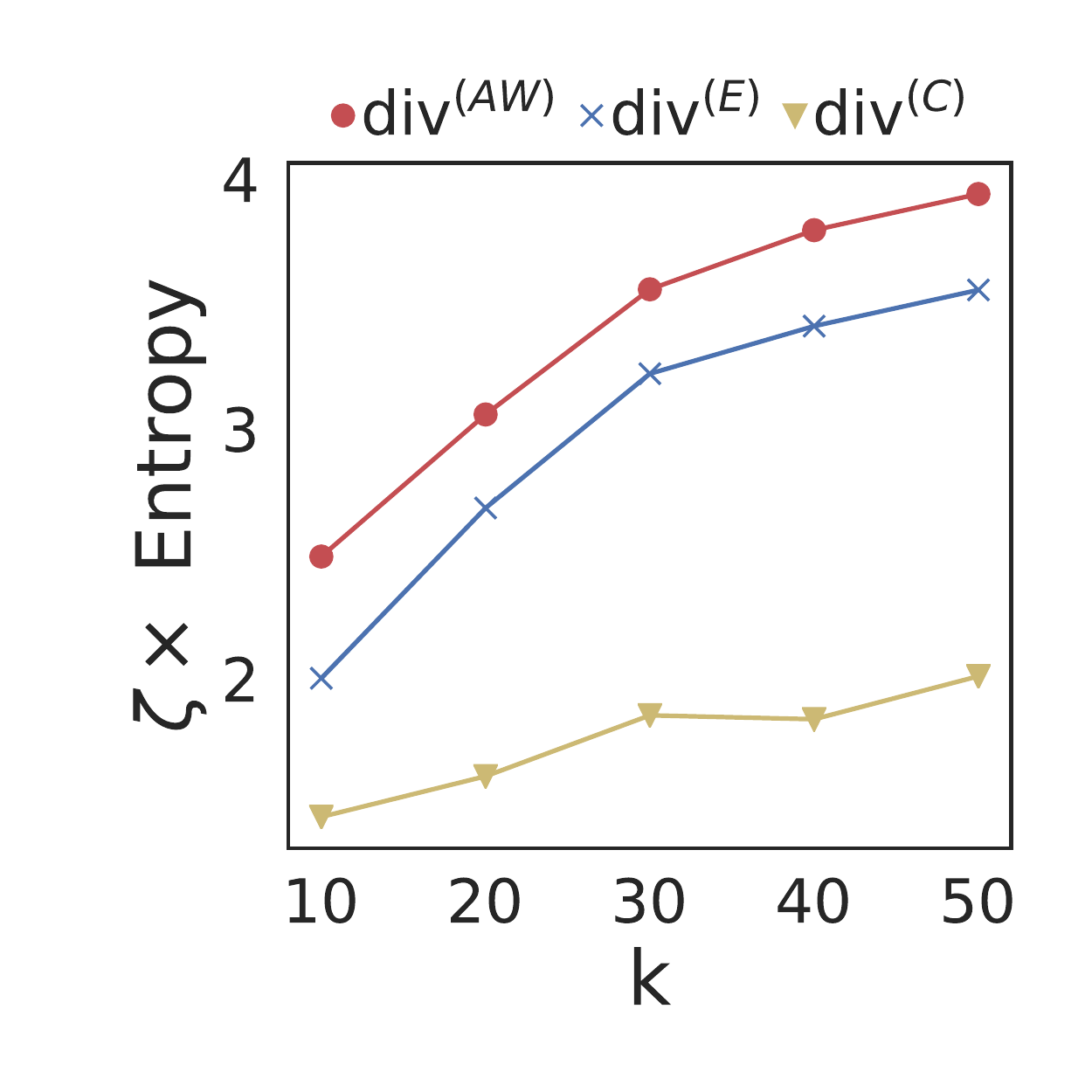} &
\includegraphics[width=.26\linewidth, height=3cm]
{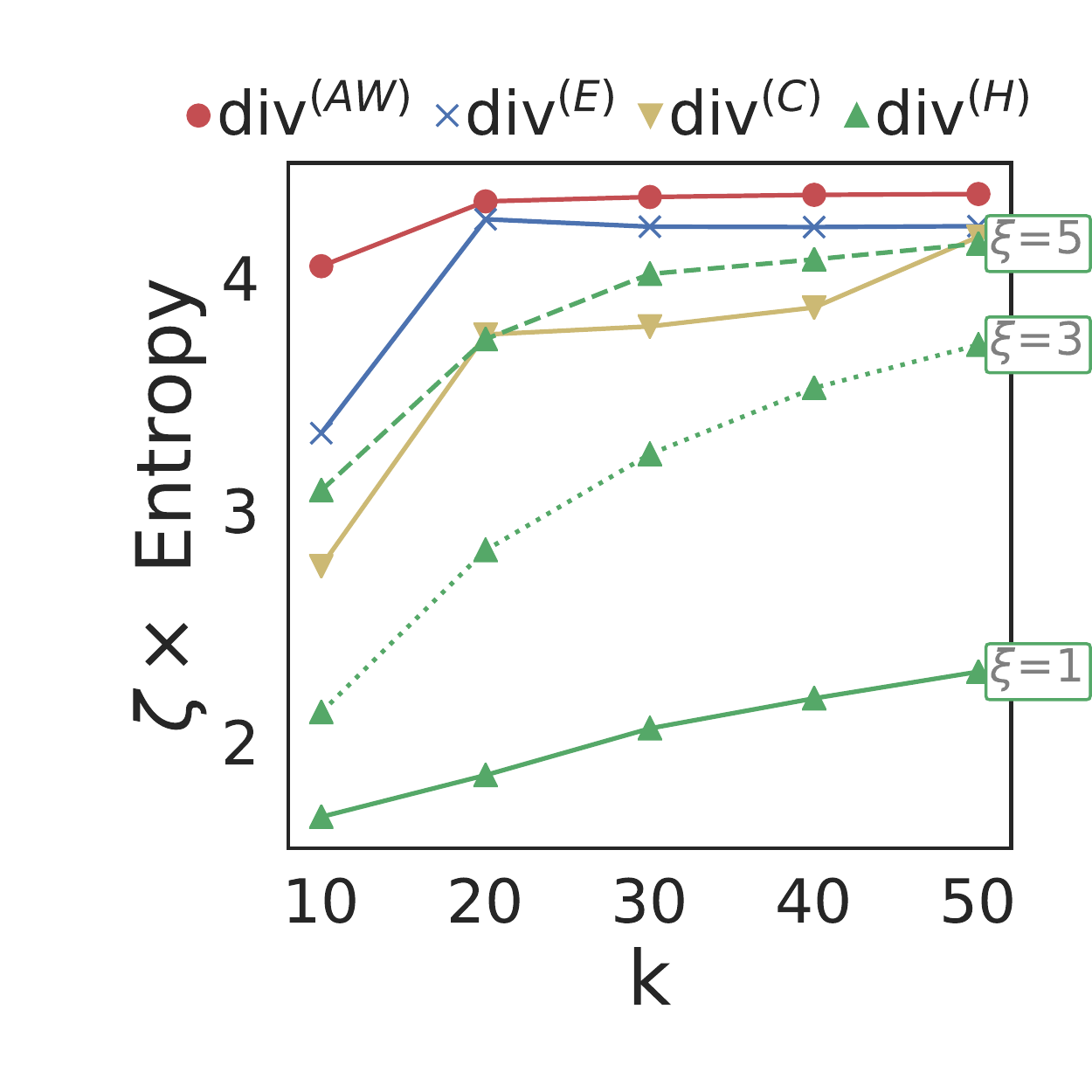} \\
 (c) Instagram  & (d) FriendFeed & (a) GooglePlus  & (b) Reddit \\ 
\end{tabular}
\caption{Entropy of the seed sets obtained by \myalgo for various diversity functions, with top-25\% target selection and
$\alpha=0$.}
\label{fig:entropy}
\end{figure}
%

\vspace{-1mm}
\subsection{Effect of the diversity functions} 
\label{sec:results:diversity}
To understand the impact of the diversity notion on the \myalgo performance, we  inspected the degree   of diversification induced by each of the functions   described in Sect.~\ref{sec:diversity}. 
 In particular,  we first measured the cross-entropy of the distribution of  attribute-values associated to the profile set of seeds, i.e., 
 $$Entropy(S) =\!\!\!\!\sum_{a \in dom(S)}  \frac{n_a}{\sum_{a' \in dom(S)} n_{a'}} \log\bigg(\frac{n_a}{\sum_{a' \in dom(S)} n_{a'}}\bigg).
 $$
Then, we multiplied the value of $Entropy(S)$  by a   factor   
 $\zeta = (1+\log(|dom|/$ $|dom(S)|))^{-1}$   that penalizes more for smaller  fraction of attribute-values covered by the profile set of $S$. 

Results shown in Fig.~\ref{fig:entropy} indicate that
 $div^{(AW)}$ generally yields seed sets with higher 
cross-entropy than the other diversity functions --- in fact,  to   maximize $div^{(AW)}$, \myalgo tends to favor a uniform distribution of the attribute-values over the seed set. 
Also, $div^{(AW)}$ achieves  higher coverage of the attribute domains (i.e., lower penalization factor). 
 The second best diversity function is the entropy-based one, $div^{(E)}$, which shows trends similar to $div^{(AW)}$. 
 
\begin{figure}[t!]
\centering
\begin{tabular}{ccc}
\includegraphics[width=.3\linewidth, height=3cm]{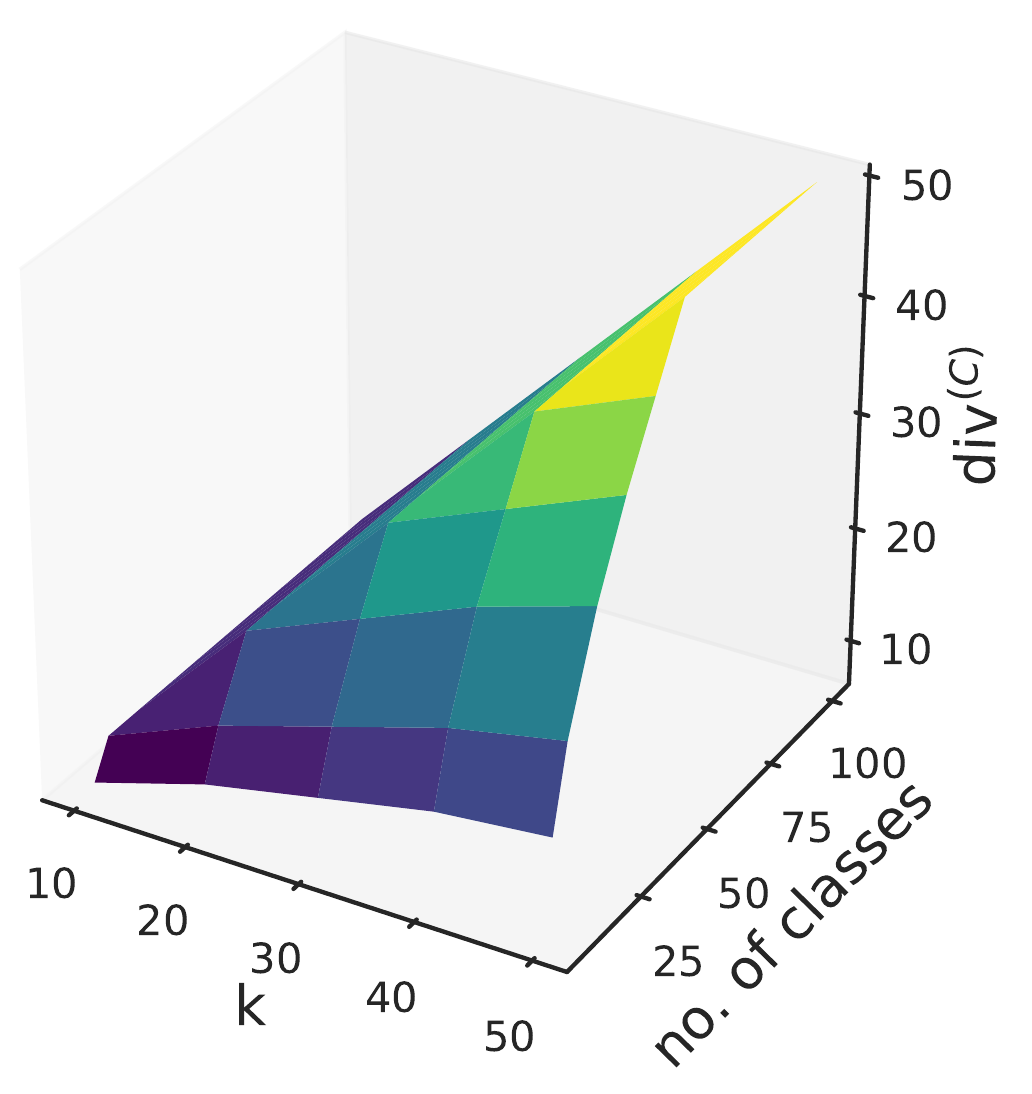} &
\includegraphics[width=.3\linewidth, height=3cm]
{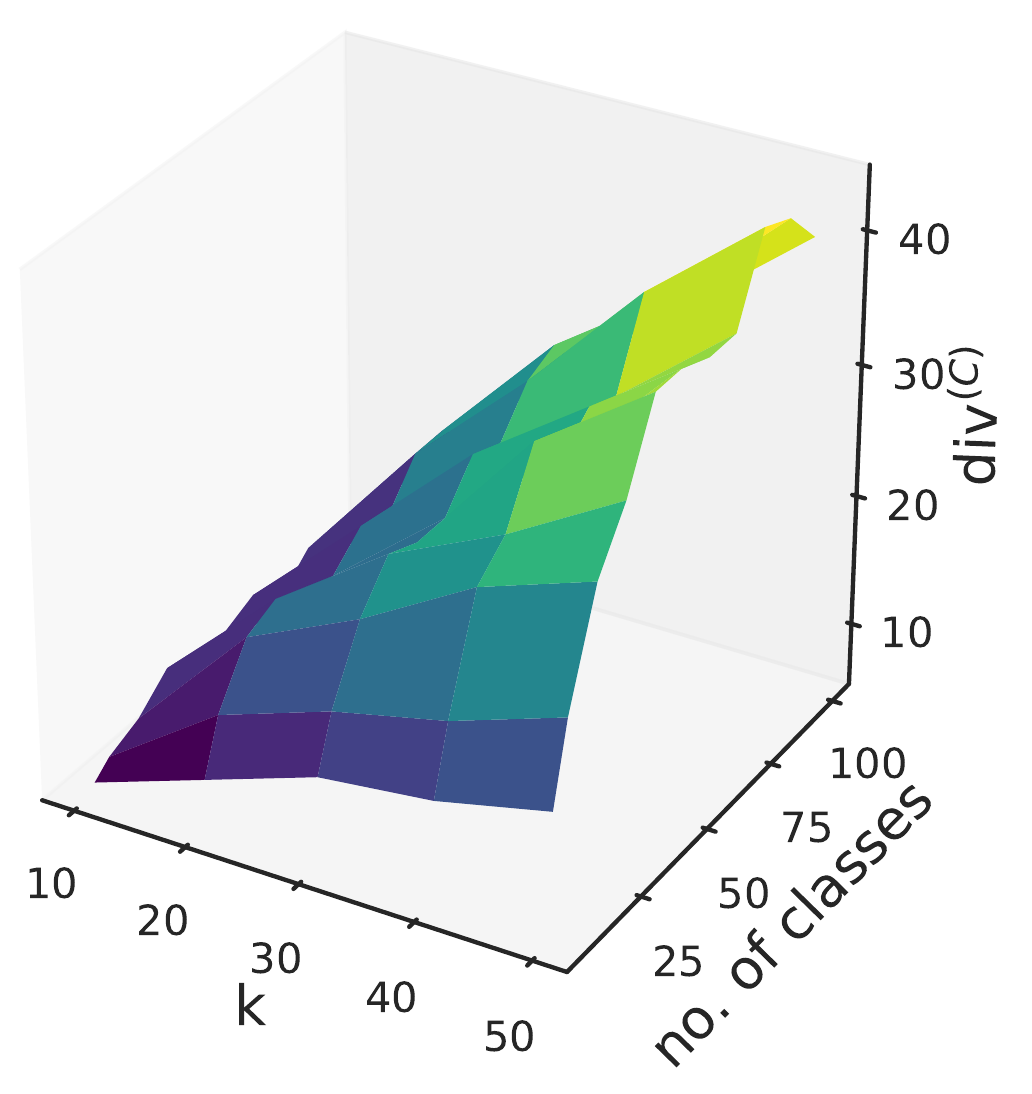} 
& 
\includegraphics[width=.3\linewidth, height=3cm]
{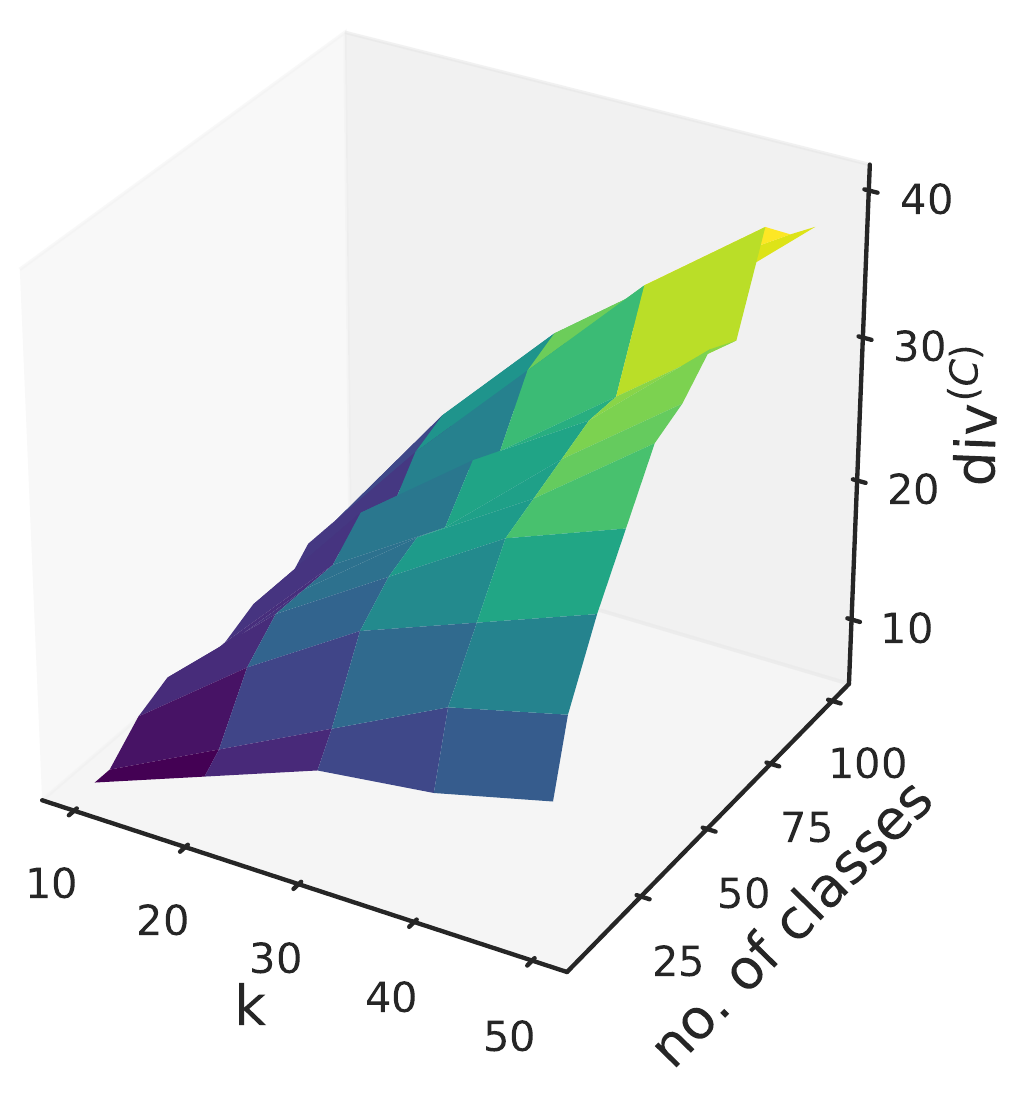}  \\ 
(a) $\alpha=0.0$ & (b) $\alpha=0.5$ & (c)  $\alpha=1.0$
\end{tabular} 
\caption{\textit{Class-based} diversity on Instagram by varying the number of
classes, $k$, and $\alpha$, with top-25\% target selection.}
\label{fig:surface}
\end{figure}

Conversely, $div^{(C)}$ and $div^{(H)}$ lead to less diversified seed sets. This is actually not surprising since the class-based notion of diversity relies on the grouping of the profiles (i.e., coarser grain than at attribute-value level) and it is maximized when all profiles in $S$ are chosen from different classes (i.e., $k\equiv h$, cf. Sect.~\ref{sec:diversity:class}), regardless of the distribution of their constituent attribute-values. 
   In this regard, we further investigated  how the combination 
of the budget $k$ and the number of classes (into which the profile set   is partitioned) affects the diversity value. 
  Fig.~\ref{fig:surface} shows that $div^{(C)}$ increases  more rapidly with the increase in the number of classes   w.r.t.  $k$. 

Also,   the Hamming-based diversity, $div^{(H)}$, 
 consistently behaves  worse 
than $div^{(AW)}$ and $div^{(E)}$, while it is comparable  to $div^{(C)}$ for higher radius. 
  Indeed,   $div^{(H)}$ strongly depends on the setting of the radius: as expected, the diversity increases for higher  values of the radius $\xi$. 
  This is explained since 
Eq.~(\ref{def:hamming-div}) increases as the union of the Hamming balls of the nodes in   the seed set grows; however, setting $\xi=1$ leads to Hamming balls containing nodes that are not really 
different from each other. As a consequence, Eq.~(\ref{def:hamming-div}) would to be deceived because
a huge Hamming ball may corresponds to a poorly diversified seed set.

In the rest of the result presentation, we will refer to the attribute-wise diversity only. Our justification is that $div^{(AW)}$   (i) has shown  effectiveness in the diversification of the seed set that is as good as or better than $div^{(E)}$, while outperforming $div^{(C)}$ and $div^{(H)}$,  
(ii) it allows marginal gain computation that is  clearly more efficient than the conditional entropy computation required in $div^{(E)}$, 
and (iii) it does not depend from additional a-priori knowledge like  $div^{(C)}$ does, or parameters like $div^{(H)}$ does.

\vspace{-1.5mm}
 \subsection{Evaluation of identified seed sets}
\label{sec:results:seedset}
\vspace{-0.5mm}
Here we discuss how the different settings of parameters in \myalgo,  particularly $\alpha$ and the attribute distributions, affect the   seed identification.  
 
\begin{figure}[t!]
\begin{tabular}{@{\hskip -1mm}c@{\hskip -0.6mm}c@{\hskip -0.6mm}c@{\hskip -0.6mm}c}
\includegraphics[width=.26\linewidth]{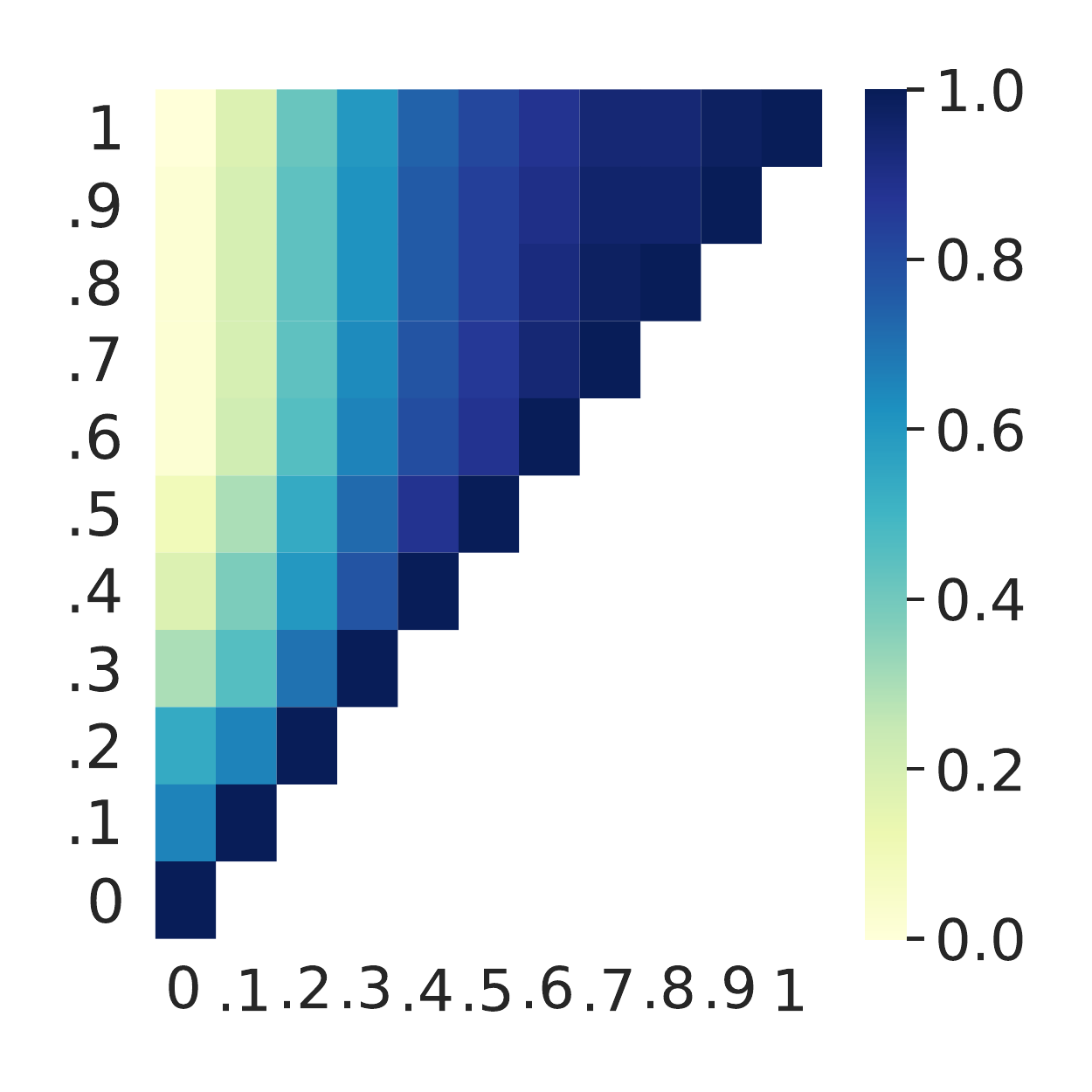} & 
\includegraphics[width=.26\linewidth]{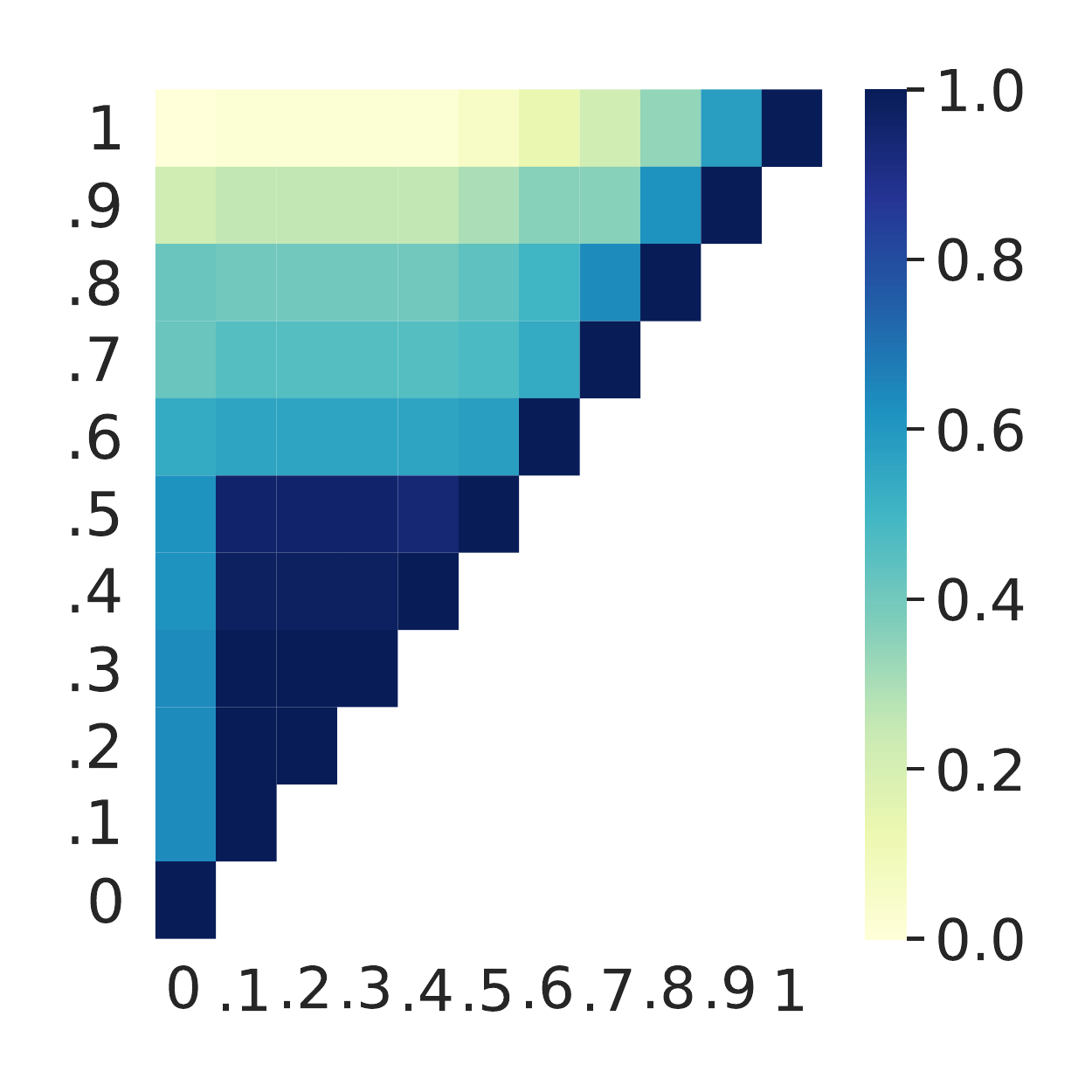} &
\includegraphics[width=.26\linewidth]{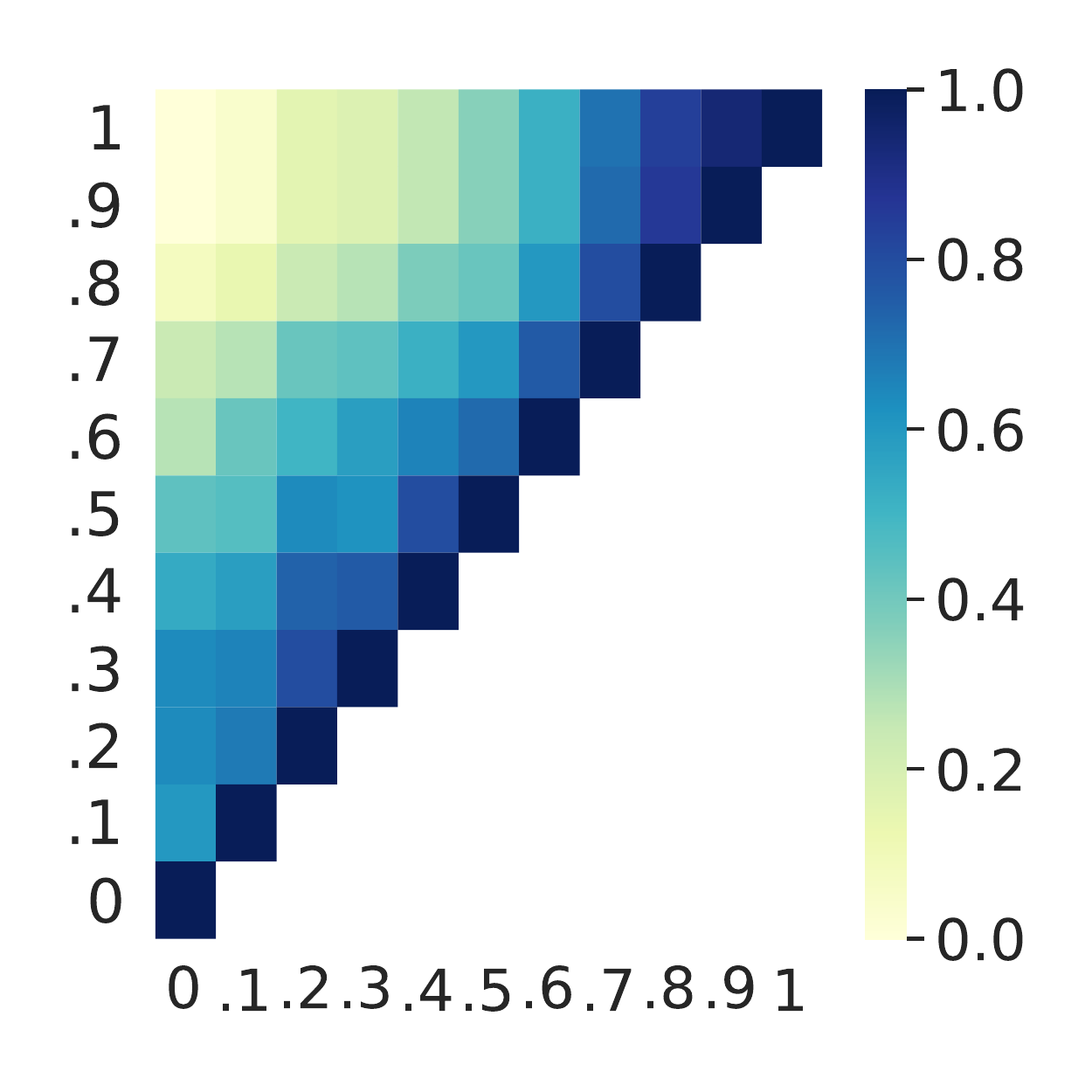}  &
\includegraphics[width=.26\linewidth]{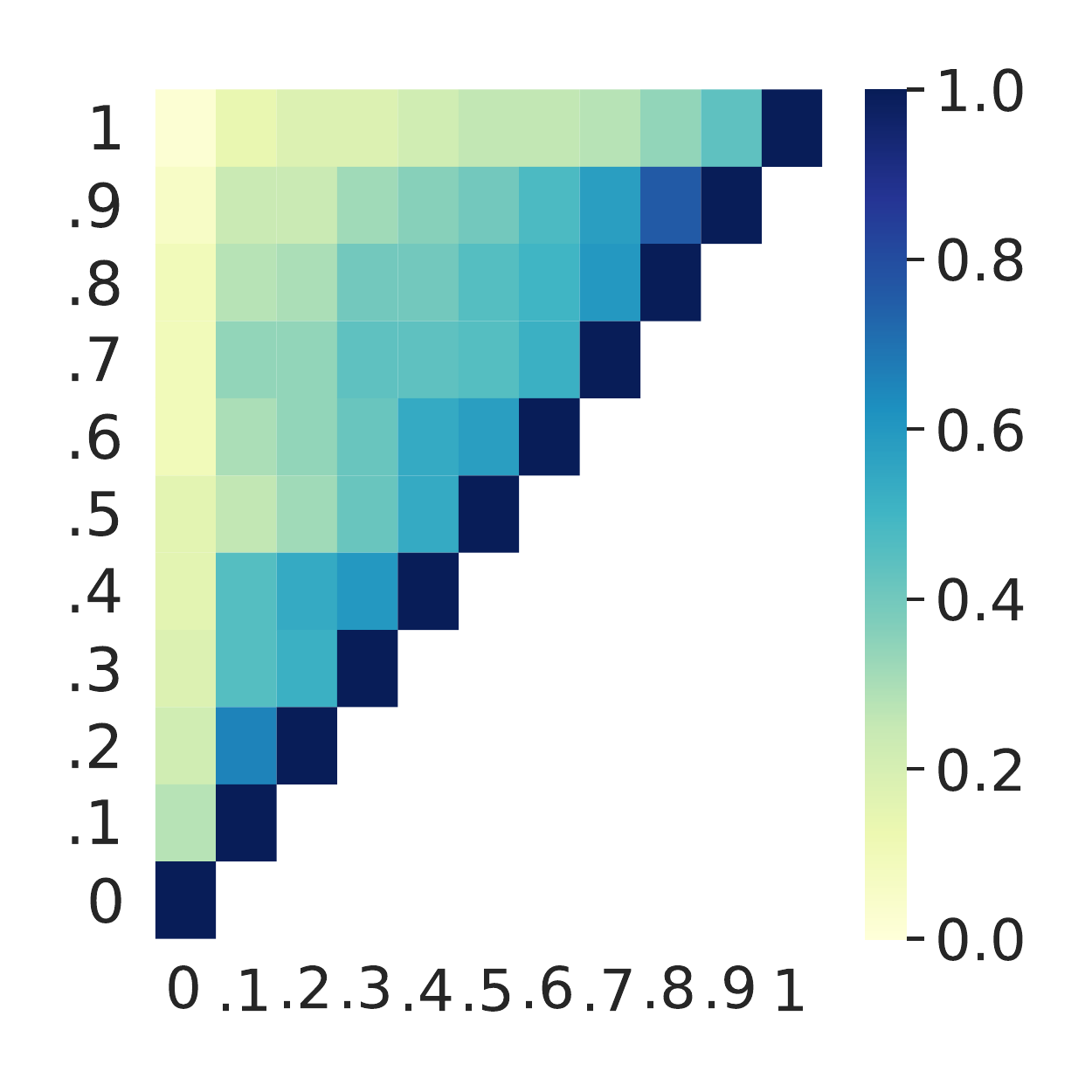} \\
(a) Instagram  & (b) FriendFeed  & (c) GooglePlus & (d) Reddit \\
\end{tabular}
\caption{Normalized overlap of seed sets, for   $\alpha \in [0,1]$ (with increments of  $0.1$),  $k=50$, top-25\%  target selection, and exponential distribution of attributes (except Reddit). }
\label{fig:seeds_overlap_exp_distribution}
\end{figure}

\emph{\underline{Sensitivity to $\alpha$}.\ }  
Heatmaps in Fig.~\ref{fig:seeds_overlap_exp_distribution} show the pairwise   overlaps of   seed sets, normalized by $k$, for varying $\alpha$.    
 %
 Focusing first on the overlaps between the seed set corresponding to $\alpha=1$ (i.e., capital contribution only) and the ones corresponding to diversity at different degrees ($\alpha < 1$), the overlap decreases rapidly for lower $\alpha$. (This trend is less evident for Instagram because of its tighter connectivity than  FriendFeed, GooglePlus and Reddit, as in fact it corresponds to the maximal strongly connected component of the original network graph~\cite{8326536}).    
 While in general overlaps always change for pairs of seed sets corresponding to different settings of $\alpha$, it appears that the fading of overlaps becomes more gradual on networks with stronger small-world characteristics (i.e., GooglePlus).  
 Moreover, results (shown in \textbf{\em Appendix}, Fig.~\ref{fig:seeds_overlap_exp_distribution-app}) obtained at top-5\% and top-10\% target selection, also confirm the variability in the seed set overlap, which is again more evident on the larger networks. 
%

\begin{figure*}[t!]
\centering 
\begin{tabular}{ccc}
\hspace{-4mm}
\includegraphics[width=.28\linewidth, height=3cm]{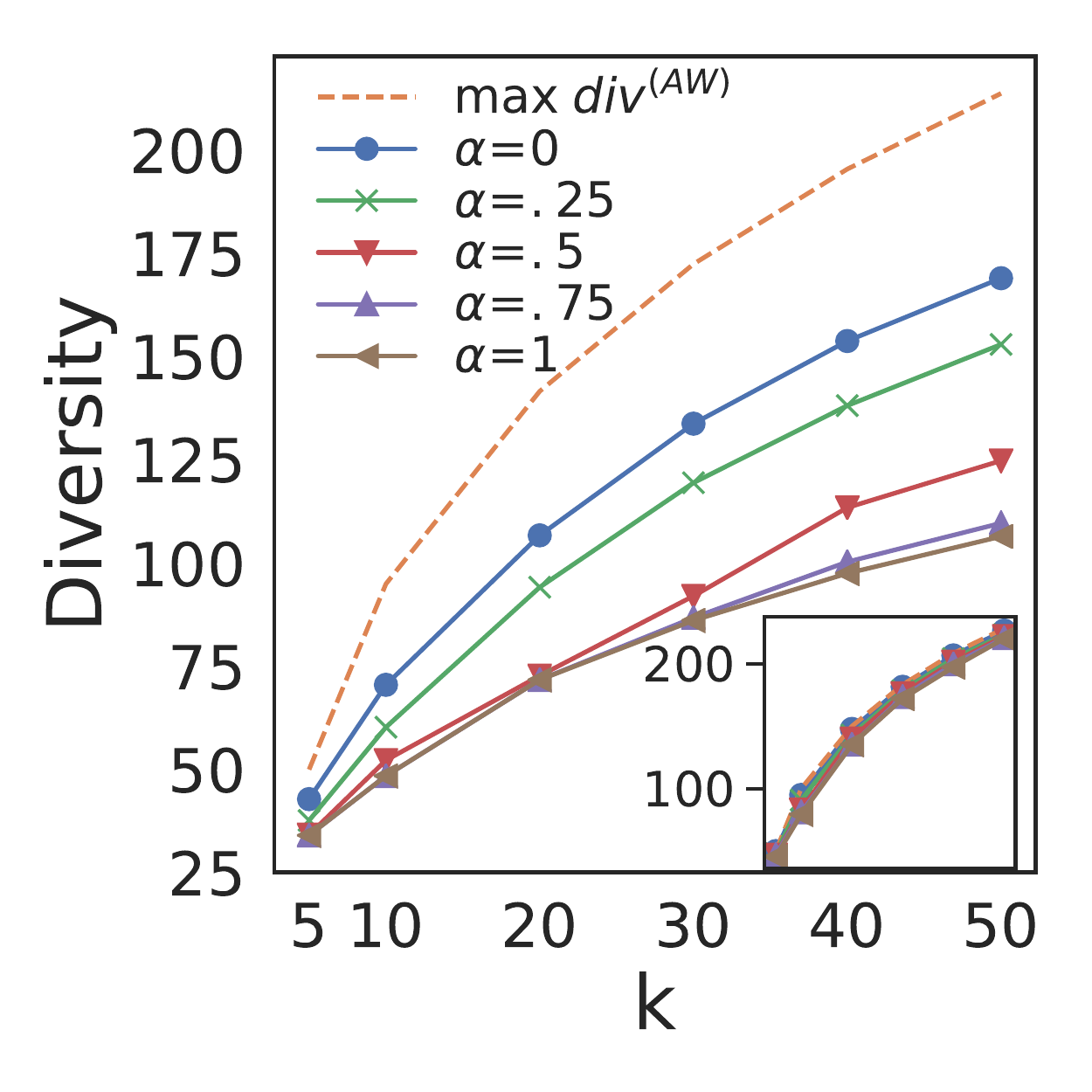} & 
\hspace{-4mm}
\includegraphics[width=.28\linewidth, height=3cm]{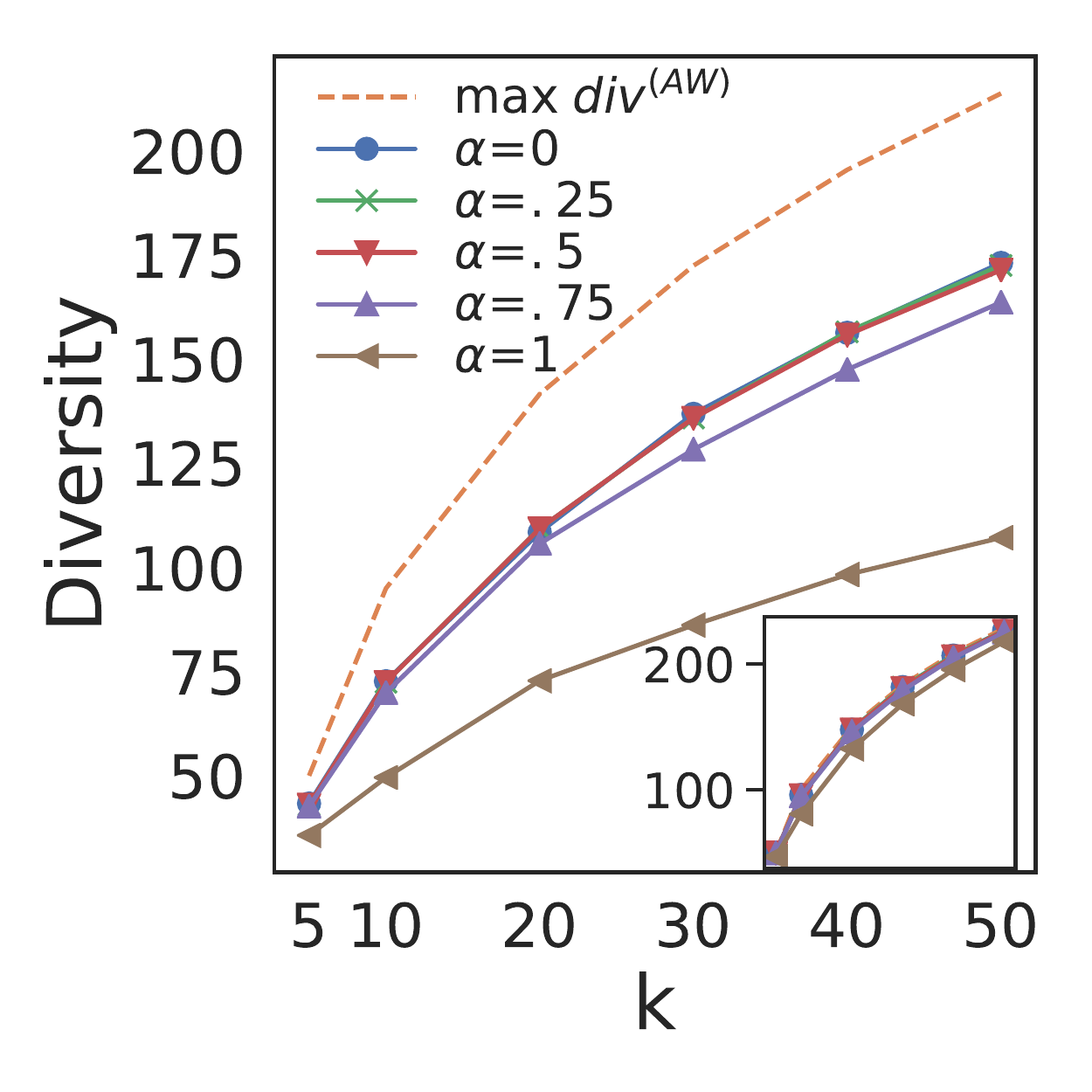} &
\hspace{-4mm}
\includegraphics[width=.28\linewidth, height=3cm]{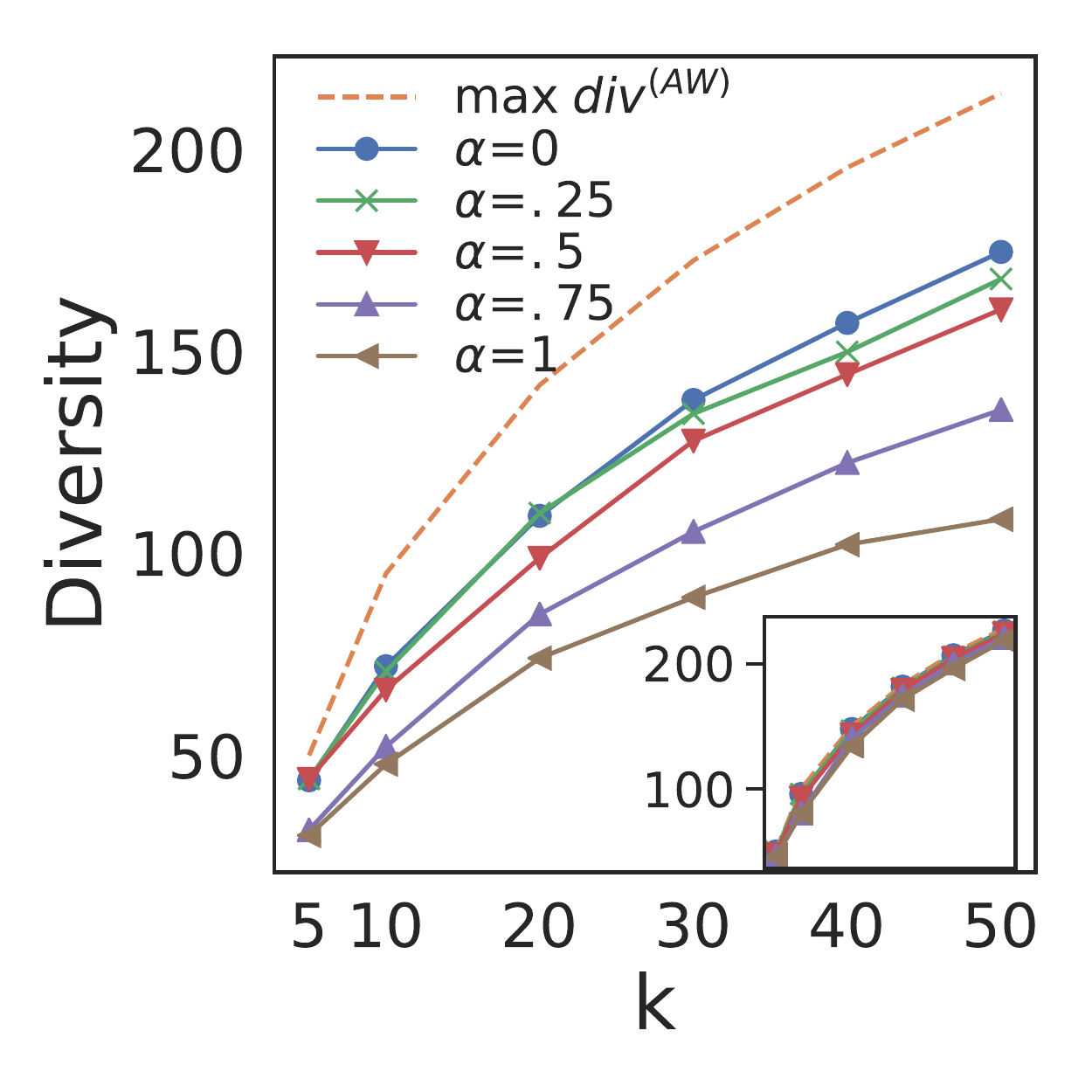} \\
  (a) Instagram &  (b) FriendFeed  &  (c) GooglePlus  
\end{tabular}
\caption{Exponential (main) vs. uniform (inset) distribution:  attribute-wise of seed set  for varying $k$ and $\alpha$, top-25\%  target selection, and  comparison to  maximum diversity value.}
\label{fig:diversity_line_plot}
\end{figure*}

\emph{\underline{Effect of the attribute distribution.\ }} 
The previous analysis refers to   exponential distribution of the   attributes. 
 We observed however that the sensitivity of \myalgo to the setting of $\alpha$ becomes much lower when a uniform distribution law is adopted.  
This prompted us to   investigate the reasons underlying this behavior. To this end, 
  we compared the diversity value associated to each seed set, 
by varying $\alpha$ and  distributions, with the maximum possible value
$div^*[k]$  (Eq.~\ref{eq:diversity_max_possible_value}); this is  achieved when all the attribute values are equally distributed over the seeds. 

Not surprisingly, looking at the insets  of Fig.~\ref{fig:diversity_line_plot} that correspond to uniform distribution,  we observe that the trends of seed-set diversity at varying $\alpha$ are all close to each other as well as to the maximum value.  By contrast, using exponential distributions (main plots  of Fig.~\ref{fig:diversity_line_plot}), it is evident that the slope of the diversity tends to decrease with higher $\alpha$, thus increasing the  gap with the maximum diversity curve.  
%
Moreover, different settings of the target selection threshold   have no significant impact on the   trends already observed for top-25\% (results   shown in \textbf{\em Appendix}, Fig.~\ref{fig:diversity_line_plot-app}).   
 In the following,   results   correspond  to exponential  distribution of the attributes, unless otherwise specified. 



\vspace{-3mm}
\subsection{Comparison with \algo{DTIM}}
\label{sec:results:comparison_DTIM}
\vspace{-2mm}

\begin{figure}[t!]
\centering
\begin{tabular}{@{\hskip -1.1mm}c@{\hskip -0.6mm}c@{\hskip -0.6mm}c@{\hskip -0.6mm}c}
\includegraphics[width=.26\linewidth, height=2.8cm]{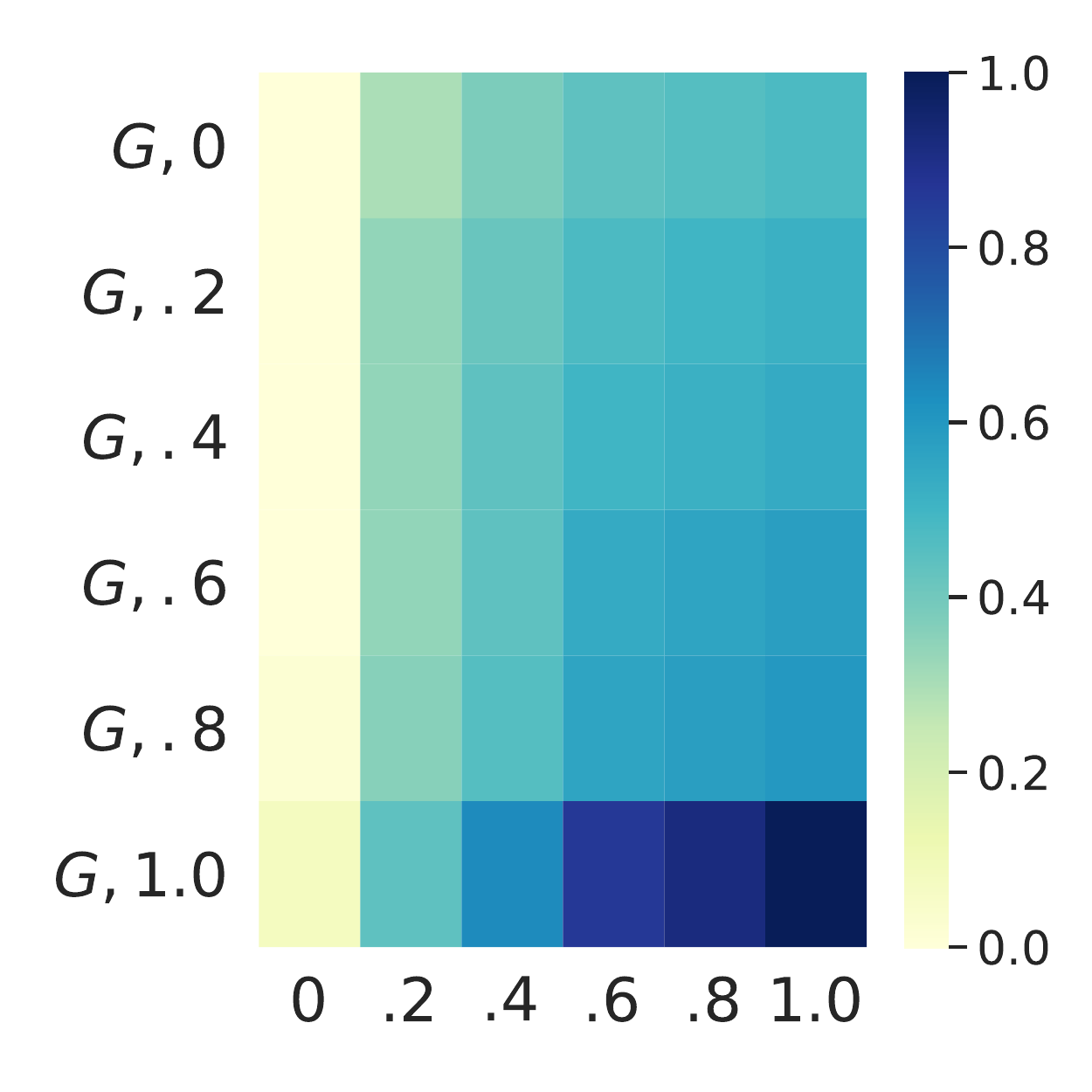} &
\includegraphics[width=.26\linewidth,  height=2.8cm]{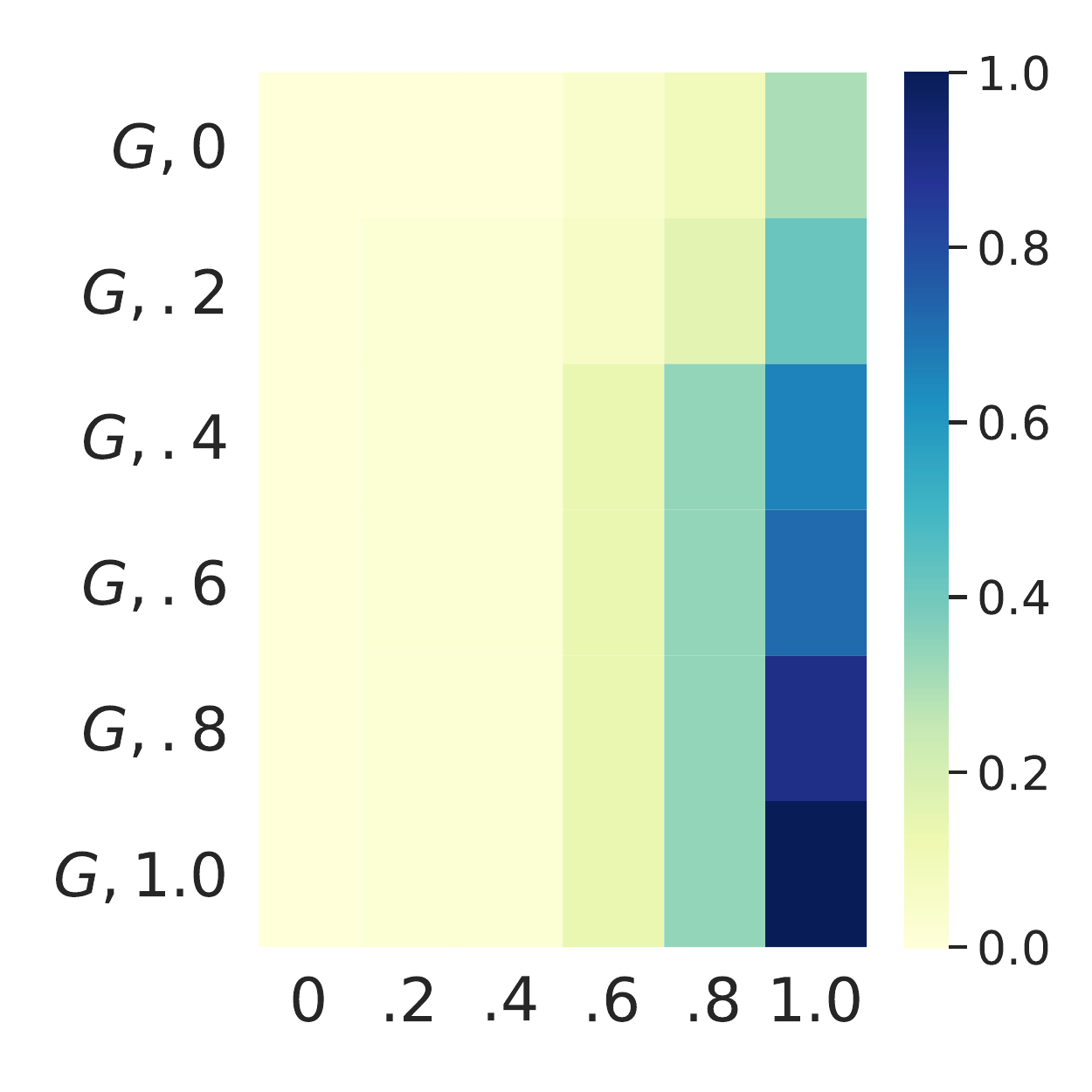} &
\includegraphics[width=.26\linewidth,  height=2.8cm]{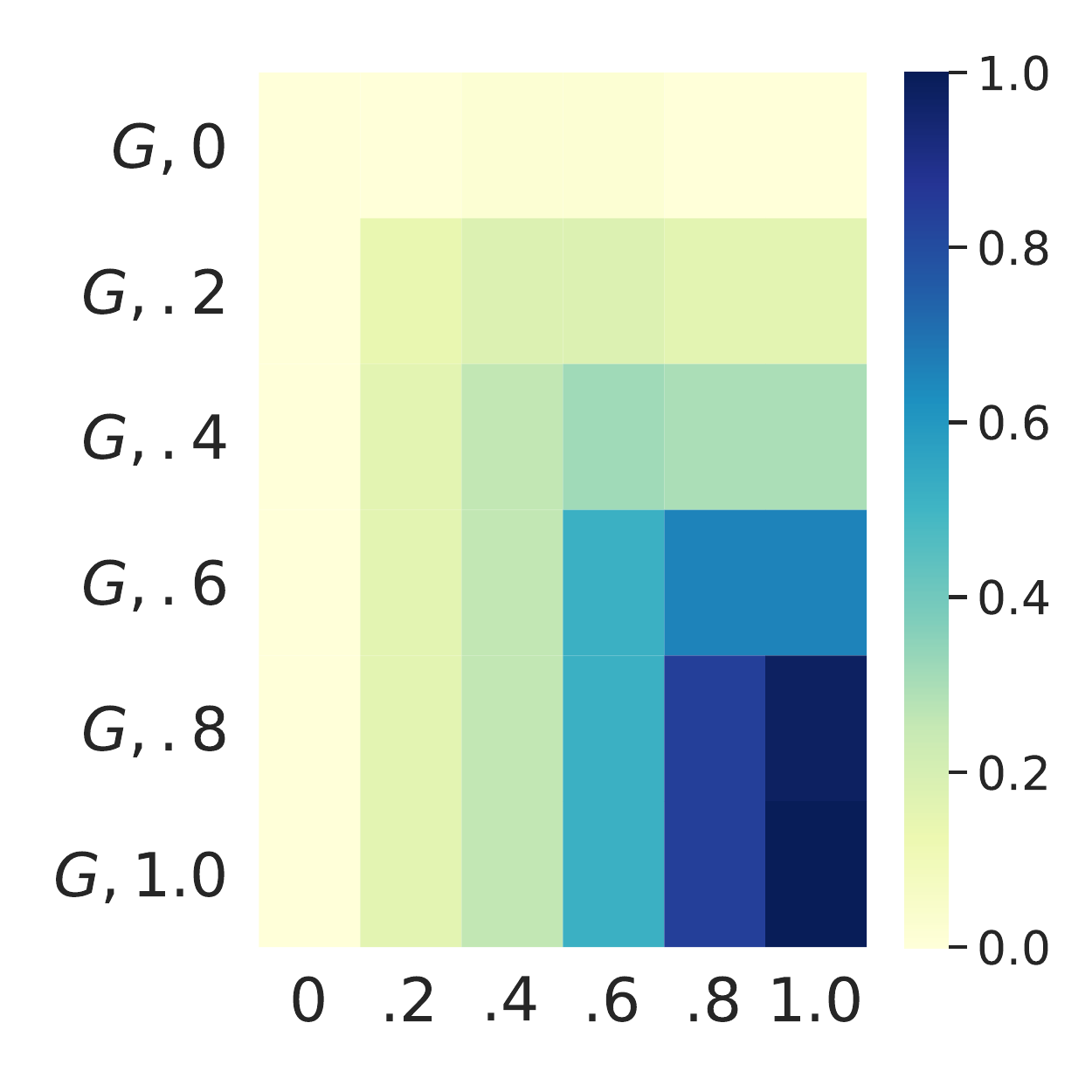} &
\includegraphics[width=.26\linewidth,  height=1.1in]{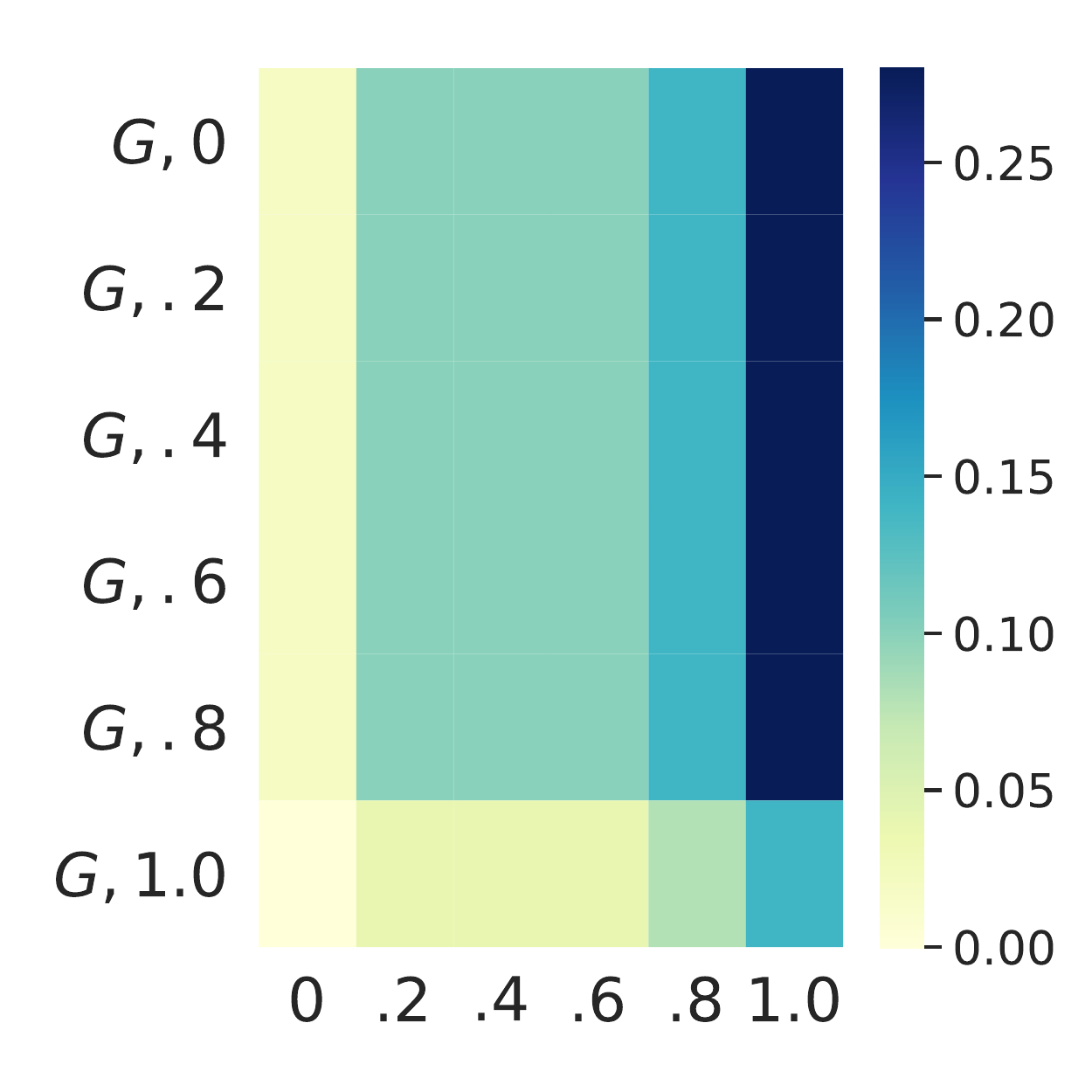} \\
  (a) Instagram  &  (b) FriendFeed & (c)  GooglePlus &  (d) Reddit \\
%
\end{tabular}
\caption{Topology-based vs. attribute-based diversity: Normalized overlap of seed sets, for selected values of $\alpha$, $k=50$, and top-25\% target selection.}
\label{fig:seeds_overlap_diff_diversity}
\vspace{-2mm}
\end{figure}

{\bf Stage 1:\ }
 We  first evaluated   the  integration of the topology-driven 
 diversity function~\cite{8326536} into our RIS-based framework.  
 We analyzed   the normalized overlap of seed sets obtained by \myalgo  and by the resulting \algo{DTIM}-based variant. 
  Figure~\ref{fig:seeds_overlap_diff_diversity} shows low-mid lack of normalized overlap  between compared seed sets;  
 in particular, overlap is  much closer to zero   for the largest networks, which are also sparser (and hence, more realistic) than Instagram network.  

\begin{figure}[t]
\centering
\begin{tabular}{@{\hskip -1mm}c@{\hskip -0.6mm}c@{\hskip -0.6mm}c}
\includegraphics[width=.28\linewidth, height=3cm]{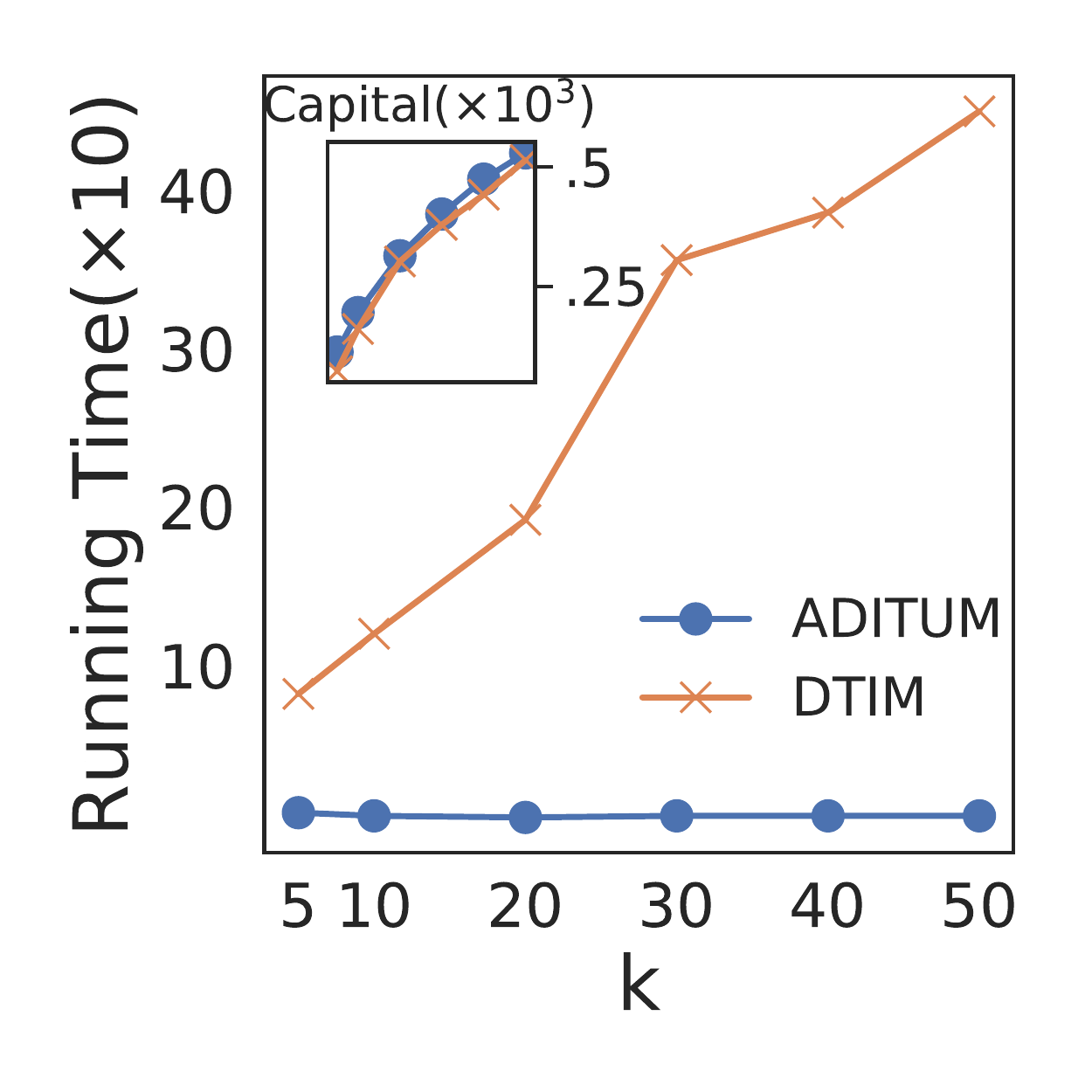} &
\includegraphics[width=.28\linewidth, height=3cm]{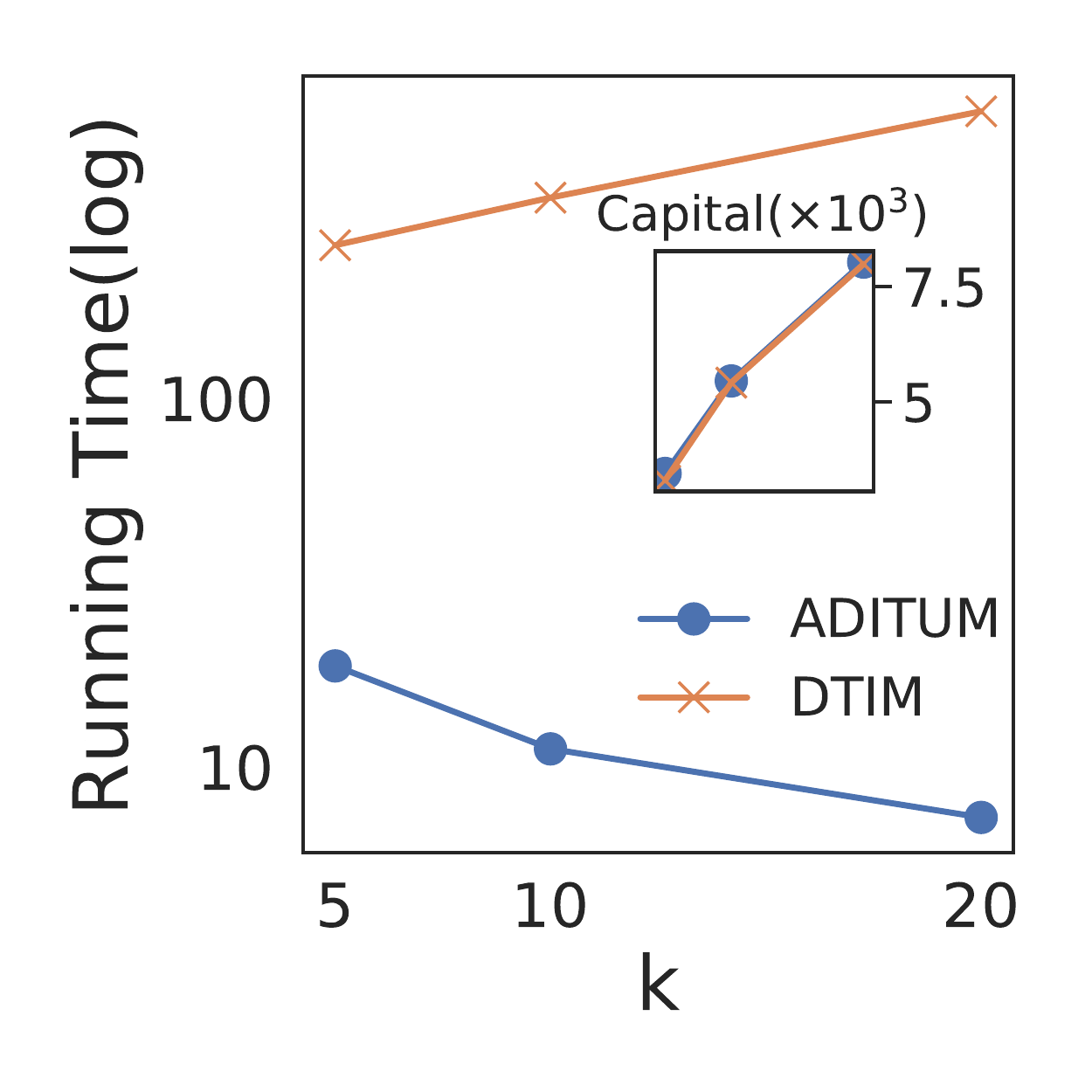} &
\includegraphics[width=.28\linewidth, height=3cm]{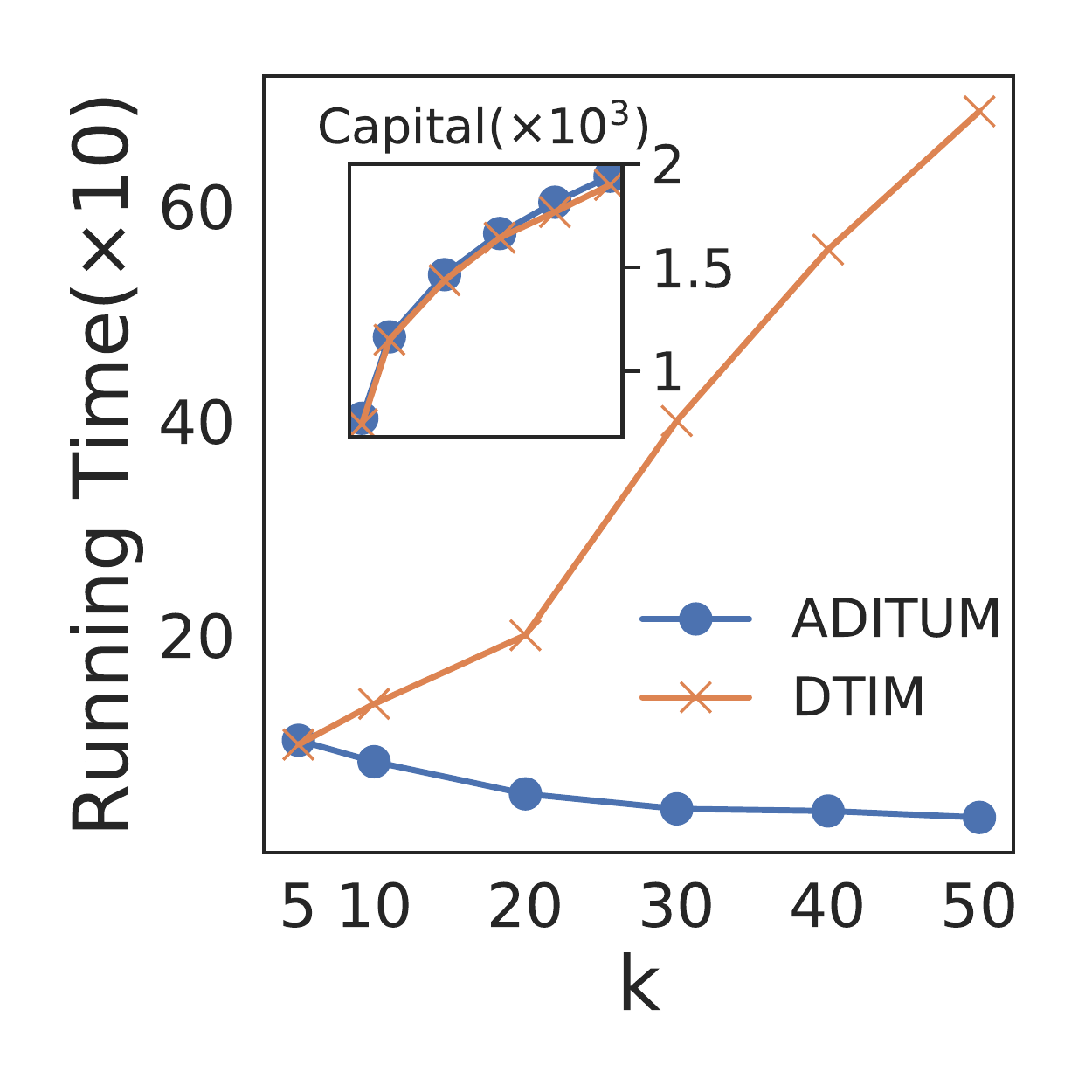} \\
  (a) Instagram  &  (b) FriendFeed &  (c) GooglePlus  \\
\end{tabular}
\caption{\myalgo ($\epsilon=0.1$)  vs. \algo{DTIM} ($\eta=10^{-4}$): Running time in seconds (main plot) and expected capital (inset) for varying $k$,
top-25\% target selection and $\alpha=1$. }
\label{fig:aditum_vs_dtim}
\vspace{-2mm}
\end{figure}

{\bf Stage 2:\ }
In the second stage of   evaluation, we compared  \myalgo and \algo{DTIM} in terms of the  expected capital. In 
Fig.~\ref{fig:aditum_vs_dtim}, the insets  show  results of a Monte Carlo simulation
 (with 10\:000 runs) for the estimation of the capital associated with the seed sets provided by each of the methods with
 $\alpha=1$ (i.e., without the diversity contribution). Also, we set  $\eta=10^{-4}$ for \algo{DTIM}, which means minimal path-pruning, and hence  highest estimation accuracy for the competitor.  
 We observe  that \myalgo keeps a relatively small advantage over \algo{DTIM} as for the estimated capital. 
 Nonetheless, it should be emphasized that, as expected from a comparison between a  RIS-based method and a greedy method, \myalgo outperforms \algo{DTIM} in terms of running time, up to   3 orders of magnitude (e.g., in FriendFeed with $k\geq10$), and this gap becomes even more evident as both $k$ and the network size increase. 
  Note   that, while the running time of \algo{DTIM} tends to increase linearly in $k$, for \myalgo it may even decrease with $k$: likewise \algo{TIM+}, this is a result of the interplay of the main factors that determine the number of random RR-Sets. 
%

\vspace{-1mm}
\subsection{Comparison with \algo{Deg-D} diversity and attribute representation}
\label{sec:results:comparison_Tang}
\vspace{-1mm}
%
As concerns the comparison with \algo{Deg-D}, we again devised two stages of evaluation: 
 (1) comparison of seed sets produced by \myalgo and by \algo{Deg-DU}/\algo{Deg-DW}, and (2) adaptation of our RIS framework to  numerical-attribute diversity   used by \algo{Deg-D} (cf. Sect.~\ref{sec:eval}-Setting). 

\begin{figure}[t]
\begin{minipage}[t]{0.49\linewidth}
\begin{tabular}{cc}
\hspace{-4mm}
	\includegraphics[width=.52\linewidth, height=1.1in]{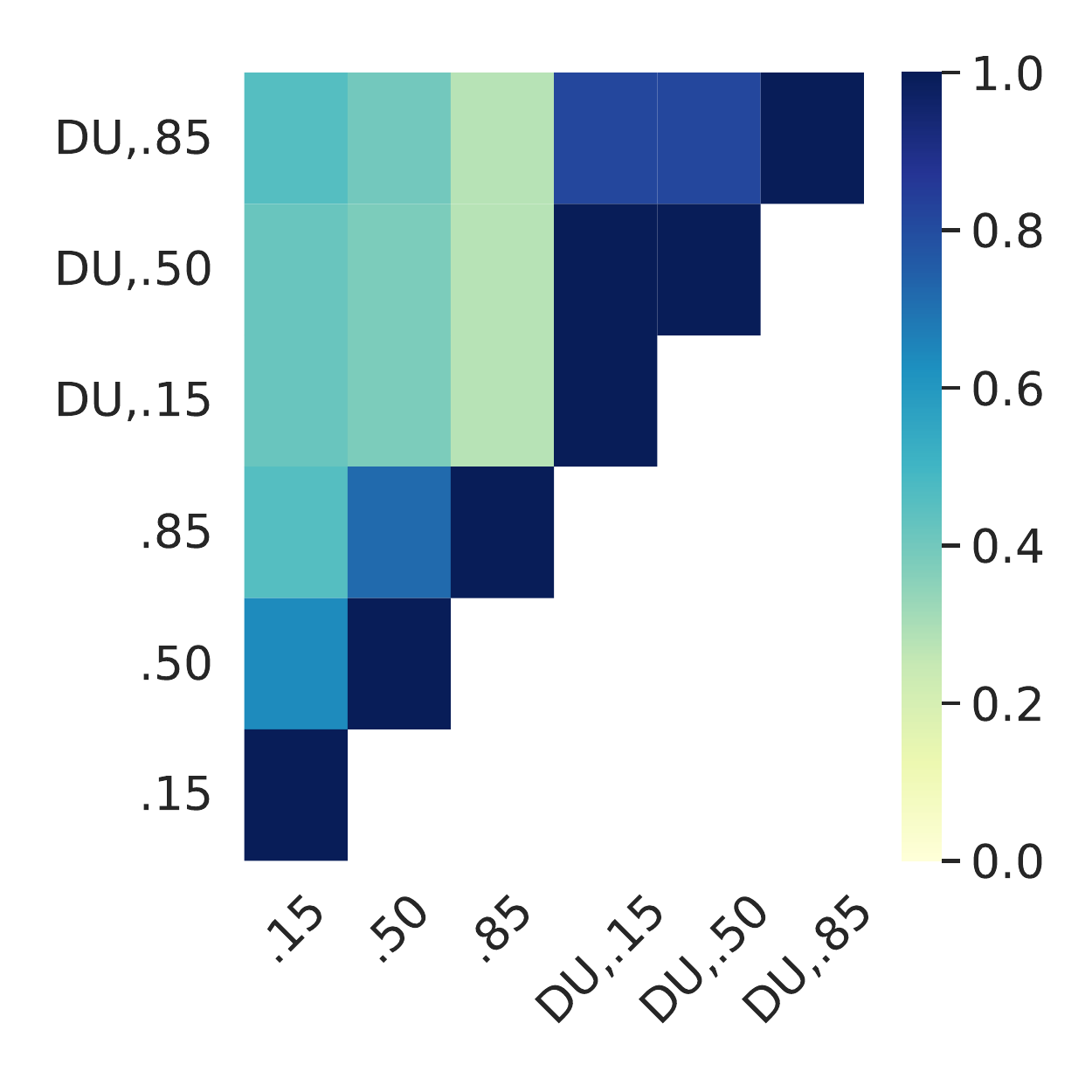} &
\hspace{-9mm}
	\includegraphics[width=.52\linewidth, height=1.1in]{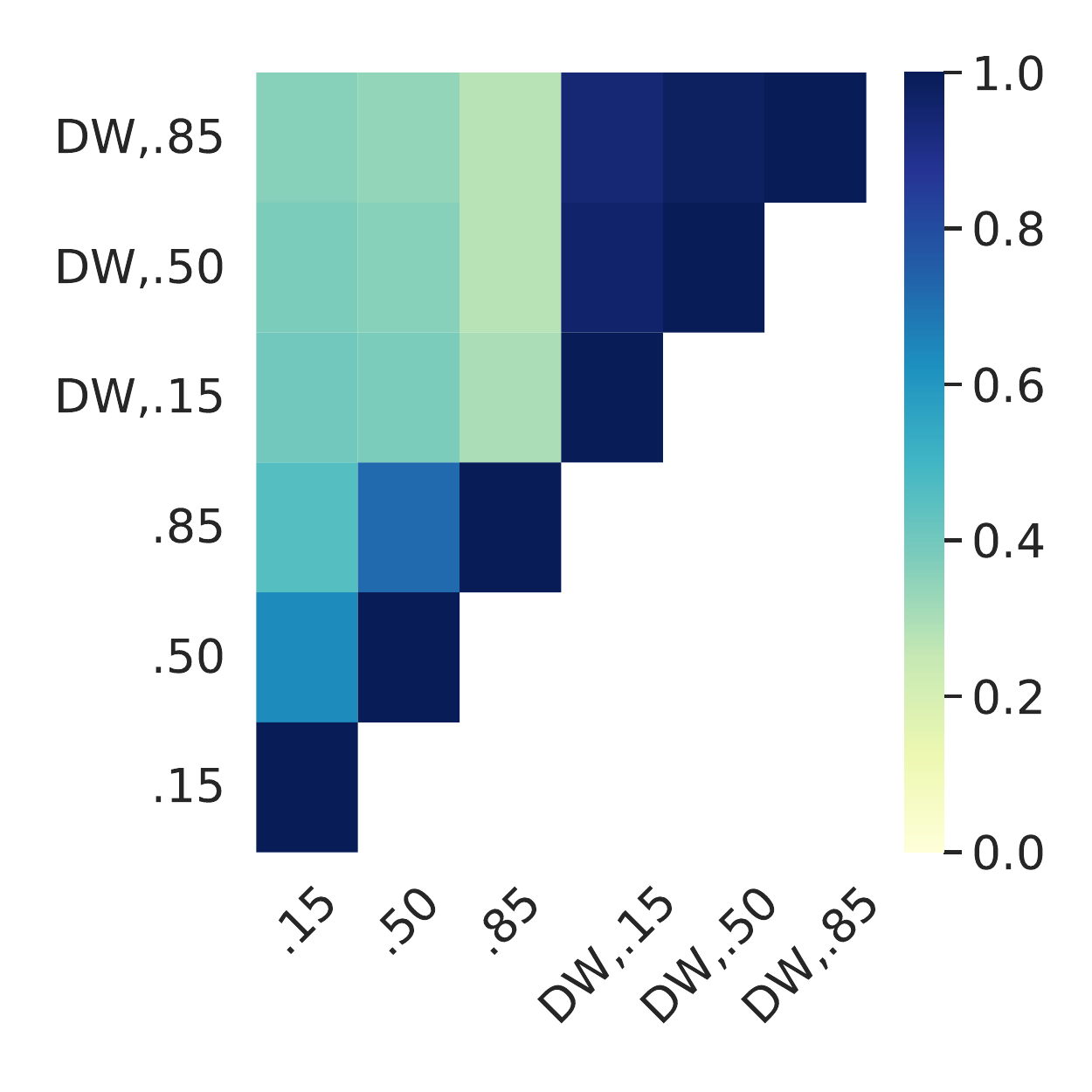} 
\end{tabular}
\vspace{-2mm}
\caption{\myalgo vs. \algo{Deg-DU} (left) and  \algo{Deg-DW} (right):  Normalized overlap of seed sets, for $\gamma \in \{ 0.15, 0.5, 0.85 \}$,  $k=50$, and top-100\% target selection, on MovieLens.}
\label{fig:tang_seeds_overlap}
\end{minipage}
    \hfill     
\begin{minipage}[t]{0.49\linewidth}
\begin{tabular}{@{\hskip -3mm}c@{\hskip -2mm}c}

	\includegraphics[width=.52\linewidth]{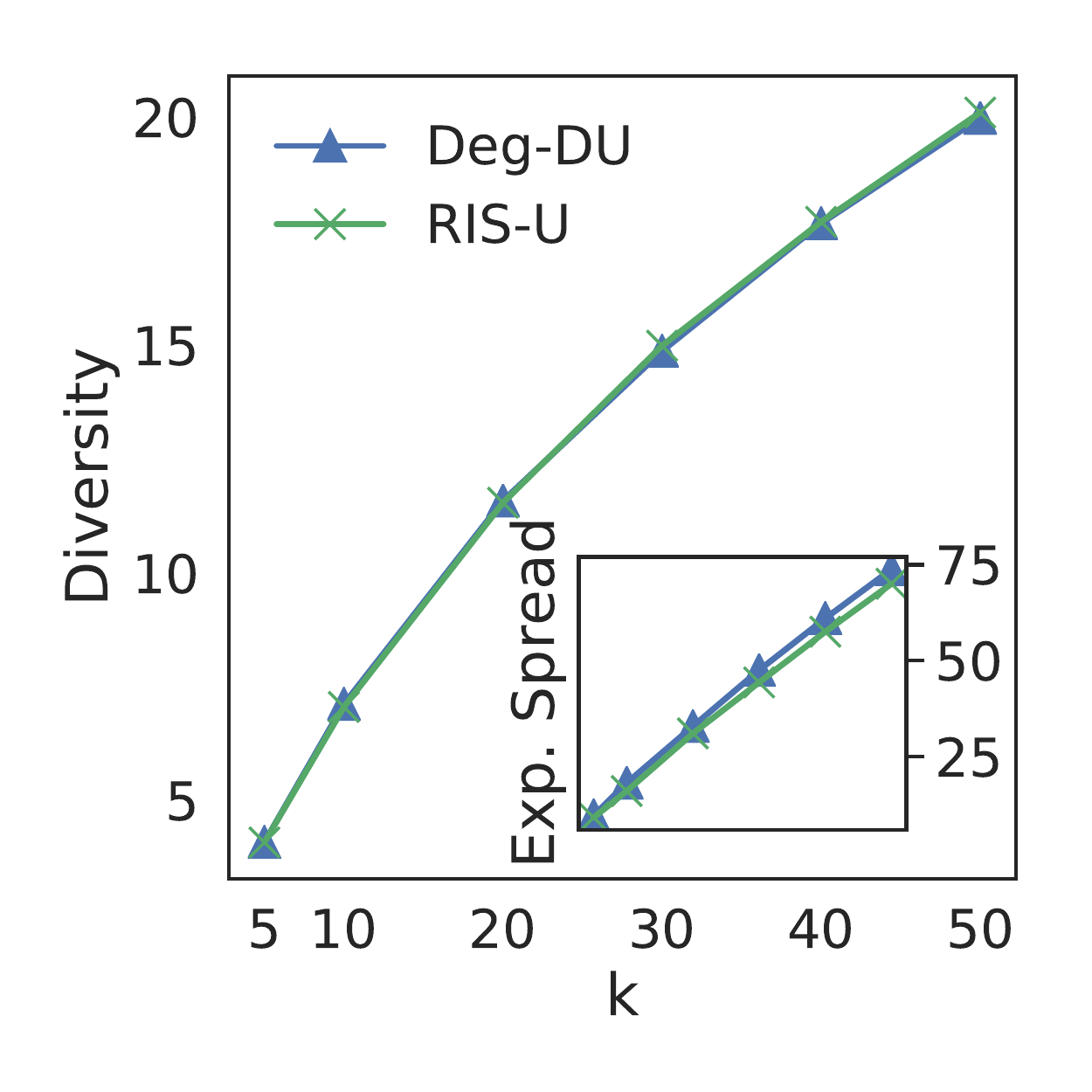} &
	\includegraphics[width=.52\linewidth]{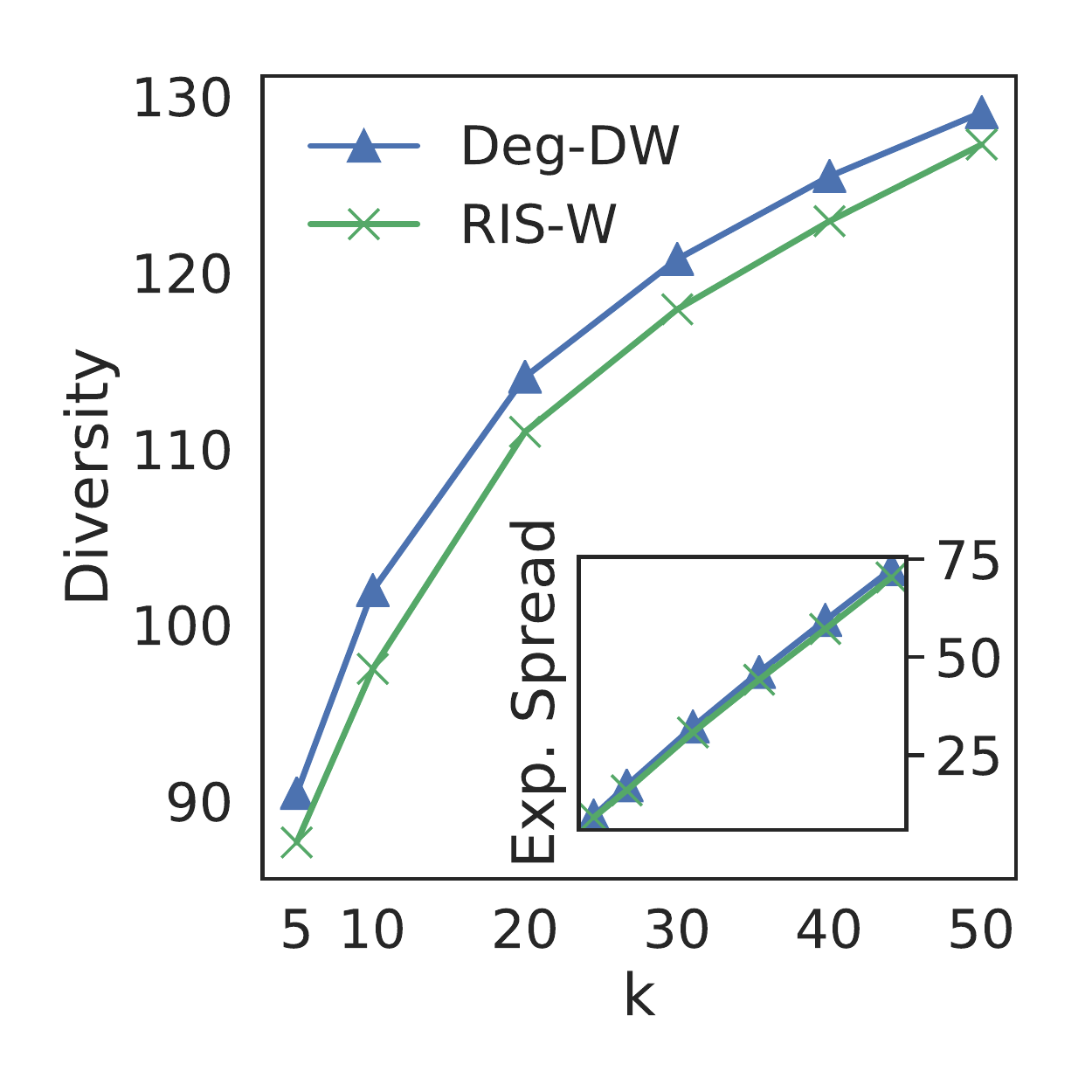} 
\end{tabular}
\vspace{-3mm}
\caption{\algo{Deg-DU} vs. \algo{RIS-U} (left) and \algo{Deg-DW} vs. \algo{RIS-W} (right):  Expected spread (inset) and seed set diversity  by varying $k$, for $\gamma=0.5$, on MovieLens. }
	\label{fig:tang_comparison_line_plot}
\end{minipage}
\end{figure}

 {\bf Stage 1:\ }
   Fig.~\ref{fig:tang_seeds_overlap} shows  the normalized overlaps of seed sets. Two main remarks can be drawn: first, 
  the overlaps between \myalgo and \algo{Deg-D} are always quite low ($0.28\sim 0.43$), and second,  the setting of $\gamma$ (i.e., $1-\alpha$) has   little effect on \algo{Deg-D}. 

 {\bf Stage 2:\ }
   Fig.~\ref{fig:tang_comparison_line_plot} refers to   numerical attribute representation and integration of \algo{Deg-DU} and \algo{Deg-DW} functions into our framework,   denoted as \textit{RIS-U} and \textit{RIS-W}.  We set $\gamma = \alpha = 0.5$ to equally balance the contributions of diversity and spread in the methods' objective function. 
  We observe that the seed-set diversity values are the same for the two methods in the uniform   setting of the numerical-attribute diversity (i.e., \algo{Deg-DU} and \textit{RIS-U}). 
   Conversely, in the weighted setting,  the RIS-based diversity curve is only slightly below the \algo{Deg-DW} curve. 
    Also, the insets show   very similar  expected spread (on average over 10\,000 Monte Carlo runs).   
  Overall, this indicates  flexibility of  our RIS-based framework, which  can also be properly adapted to   integrate numerical-based diversity functions.

\section{Conclusions}
\label{sec:conclusions}

We proposed a novel targeted influence maximization  problem  
 which accounts for the diversification of the seeds according to    side-information  available at node level in the general form of 
  categorical attribute values.  
 We also design a class of nondecreasing monotone and submodular functions to determine diversity of the categorical profiles associated to seed nodes. 
 Our developed RIS-based \myalgo algorithm was compared to  two IM methods, the one exploiting topology-driven diversity  and the other one accounting for numerical-based diversity in IM. 
 While showing different and more flexible behavior than the competitors,  \myalgo takes the advantages of ensuring the RIS-typical theoretical-guarantee  and computational complexity under a general, side-information-based setting of node diversity. 
 A further strength point of our diversity-sensitive framework lays on its versatility since \myalgo can easily be extended to incorporate other definitions of node diversity. 
 In this regard, we plan to define diversity notions based on representation learning techniques, including network embedding methods.    

%


\newpage 
\section*{Appendix}
\addcontentsline{toc}{section}{Appendices}
\renewcommand{\thesubsection}{\Alph{subsection}}

 \setcounter{lemma}{0} 
   \setcounter{proposition}{0}
\subsection{Proofs of theoretical results in Section 4}

\begin{proposition} 
The attribute-wise   diversity function defined in Eq.~(\ref{eq:diversity}) is monotone and submodular.
\end{proposition}
{\em Proof.\ } 
Function $div(S)$ in Eq.~(\ref{eq:diversity}) is monotone and submodular provided that  $div_A(S)$ in Eq.~(\ref{eq:divA}) is such as well, for any choice of $A \in \A$ and  setting of coefficients $\omega$, since $div(S)$ is a linear combination of functions $div_A(S)$ with nonnegative weights.  
 Monotonicity of Eq.~(\ref{eq:divA}) is trivially satisfied. 
 As concerns submodularity, let us assume $\lambda =1$ without loss of generality. 
Note that the inclusion of a node $u$ into $S$  corresponds to $1/k_1$, with $k_1$ equal to the size of  $S' \subseteq S$ such that, for each $v \in S'$, it holds that $val_A(v) \equiv val_A(u)$; moreover, the inclusion of   node $u$ into $T$ ($S \subseteq T$)  is $1/k_2$, with $k_2$ equal to the size of  $T' \subseteq T$ such that for each $v \in T'$, $val_A(v) \equiv val_A(u)$. Since $S \subseteq T$, it holds that $k_2 \geq k_1$, or   $1/k_1 \geq 1/k_2$, which concludes the proof.
\hfill~$\blacksquare$

\begin{lemma} \label{lemma:max_div_value}
Given a set $S$ and a categorical attribute $A$,  \ 
 consider $M_A = \max_{a \in dom_A(S)} n_a$ and 
    $m_A = \min_{a \in dom_A(S)} n_a$.   For any $$S = \argmax_{S' \subseteq \mathcal{V} \ s.t. \ |S'| \leq k} div_A(S'),$$ it holds that   $M_A - m_A <= 1.$
\end{lemma}
{\em Proof.\ } 
Assume by contradiction that there exists a set $S$ that maximizes $div_A$ (for any $A \in \A$) such  that  $M_A - m_A > 1$. Without loss of generality,    assume $M_A = m_A+2$  and $\lambda = 1$. 
Let $a^{(M)}$ and $a^{(m)}$ denote the categorical values corresponding 
to $M_A$ and $m_A$, respectively. It is easy to note that, if we remove 
a node with profile containing  $a^{(M)}$, resp. we add a node with profile containing $a^{(m)}$, then $div_A$ will  decrease by a quantity $\delta^- = 1/(M_A)$, resp.  increase
by a  quantity $\delta^+=1/(m_A+1)$. Since $\delta^- < \delta^+$, the diversity value is increased, therefore $S$ cannot be the optimal solution, which proves our statement.
\hfill~$\blacksquare$

\begin{proposition} 
\label{def:diversity_max_possible_value}
 Given the set of categorical attributes $\mathcal{A} = \{A_1, \ldots, A_m \}$,  $m$-real valued coefficients $\omega_j \in [0,1]$ ($j=[1..m]$), and a budget $k$, the theoretical maximum value     for Eq.~(\ref{eq:diversity}) is function of $k$ and determined as ($d_j \triangleq |dom_{A_j}|$):
\begin{equation}
\label{eq:diversity_max_possible_value}
div^*[k] = \sum_{j=1}^{m} \omega_j \left(d_j \sum_{i=1}^{k/d_j} \frac{1}{i^{\lambda}} +  \frac{k~\mathrm{mod}~d_j}{\big(1+ \frac{k}{d_j}\big)^{\lambda}} \right)
\end{equation}
\end{proposition}

{\em Proof sketch.\ } 
Equation~(\ref{eq:diversity_max_possible_value}) can be derived   based on the observation that the maximum possible value achievable w.r.t. a budget $k$ is obtained when the categorical values
are equally distributed among the $k$ nodes. 
Without loss of generality, let us  consider the case with 
one categorical attribute $A$. If we need to select $k$ nodes, one at a time, the best choice corresponds to select the node with  value $a^*= \argmin_{a \in dom_A(S)}$ $n_a$,
as it yields the maximum marginal gain. 
It straightforwardly follows that, by adopting this strategy,   a set $S$ can be   produced  to satisfy 
the requirement stated in Lemma~\ref{lemma:max_div_value} for the  maximization of Eq.~(\ref{eq:diversity}).
\hfill~$\blacksquare$

\begin{proposition} 
The Hamming-based diversity function defined in Eq.~(\ref{eq:hamming-div}) is monotone and submodular.
\end{proposition}
{\em Proof sketch.\ }
Monotonicity of Eq.~(\ref{eq:hamming-div}) is trivial. 
In fact, since the equation takes into account the union
of the Hamming balls associated with any node in the set, 
greater sets can only lead to greater Hamming balls, 
thus Eq.~(\ref{eq:hamming-div}) is only allowed to increase. 
 
As concerns the submodularity, it should be noted that for
any $S \subseteq T \subseteq V$, it holds that 
$B^{\xi}_S  \subseteq B_T^{\xi}$.  
In light of   Fact~\ref{fact:hamming-inc}, we can write the
inequality between the marginal gain of any node $v$ with respect to
$S$ and $T$ as:
\[
\begin{array}{lcl}
\cancel{div(S)} + \mid B_v^{\xi} \setminus B_S^{\xi} \mid - \cancel{div(S)} & \geq & \cancel{div(T)} + \mid B_v^{\xi} \setminus B_T^{\xi} \mid  - \cancel{div(T)}\\
\end{array}
\]
In order to prove the submodularity, we can proceed by contradiction.
 Suppose there exists a node $v$ such that the following inequality
is strictly satisfied: 

\[
\begin{array}{lcl}
	\cancel{\mid B_v^{\xi} \mid} - \mid B_v^{\xi}  \cap B_S^{\xi} \mid  & < & \cancel{\mid B_v^{\xi} \mid} -  \mid B_v^{\xi}  \cap B_T^{\xi} \mid \\
	\mid B_v^{\xi}  \cap B_S^{\xi} \mid  & > & \mid B_v^{\xi}  \cap B_T^{\xi} \mid
\end{array}
\]

It is easy to verify that the above inequality is a contradiction, in fact since $B_S^{\xi} \subseteq B_T^{\xi}$, there cannot exist
any node $u$ belonging to the intersection in the leftmost side of the equation that does not belong to the 
intersection in the rightmost side. 
\hfill~$\blacksquare$

\begin{proposition} 
The entropy-based diversity function defined in Eq.~(\ref{eq:entropy-div}) is monotone and submodular.
\end{proposition}
{\em Proof sketch.\ }
Monotonicity and submodularity are ensured given the strict relation between the  joint entropy function and a polymatroid~\cite{Fujishige78}. Moreover, as concerns submodularity in particular, note that in the inequality $H(\mathcal{X}_S,X) - H(\mathcal{X}_S) \geq H(\mathcal{X}_T,X) - H(\mathcal{X}_T)$ (with $\mathcal{X}_S \subset \mathcal{X}_T \subseteq \mathcal{X}$ and $X \in \mathcal{X}, X \notin \mathcal{X}_T$), each of the two terms is just the conditional entropy of variable $X$ given $\mathcal{X}_S$ and $\mathcal{X}_T$, respectively. Therefore,  $H(X | \mathcal{X}_S)  \geq H(X | \mathcal{X}_T)$ holds since conditioning cannot increase entropy. 
\hfill~$\blacksquare$

\begin{proposition}
Given a budget $k$  and $h$ classes, 
the function in Eq.~(\ref{eq:partition-div}), equipped with $\mathrm{f}(x)=\log(1+x)$, with $r_j = 1, \forall v_j \in \V$, achieves the minimum value of $\log(1+k)$ when all $k$ nodes belong to the same class (i.e.,  1 class  covered),  and the maximum value of $k$ when all $k$ nodes belong to different classes (i.e.,  $k$ classes  covered). 
\end{proposition}
%
{\em Proof sketch.\ }
The values of $\log(1+k)$ and $k$ are immediately derived by evaluating  Eq.~(\ref{eq:partition-div}) for the cases $h=1$ and $h=k$, respectively. The proof of $k$ as upper bound is immediate. To prove that $\log(1+k)$ is the lower bound of Eq.~(\ref{eq:partition-div}), consider without loss of generality a uniform class distribution, i.e., there are $k/h$ (with $h<k$) nodes that belong to each class. In this case, it holds that  $div(S)= h \log(1+k/h)$, for any size-$k$   $S$. It follows that   the inequality $\log(1+k) \leq h \log(1+k/h)$ must be verified (with equality iff $h=1$). This is immediately derived by observing that, after algebraic manipulation, the above inequality holds iff $(1+k)h^h \leq (h+k)^h$, which is true since the   terms on the left side are contained in the  polynomial $(h+k)^h$. 
\hfill~$\blacksquare$

\begin{proposition} 
The partition-based diversity function defined in Eq.~(\ref{eq:partition-div}) is monotone and submodular.
\end{proposition}
{\em Proof sketch.\ }
Monotonicity and submodularity of the function in Eq.~(\ref{eq:partition-div})   can directly be derived from the  mixture property  of submodular functions and the  composition property  of  submodular with nondecreasing concave functions~\cite{Lovasz83}, respectively. 
In fact,  the   summation  argument  of $\mathrm{f}$ is a collection of modular functions with nonnegative weights (and hence is monotone),  
 the  application of $\mathrm{f}$  yields a submodular function, and  finally summing up over the groups retains monotonicity and submodularity. 
 \hfill~$\blacksquare$

\subsection{Inappropriate set-diversity functions}  
We report details about a number of functions that, despite their  simplicity,  were demonstrated to be   unsuitable as diversity functions for our problem (cf. Sect.~\ref{sec:negativefunctions}). 
 %

Concerning   attribute-wise functions, we discussed that a simple approach would be to aggregate \textit{pairwise distances} of the node profiles w.r.t.  a given attribute   $A$. 
  We consider in particular the following definition based on pairwise attribute-value mismatchings: 
  $$
 f_1(S,A) = \frac{1}{|S|}  \sum_{u,v \in S}  \mathbbm{1}[val_{A}(u) \neq val_{A}(v)],$$   where $\mathbbm{1}[\cdot]$ denotes the indicator function.\footnote{For any nodes $u$ and $v$, we assume that if either $u$'s or $v$'s profile  is not associated with a value in the domain of $A$ (i.e., missing value for $A$),  then the indicator function will be evaluated as 1.}   
  It is easy to prove that this function is non-submodular; to give empirical evidence of this fact,  consider the following example. We are given $S = \{u,v,x\}$ with 
$val_{A}(u) = val_{A}(v) = a_1$ and $val_{A}(x)= a_2$, and $T = \{u,v,x,y\}$ with 
$val_{A}(y) = a_1$. Suppose that node $z$, with $val_{A}(z)= a_2$, is inserted into $S$ and $T$, then it holds that:  $f_1(S,A) = \frac{2}{3}$, $f_1(S \cup \{z\}, A) = \frac{4}{4}$, $f_1(T,A) = \frac{3}{4}$, and $f_1(T \cup \{z\},A) = \frac{6}{5}$. It follows that 
$f_1(S \cup \{z\}, A) -  f_1(S,A)  \not\geq f_1(T \cup \{z\},A) - f_1(T,A)$. 
Note also that the property of submodularity still does not hold if the normalization term (i.e., $|S|$) is discarded in $f_1(\cdot)$.

Let us now extend to computing  pairwise distances  of the node profiles in their entirety, focusing on the \textit{Hamming distance}, as defined in Eq.~(\ref{eq:hamming}).  
Upon this, let us define $f_2(S) =  \sum_{u, v \in S, u\neq v} dist^H(u,v)$, and two normalized versions:   $\widehat{f_2}(S) = (1/(2|S|))f_2(S)$ and $\widehat{\widehat{f_2}}(S) = (1/|S|(|S|-1))f_2(S)$.
 It is easy to check that none of such functions is appropriate. Let us consider the following example. We are given a schema with three attributes ($m=3$) and sets $S = \{u,v\}$, such that
$\A[u] = \langle a_1, \bot, \bot\rangle$, 
$\A[v] = \langle a_2, \bot, \bot\rangle$,  
 and 
 $T= \{u,v,x\}$, such that $\A[x] = \langle a_3, b_1, c_1\rangle$. Suppose that node $z$, with 
 $\A[z] = \langle a_4, \bot, \bot\rangle$, 
 is inserted into $S$ and $T$, then it holds that: 
 $f_2(S) = 2$, $f_2(T) = 14$,  $f_1(S \cup \{z\}) = 6$, and $f_1(T \cup \{z\}) = 24$. It follows that 
  $f_2(S \cup \{z\}) -  f_2(S)  \not\geq f_2(T \cup \{z\}) - f_2(T)$. 
Considering   $\widehat{f_2}(\cdot)$, we have:  $\widehat{f_2}(S) = \frac{1}{2}$, $\widehat{f_2}(T) = \frac{7}{3}$,  $\widehat{f_2}(S \cup \{z\}) = 1$, and $\widehat{f_2}(T \cup \{z\}) = 3$; thus, again 
$\widehat{f_2}(S \cup \{z\}) -  \widehat{f_2}(S)  \not\geq \widehat{f_2}(T \cup \{z\}) - \widehat{f_2}(T)$. 
Yet, when using $\widehat{\widehat{f_2}}(\cdot)$, we have: 
$\widehat{\widehat{f_2}}(S) = 1$, $\widehat{\widehat{f_2}}(T) = \frac{7}{3}$,  $\widehat{\widehat{f_2}}(S \cup \{z\}) = 1$, and $\widehat{\widehat{f_2}}(T \cup \{z\}) = 2$; in this case, mononicity is not even satisfied (since $\widehat{\widehat{f_2}}(T \cup \{z\}) \not\geq \widehat{\widehat{f_2}}(T)$).

Alternatively, we considered \textit{Jaccard distance}, i.e., 
 given the profiles of any two nodes $u,v$:   
$$
dist^J(u,v) = 1 - \frac{\sum_{j=1}^m \mathbbm{1}[val_{A_j}(u) = val_{A_j}(v)]}{|\A[u]| + |\A[v]|- \sum_{j=1}^m \mathbbm{1}[val_{A_j}(u) = val_{A_j}(v)]}.  
$$

Upon this, let us define $f_3(S) =  \sum_{u, v \in S, u\neq v} dist^J(u,v)$, and normalized version:   $\widehat{f_3}(S) = (1/(2|S|))f_2(S)$. 
Like previous functions, it can be empirically shown that $f_3(\cdot)$ and $\widehat{f_3}(\cdot)$ are not appropriate for our purposes.  
Suppose we are given a schema with five attributes ($m=5$) and sets $S = \{u,v\}$, such that
$\A[u] = \langle a, b, c, \bot, \bot\rangle$, 
$\A[v] = \langle a, b, \bot, d, \bot\rangle$,   
 and 
 $T= \{u,v,x\}$, such that $\A[x] = \A[v]$. Suppose that node $z$, with 
 $\A[z] = \langle a, \bot, \bot, d, e\rangle$, 
 is inserted into $S$ and $T$, then it holds that: 
 $f_3(S) = 1, f_3(T) = 2, f_3(S \cup \{z\}) = \frac{18}{5}, 
 f_3(T \cup \{z\}) = \frac{28}{5}$. 
 It follows that 
  $f_3(S \cup \{z\}) -  f_3(S)  \not\geq f_3(T \cup \{z\}) - f_3(T)$. 
Considering   $\widehat{f_3}(\cdot)$, we have:  $\widehat{f_3}(S) = \frac{1}{4}$, $\widehat{f_2}(T) = \frac{1}{3}$,  $\widehat{f_2}(S \cup \{z\}) = \frac{3}{5}$, and $\widehat{f_2}(T \cup \{z\}) = \frac{7}{10}$; thus, again 
$\widehat{f_3}(S \cup \{z\}) -  \widehat{f_3}(S)  \not\geq \widehat{f_3}(T \cup \{z\}) - \widehat{f_3}(T)$.

 The above Jaccard distance function could also   be exploited to allow for measuring the dissimilarity of all profiles in any set $S=\{v_1, \ldots, v_k\} \subseteq \V$: 
 $$
f_4(S) = 1 - \frac{\sum_{j=1}^m \mathbbm{1}[val_{A_j}(v_1) = \ldots = val_{A_j}(v_k)]}{\sum_{j=1}^m |\bigcup_{v \in S} \{val_{A_j}(v)\}|}. 
$$

However, it is straightforward to show that the above function can easily yield useless results; e.g., referring to the previous example, the marginal gains of $z$ w.r.t. $S$ and $T$ are the same. Even worse, a normalization of $f_4(S)$ by set-size does not even ensure monotonicity.

\subsection{Complexity aspects of \myalgo}

\begin{proposition}
\myalgo   runs in $O((k+l)(|\E|+|\V|)$ $\log |\V| /\epsilon ^2)$ time and returns  a $(1-1/e-\epsilon)$-approximate solution 
with at least $1-|\V|^{-l}$ probability.
\end{proposition}

{\em Proof sketch.\ } 
\myalgo is developed under the RIS framework and follows the typical two-phase schema of \algo{TIM}/ \algo{TIM+} methods, i.e., parameter estimation and (seed) node selection, for which the theoretical results in the Proposition hold. 
 Due to the targeted nature of the problem under consideration,  the expected capital must be computed in place of the expected spread; however, this only implies to choose a distribution over the roots of the RR-Sets, which depends on the target scores of the nodes in the network. Thus, the  asymptotic complexity of \algo{TIM}/\algo{TIM+} is not increased. 
 Moreover, two major differences occur in the
seeds selection phase of 
 \myalgo w.r.t.  \algo{TIM}/\algo{TIM+}, i.e.,
the lazy forward approach and the computation of the
marginal gain w.r.t. the diversity function.
However, both aspects do not affect the asymptotic complexity, since  the former allows saving runtime only and the latter does not represent any overhead (computing a node's marginal gain is made in nearly constant time, for each   of the diversity functions). 
Therefore, we can conclude that \myalgo has the same asymptotic complexity of \algo{TIM}/\algo{TIM+}.
\hfill~$\blacksquare$

\vspace{3mm}
\subsection{Sensitivity to $\alpha$}
\begin{figure}[t]
\begin{tabular}{@{\hskip -2mm}ccc}
\includegraphics[width=.33\linewidth, height=3.5cm]{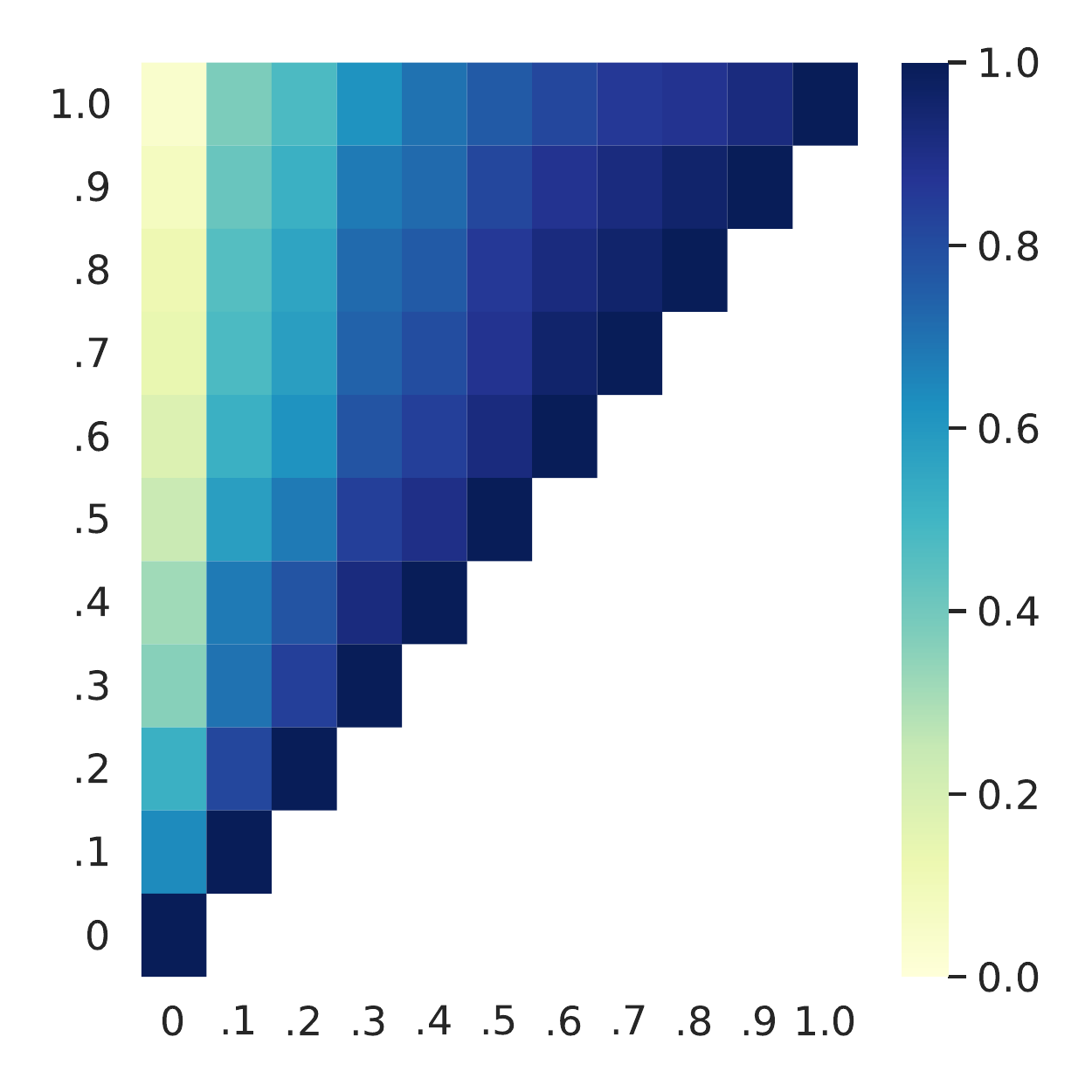} &
\includegraphics[width=.33\linewidth, height=3.5cm]{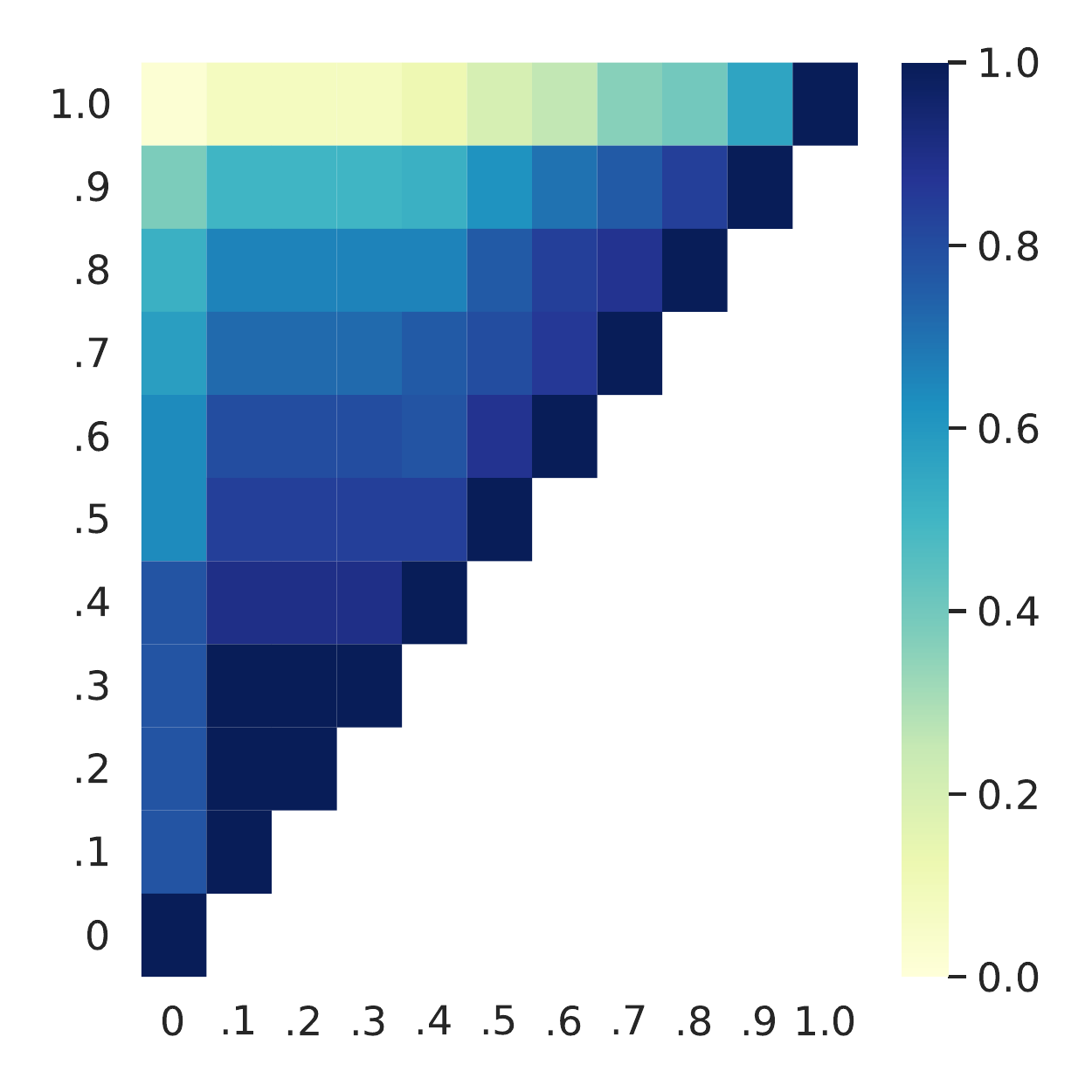} &
\includegraphics[width=.33\linewidth, height=3.5cm]{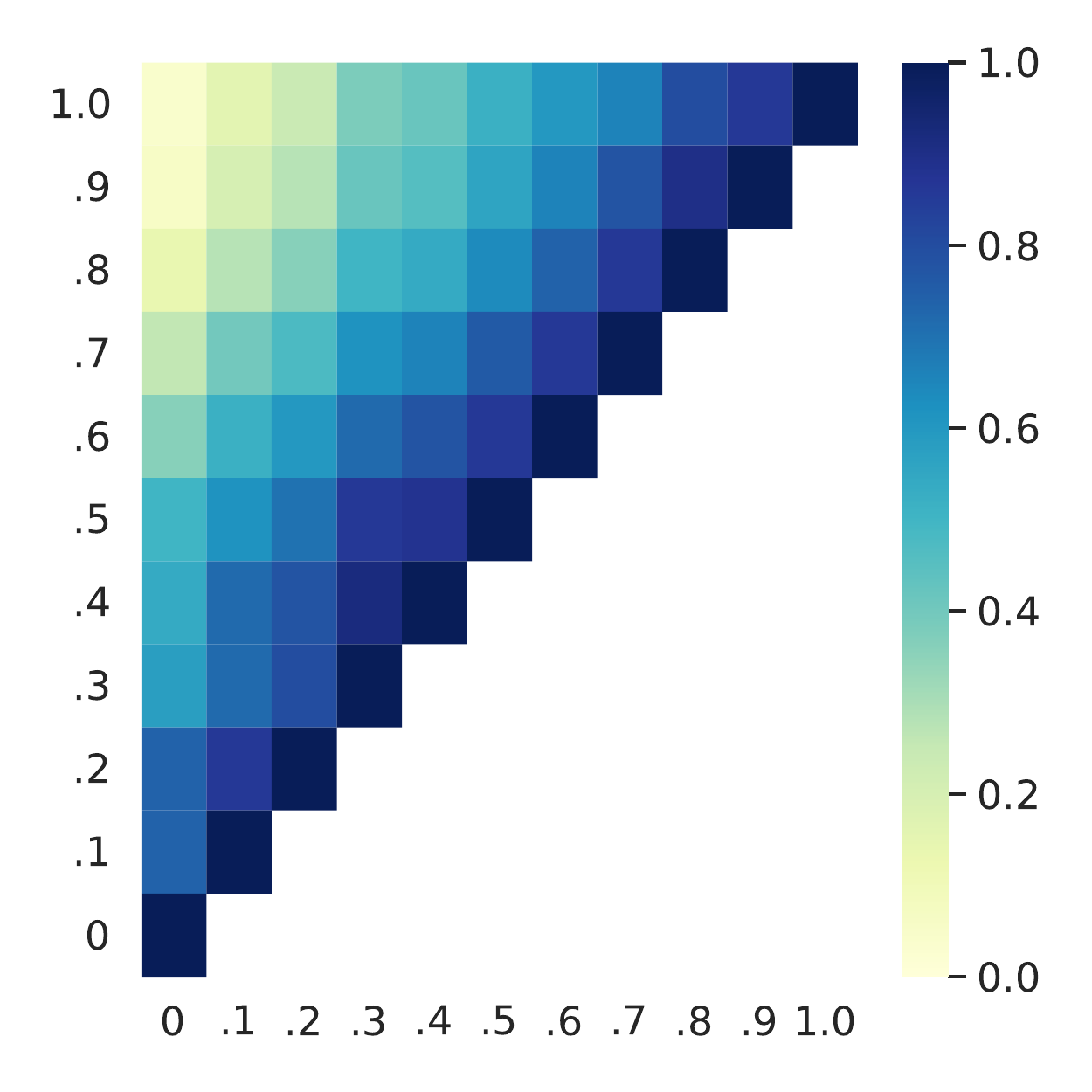} \\
(a) Instagram  & (b) FriendFeed  & (c) GooglePlus  \\
\includegraphics[width=.33\linewidth, height=3.5cm]{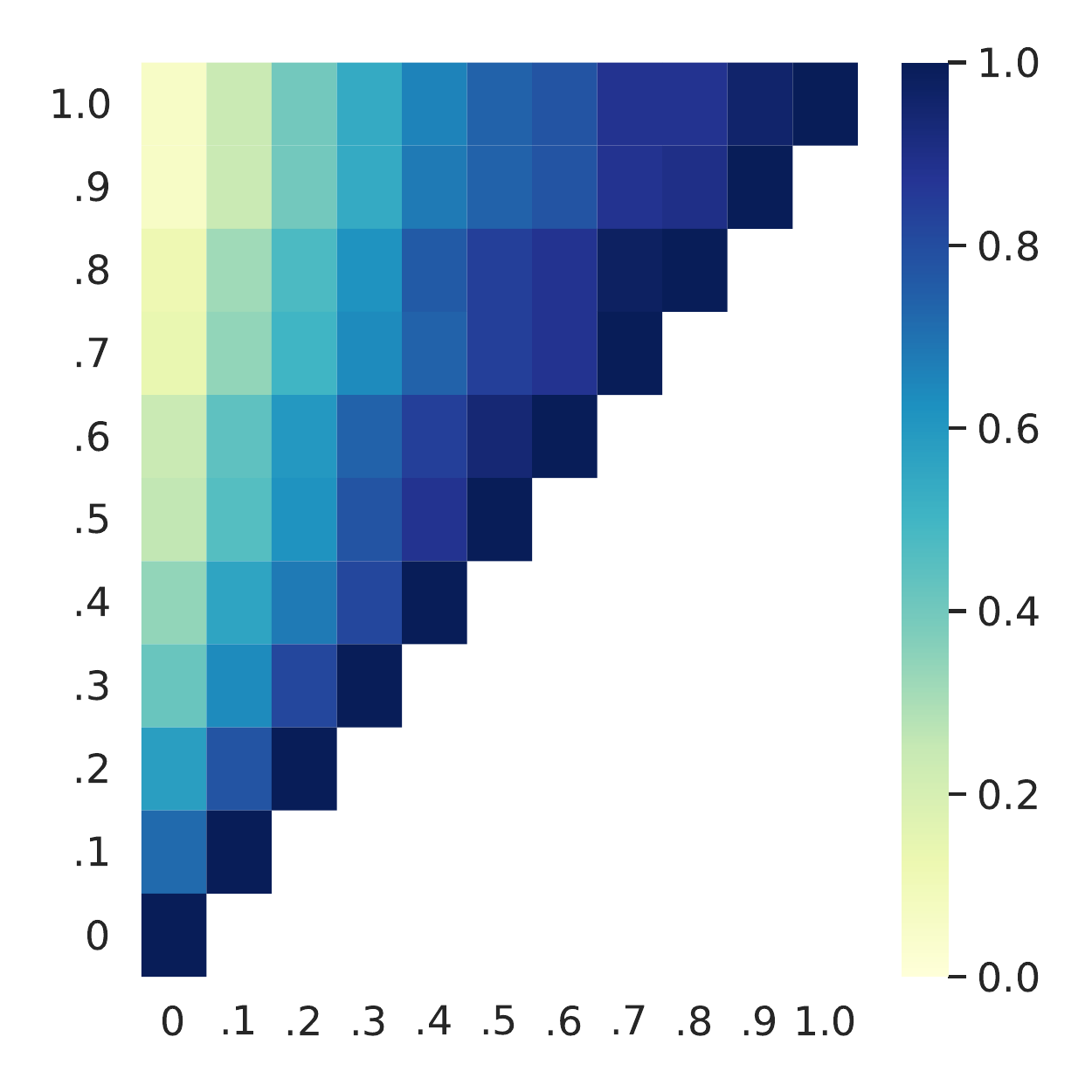} &
\includegraphics[width=.33\linewidth, height=3.5cm]{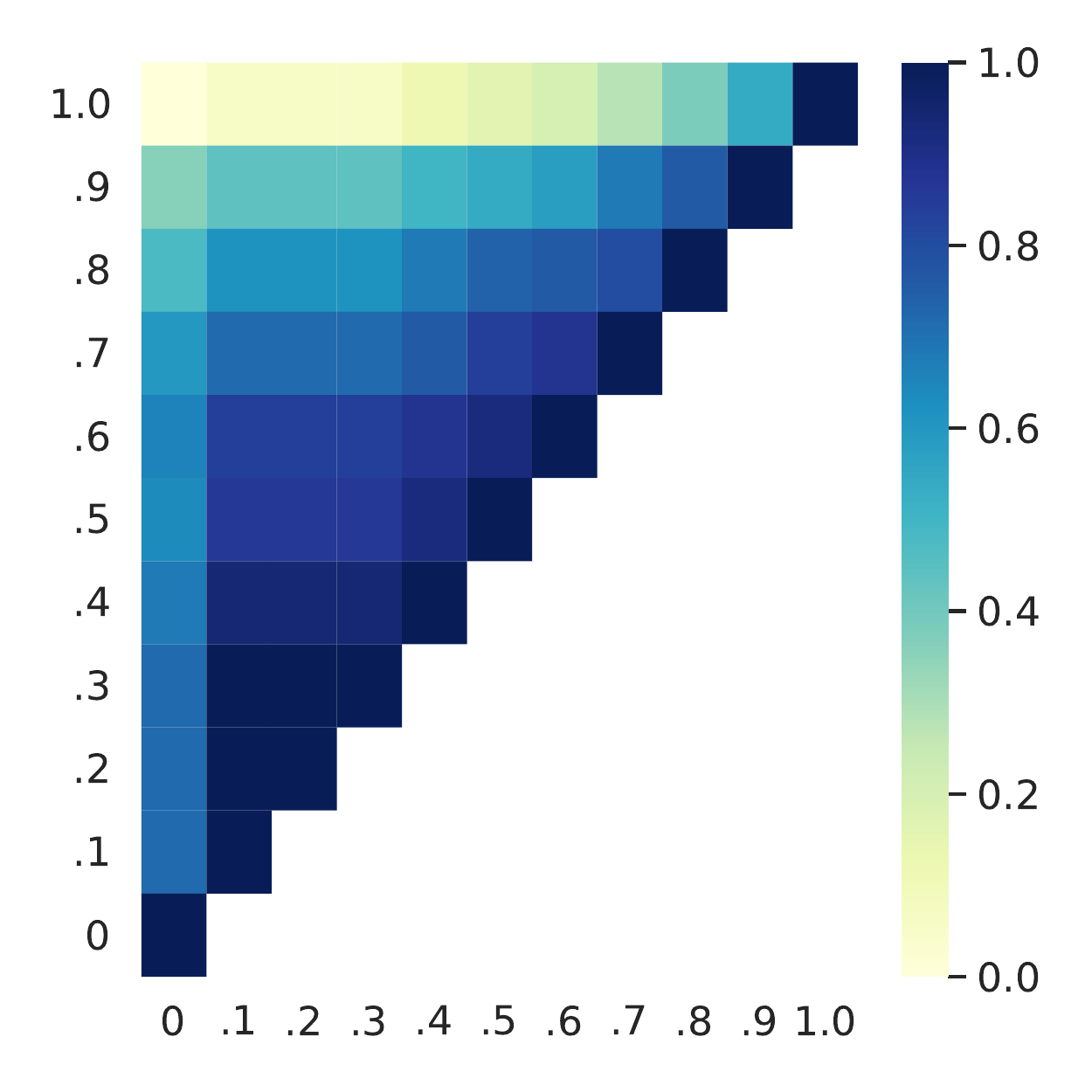} &
\includegraphics[width=.33\linewidth, height=3.5cm]{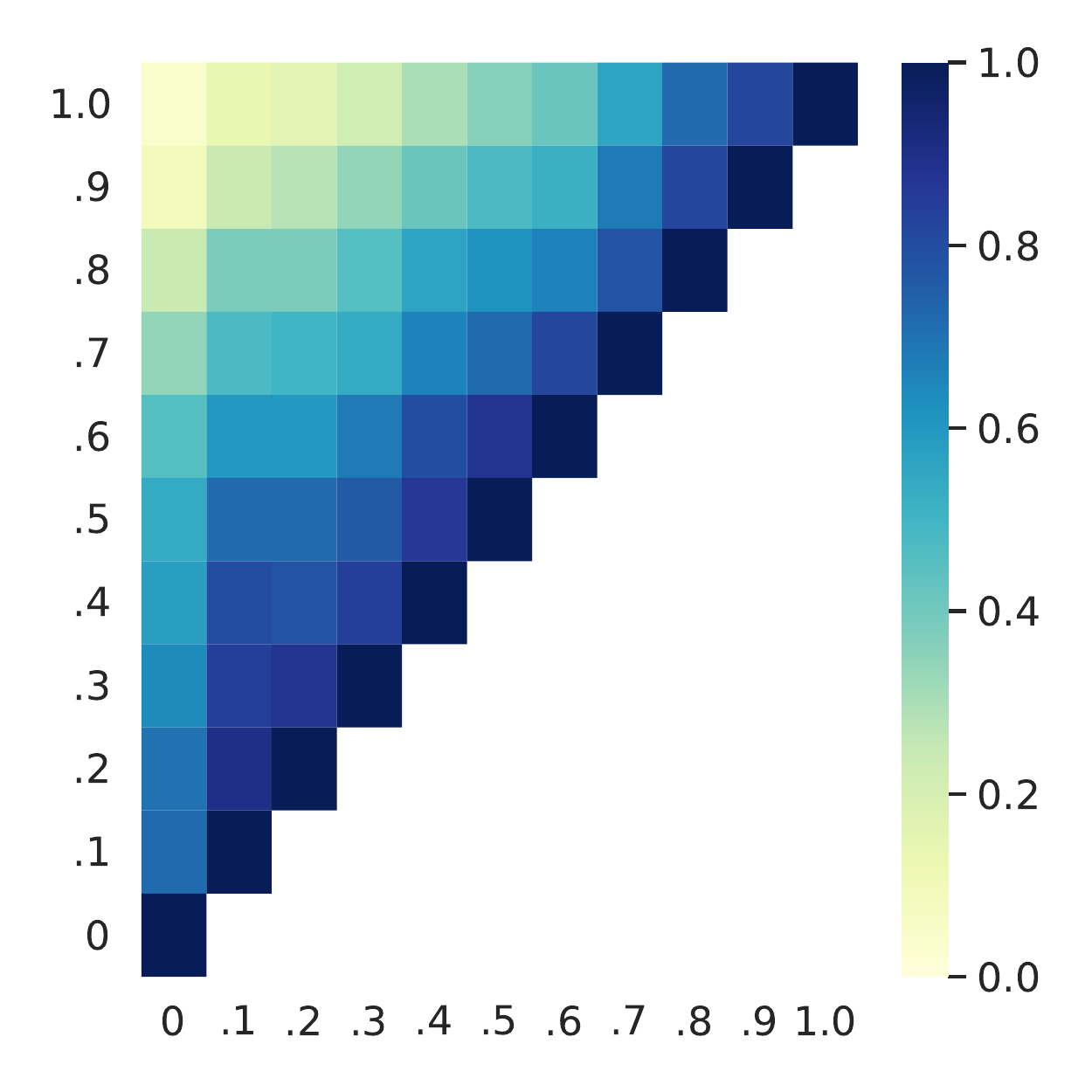} \\
(d) Instagram  & (e) FriendFeed  & (f) GooglePlus  \\
\end{tabular}
\caption{Normalized overlap of seed sets, for   $\alpha \in [0,1]$ (with increments of  $0.1$),  $k=50$, top-5\% (top) and top-10\% (bottom) target selection, and exponential distribution of attributes.}
\label{fig:seeds_overlap_exp_distribution-app}
\end{figure}

Figure~\ref{fig:seeds_overlap_exp_distribution-app} shows further results on normalized overlap of seed sets, for top-5\% and top-10\% target selection threshold.

  \vspace{3mm}
\subsection{Effect of the attribute distribution}
\begin{figure}[h!]
\centering 
\begin{tabular}{@{\hskip -2mm}ccc}
\includegraphics[width=0.315\linewidth]{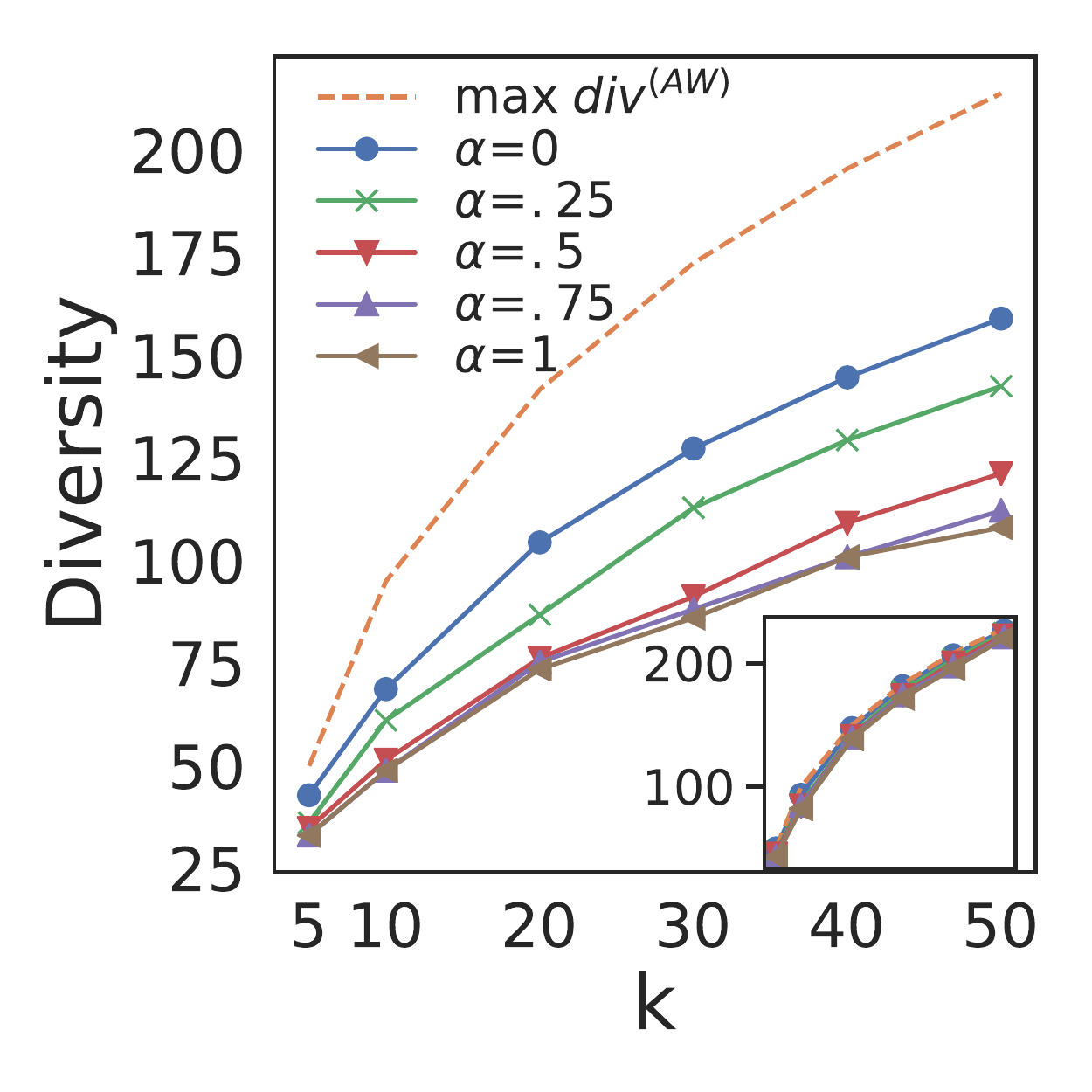} &
\includegraphics[width=0.315\linewidth]{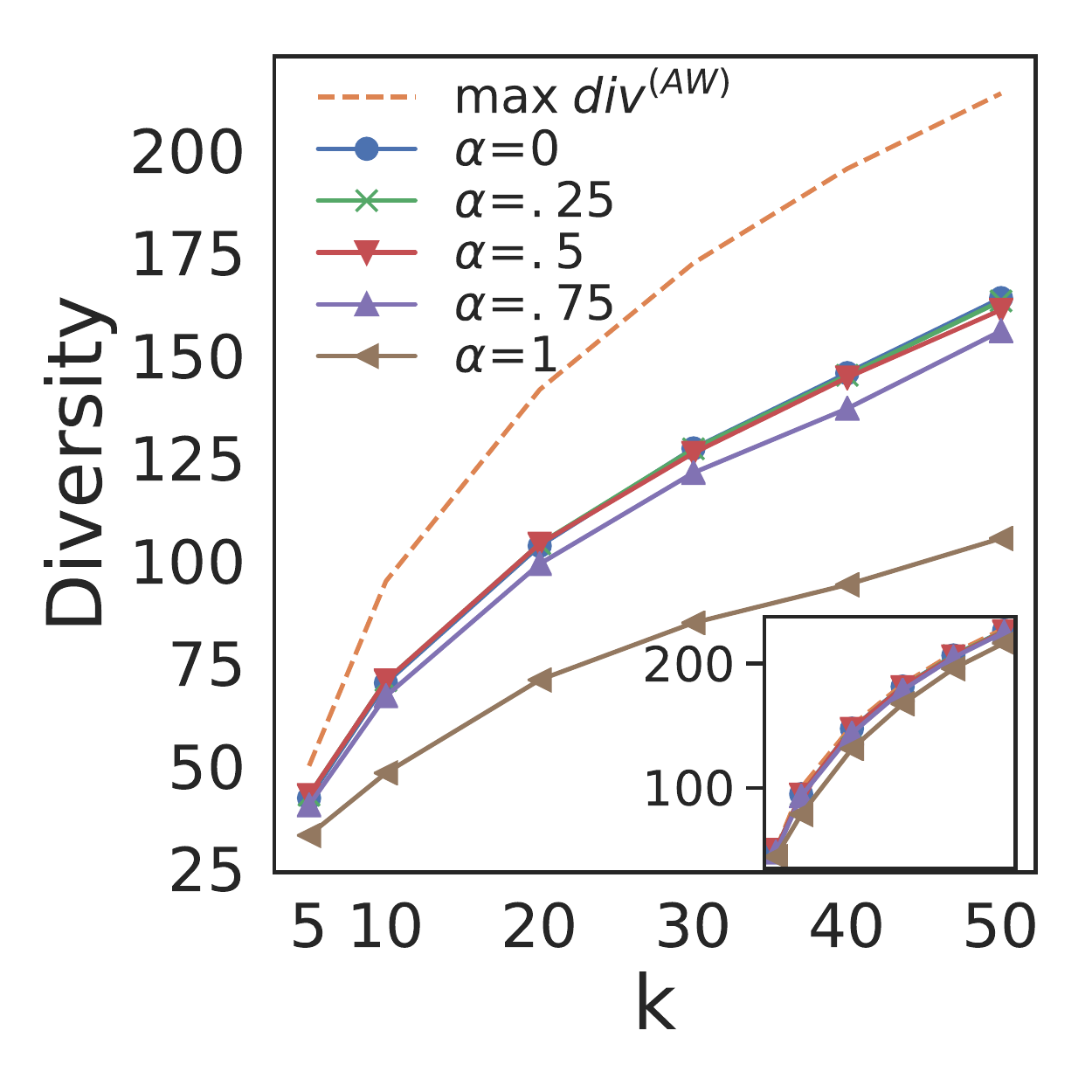} &
\includegraphics[width=0.31\linewidth]{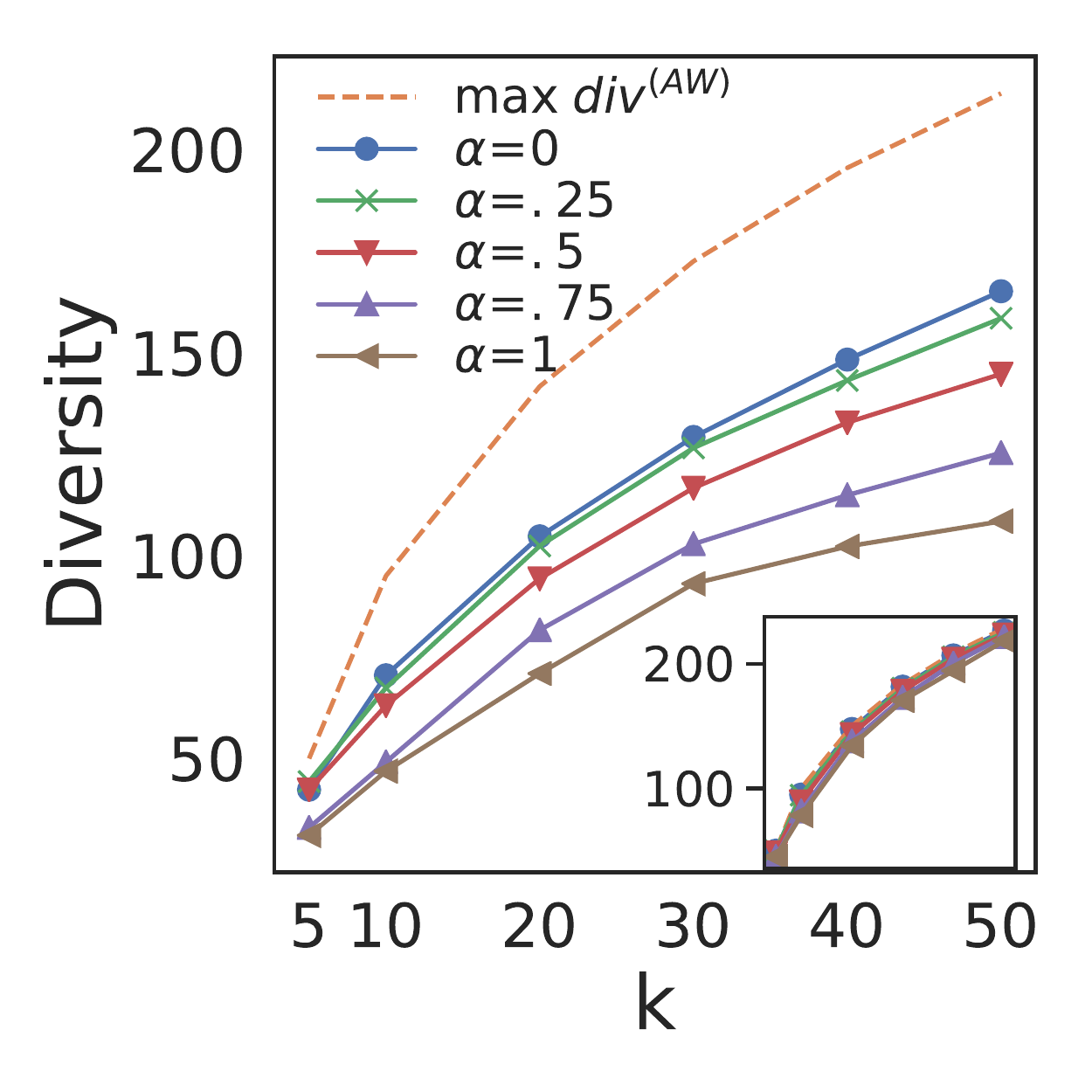} \\
(a) Instagram  & (b) FriendFeed  & (c) GooglePlus \\

\includegraphics[width=0.315\linewidth]{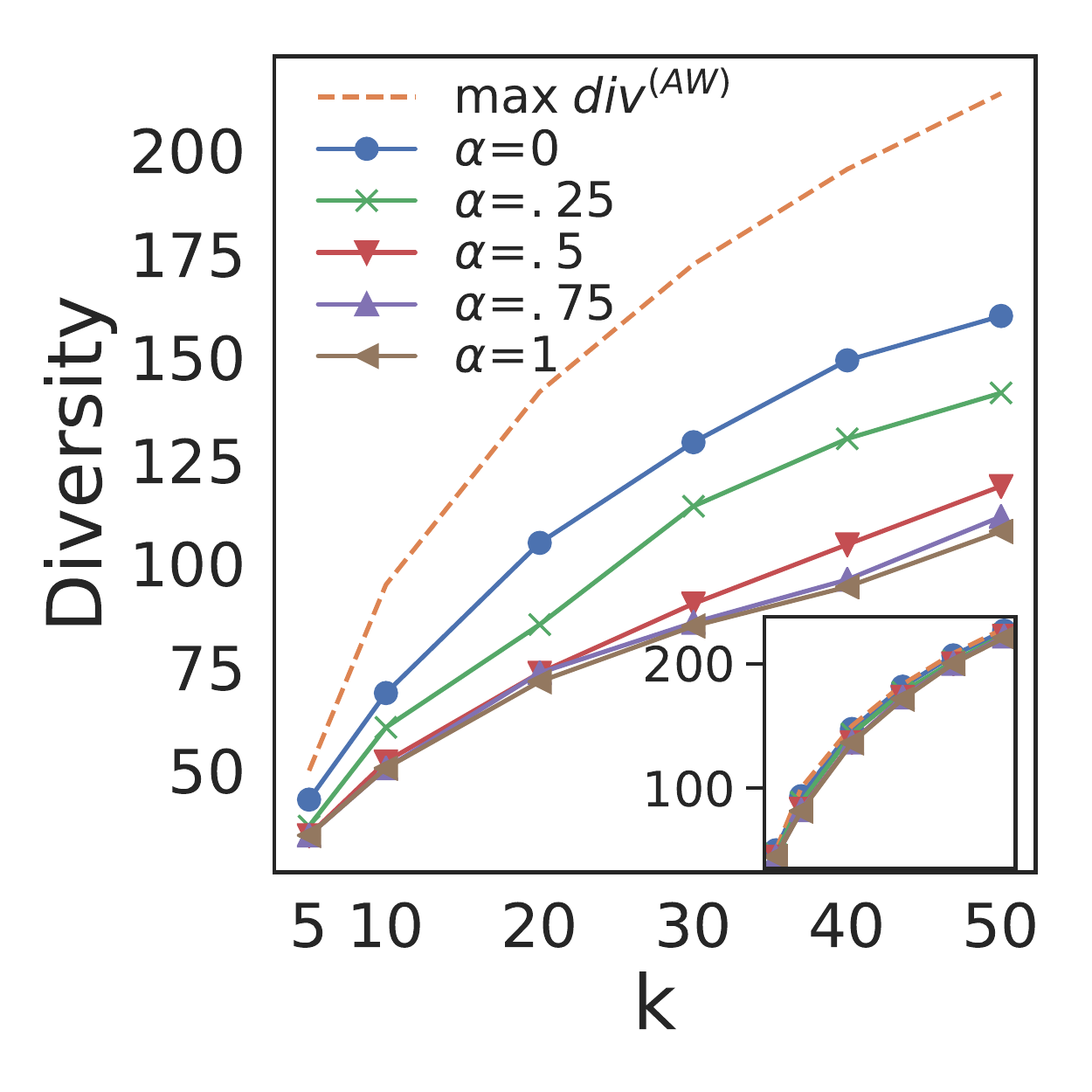} &
\includegraphics[width=0.315\linewidth]{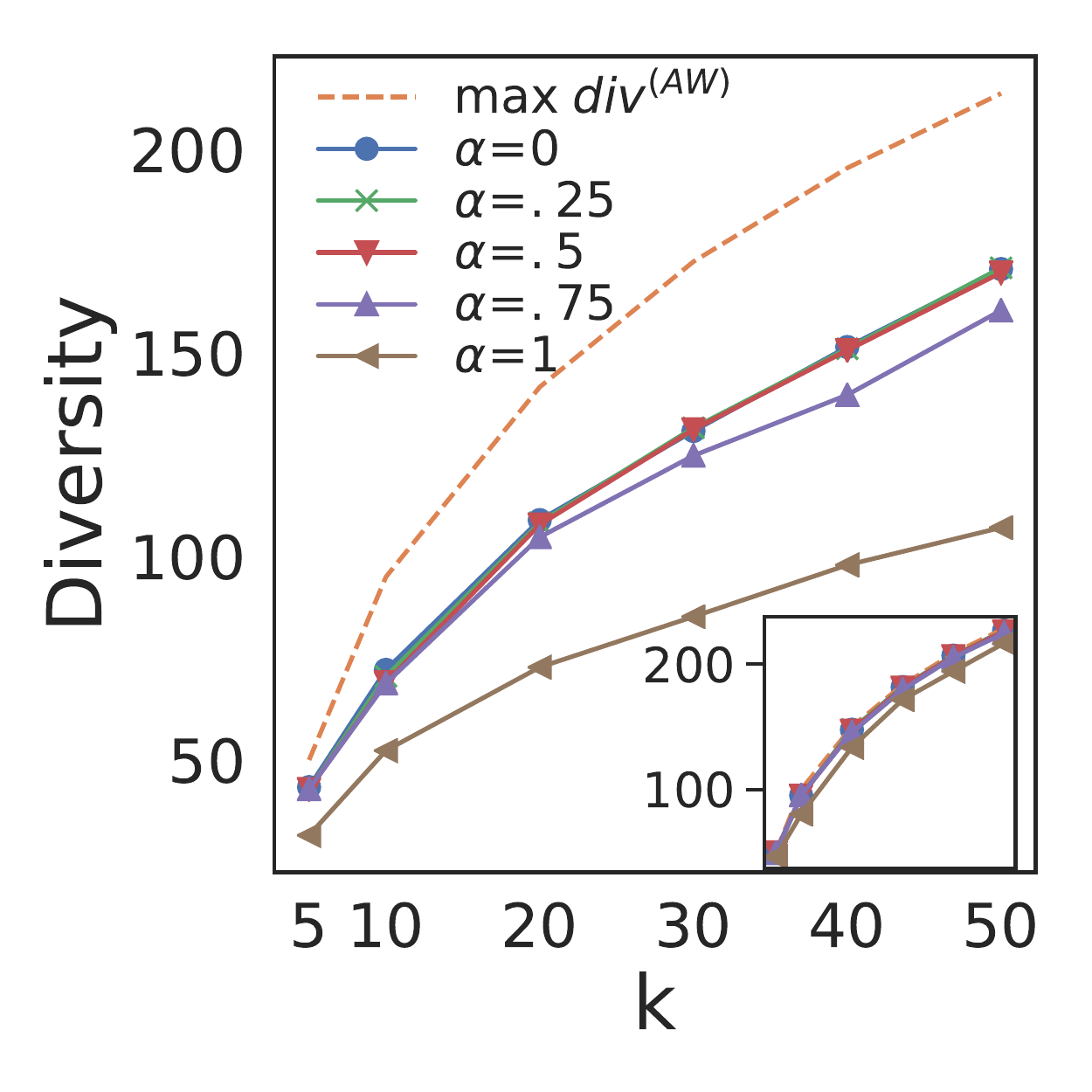} &
\includegraphics[width=0.315\linewidth]{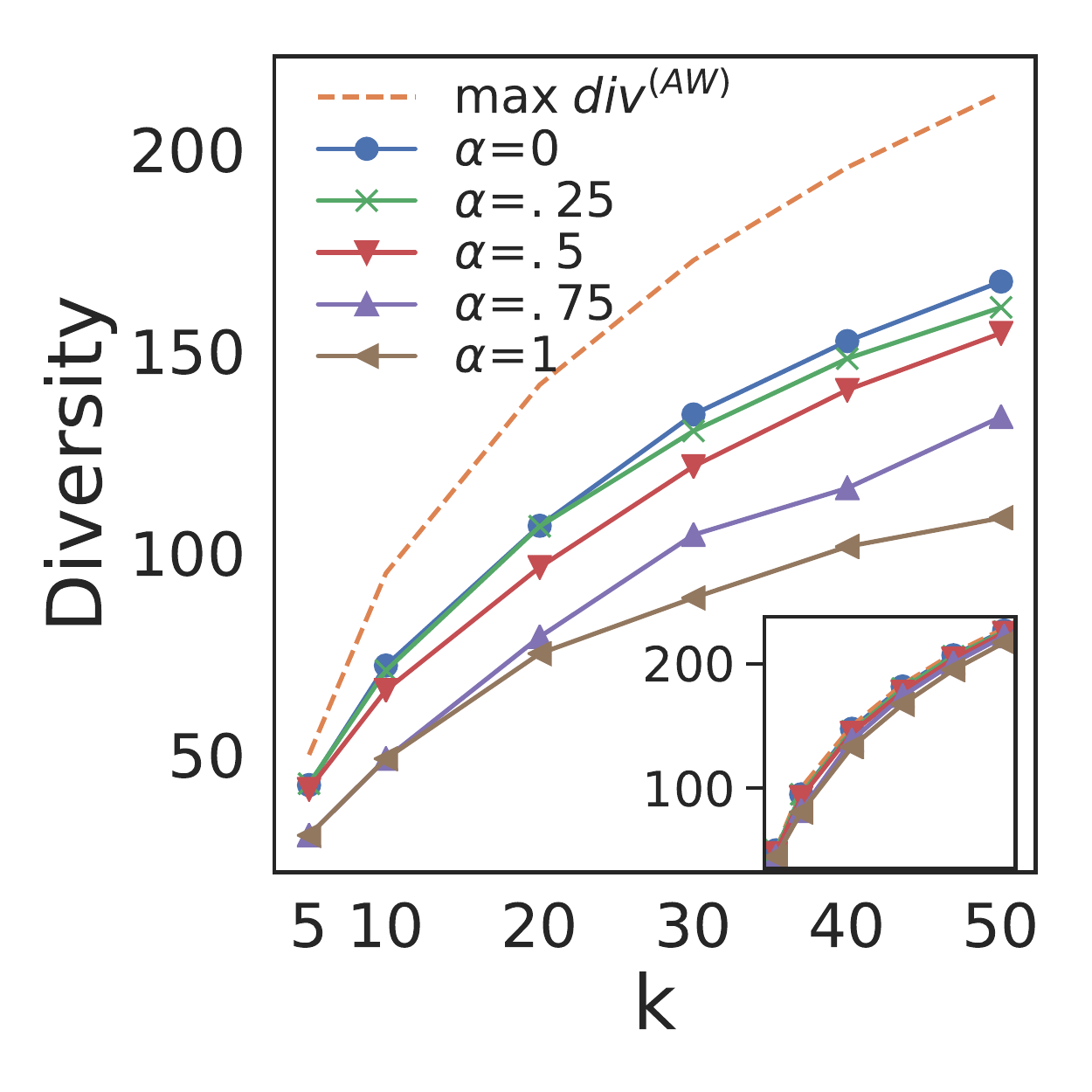} \\
(d) Instagram  &   (e) FriendFeed  & (f) GooglePlus  \\
\end{tabular}
\caption{Exponential (main) vs. uniform (inset) distribution:  seed-set diversity for varying $k$ and $\alpha$, top-5\% (a--c) and top-10\% (d--f)  target selection, and  comparison to  maximum diversity value.}
\label{fig:diversity_line_plot-app}
\end{figure}

Figure~\ref{fig:diversity_line_plot-app} shows further results on comparison between exponential and uniform distributions, for top-5\% and top-10\% target selection threshold.

\begin{thebibliography}{10}
\providecommand{\url}[1]{{#1}}
\providecommand{\urlprefix}{URL }
\expandafter\ifx\csname urlstyle\endcsname\relax
  \providecommand{\doi}[1]{DOI~\discretionary{}{}{}#1}\else
  \providecommand{\doi}{DOI~\discretionary{}{}{}\begingroup
  \urlstyle{rm}\Url}\fi

\bibitem{Suman+19}
Banerjee, S., Jenamani, M., Pratihar, D.K.: {ComBIM: A community-based solution
  approach for the Budgeted Influence Maximization Problem}.
\newblock Expert Systems with Applications \textbf{125}, 1--13 (2019)

\bibitem{BaoCZ13}
Bao, Q., Cheung, W.K., Zhang, Y.: Incorporating structural diversity of
  neighbors in a diffusion model for social networks.
\newblock In: Proc. {IEEE/WIC/ACM} Int. Conf. on Web Intelligence, pp. 431--438
  (2013)

\bibitem{doi:10.1137/1.9781611973402.70}
Borgs, C., Brautbar, M., Chayes, J., Lucier, B.: Maximizing social influence in
  nearly optimal time.
\newblock In: Proc. ACM-SIAM Symp. on Discrete Algorithms (SODA), pp. 946--957
  (2014)

\bibitem{8326536}
Cali\`{o}, A., Interdonato, R., Pulice, C., Tagarelli, A.: {Topology-driven
  Diversity for Targeted Influence Maximization with Application to User
  Engagement in Social Networks}.
\newblock IEEE Trans. Knowl. Data Eng. \textbf{30}(12), 2421--2434 (2018)

\bibitem{Chenbook}
Chen, W., Lakshmanan, L.V.S., Castillo, C.: Information and Influence
  Propagation in Social Networks.
\newblock Morgan \& Claypool (2013)

\bibitem{CoverThomas2006}
Cover, T.M., Thomas, J.A.: Elements of Information Theory.
\newblock John Wiley \& Sons, Inc. (2006)

\bibitem{Deng+16}
Deng, X., Pan, Y., Shen, H., Gui, J.: Credit distribution for influence
  maximization in online social networks with node features.
\newblock J Intell Fuzzy Syst \textbf{31}(2), 979--990 (2016)

\bibitem{Fujishige78}
Fujishige, S.: Polymatroid dependence structure of a set of random variables.
\newblock Inform. Contr. \textbf{39}, 55--72 (1978)

\bibitem{Simpath}
Goyal, A., Lu, W., Lakshmanan, L.V.S.: Simpath: An efficient algorithm for
  influence maximization under the linear threshold model.
\newblock In: 2011 IEEE 11th International Conference on Data Mining, pp.
  211--220 (2011)

\bibitem{GulerVTNZYO14}
Guler, B., Varan, B., Tutuncuoglu, K., Nafea, M.S., Zewail, A.A., Yener, A.,
  Octeau, D.: Optimal strategies for targeted influence in signed networks.
\newblock In: Proc. Int. Conf. on Advances in Social Networks Analysis and
  Mining (ASONAM), pp. 906--911 (2014)

\bibitem{GuoZZCG13}
Guo, J., Zhang, P., Zhou, C., Cao, Y., Guo, L.: Personalized influence
  maximization on social networks.
\newblock In: Proc. ACM Conf. on Information and Knowledge Management (CIKM),
  pp. 199--208 (2013)

\bibitem{HuangLCC13}
Huang, P., Liu, H., Chen, C., Cheng, P.: The impact of social diversity and
  dynamic influence propagation for identifying influencers in social networks.
\newblock In: Proc. {IEEE/WIC/ACM} Int. Conf. on Web Intelligence, pp. 410--416
  (2013)

\bibitem{Devotion}
Interdonato, R., Pulice, C., Tagarelli, A.: "got to have faith!": The devotion
  algorithm for delurking in social networks.
\newblock In: 2015 IEEE/ACM International Conference on Advances in Social
  Networks Analysis and Mining (ASONAM), pp. 314--319 (2015)

\bibitem{Kempe:2003}
Kempe, D., Kleinberg, J.M., Tardos, {\'{E}}.: Maximizing the spread of
  influence through a social network.
\newblock In: Proc. ACM SIGKDD Int. Conf. on Knowledge Discovery and Data
  Mining (KDD), pp. 137--146 (2003)

\bibitem{kumar2018community}
Kumar, S., Hamilton, W.L., Leskovec, J., Jurafsky, D.: Community interaction
  and conflict on the web.
\newblock In: Proceedings of the 2018 World Wide Web Conference on World Wide
  Web, pp. 933--943 (2018)

\bibitem{LagnierDGG13}
Lagnier, C., Denoyer, L., Gaussier, E., Gallinari, P.: Predicting information
  diffusion in social networks using content and user's profiles.
\newblock In: Proc. European Conf. on Information Retrieval (ECIR), pp. 74--85
  (2013)

\bibitem{LeeChung15}
Lee, J., Chung, C.: A query approach for influence maximization on specific
  users in social networks.
\newblock IEEE Trans Knowl Data Eng \textbf{27}(2), 340--353 (2015)

\bibitem{LiBSC15}
Li, H., Bhowmick, S., Sun, A., Cui, J.: Conformity-aware influence maximization
  in online social networks.
\newblock The VLDB Journal \textbf{24}, 117--141 (2015)

\bibitem{M.thai2}
Li, X., Smith, J.D., Dinh, T.N., Thai, M.T.: {Why approximate when you can get
  the exact? Optimal targeted viral marketing at scale}.
\newblock In: Proc. {IEEE} Conf. on Computer Communications (INFOCOM), pp. 1--9
  (2017)

\bibitem{lifan}
{Li}, Y., {Fan}, J., {Wang}, Y., {Tan}, K.: Influence maximization on social
  graphs: A survey.
\newblock IEEE Transactions on Knowledge and Data Engineering \textbf{30}(10),
  1852--1872 (2018)

\bibitem{kbtim}
Li, Y., Zhang, D., Tan, K.L.: Real-time targeted influence maximization for
  online advertisements.
\newblock Proc. VLDB Endow. \textbf{8}(10), 1070--1081 (2015)

\bibitem{Lovasz83}
Lov\'{a}sz, L.: Submodular functions and convexity.
\newblock In: A.~Bachem, B.~Korte, M.~Grötschel (eds.) Mathematical
  Programming: The State of the Art, pp. 235--257. Springer-Verlag Berlin
  Heidelberg (1983)

\bibitem{Lu:2015:CCC:2850578.2850581}
Lu, W., Chen, W., Lakshmanan, L.V.S.: From competition to complementarity:
  Comparative influence diffusion and maximization.
\newblock Proc. VLDB Endow. \textbf{9}(2), 60--71 (2015)

\bibitem{M.thai1}
Nguyen, H.T., Dinh, T.N., Thai, M.T.: {Cost-aware Targeted Viral Marketing in
  billion-scale networks}.
\newblock In: Proc. {IEEE} Conf. on Computer Communications (INFOCOM), pp. 1--9
  (2016)

\bibitem{Peng+18}
Peng, S., Zhou, Y., Cao, L., Yu, S., Niu, J., Jia, W.: Influence analysis in
  social networks: A survey.
\newblock Journal of Network and Computer Applications \textbf{106}, 17--32
  (2018)

\bibitem{hammingball}
Prasad, A., Jegelka, S., Batra, D.: Submodular meets structured: Finding
  diverse subsets in exponentially-large structured item sets.
\newblock arXiv CoRR \textbf{abs/1411.1752} (2014)

\bibitem{PHG}
{Qiu}, L., {Jia}, W., {Yu}, J., {Fan}, X., {Gao}, W.: {PHG: A Three-Phase
  Algorithm for Influence Maximization Based on Community Structure}.
\newblock IEEE Access \textbf{7}, 62511--62522 (2019)

\bibitem{Santos:2015:SRD:2802186.2802187}
Santos, R.L.T., Macdonald, C., Ounis, I.: Search result diversification.
\newblock Found. Trends Inf. Retr. \textbf{9}(1), 1--90 (2015)

\bibitem{annappa}
Sumith, N., Annappa, B., Bhattacharya, S.: Influence maximization in large
  social networks: Heuristics, models and parameters.
\newblock Future Generation Computer Systems \textbf{89}, 777--790 (2018)

\bibitem{6921625}
Tang, F., Liu, Q., Zhu, H., Chen, E., Zhu, F.: Diversified social influence
  maximization.
\newblock In: Proc. IEEE/ACM Int. Conf. on Advances in Social Networks Analysis
  and Mining (ASONAM), pp. 455--459 (2014)

\bibitem{IMM}
Tang, Y., Shi, Y., Xiao, X.: {Influence Maximization in Near-Linear Time: A
  Martingale Approach}.
\newblock In: Proc. ACM SIGMOD Int. Conf. on Management of Data (SIGMOD), pp.
  1539--1554 (2015)

\bibitem{Tang:2014:IMN:2588555.2593670}
Tang, Y., Xiao, X., Shi, Y.: {Influence Maximization: Near-optimal Time
  Complexity Meets Practical Efficiency}.
\newblock In: Proc. ACM SIGMOD Int. Conf. on Management of Data (SIGMOD), pp.
  75--86 (2014)

\bibitem{Wu:2016:RMC:2885506.2700496}
Wu, L., Liu, Q., Chen, E., Yuan, N.J., Guo, G., Xie, X.: Relevance meets
  coverage: A unified framework to generate diversified recommendations.
\newblock ACM Trans. Intell. Syst. Technol. \textbf{7}(3), 39:1--39:30 (2016)

\bibitem{YangHLC13}
Yang, D., Hung, H., Lee, W., Chen, W.: Maximizing acceptance probability for
  active friending in online social networks.
\newblock In: Proc. ACM SIGKDD Int. Conf. on Knowledge Discovery and Data
  Mining (KDD), pp. 713--721 (2013)

\bibitem{8031449}
{Zhang}, K., {Zhang}, Z., {Wu}, Y., {Xu}, J., {Niu}, Y.: A core theory based
  algorithm for influence maximization in social networks.
\newblock In: 2017 IEEE International Conference on Computer and Information
  Technology (CIT), pp. 31--36 (2017)

\bibitem{ZhouZC14}
Zhou, J., Zhang, Y., Cheng, J.: Preference-based mining of top-$k$ influential
  nodes in social networks.
\newblock Future Generation Comp. Syst. \textbf{31}, 40--47 (2014)

\end{thebibliography}
\end{document}